\documentclass{amsart}
\usepackage{graphicx,amssymb}
\newtheorem{thm}{Theorem}[section]

\newtheorem{lem}[thm]{Lemma}

\newtheorem{con}[thm]{Conjecture}
\theoremstyle{definition}

\theoremstyle{remark}

\numberwithin{equation}{section}

\newcommand{\av}[1]{\left\langle{#1}\right\rangle}
\newcommand{\mbold}[1]{\mbox{\boldmath${#1}$}}
\def\beq{\begin{eqnarray}}
\def\eeq{\end{eqnarray}}

\begin{document}

\title[Future Asymptotic Behaviour of Tilted Bianchi models of type IV and VII$_h$]{Future Asymptotic Behaviour of Tilted Bianchi models \\ of type IV and VII$_h$}
\author[Sigbj{\o}rn Hervik, Robert van den Hoogen and Alan Coley]{Sigbj{\o}rn Hervik$^{1}$, Robert van den Hoogen$^{2, 1}$ and Alan Coley$^{1}$}%
\address{$^{1}$ Department of Mathematics \& Statistics, Dalhousie University,
Halifax, Nova Scotia,
Canada B3H 3J5}%
\address{$^2$ Department of Mathematics, Statistics and Computer Science,
St. Francis Xavier University, Antigonish, Nova Scotia, Canada B2G 2W5}
\email{ herviks@mathstat.dal.ca,rvandenh@stfx.ca,aac@mathstat.dal.ca}%


\date{\today}%
\begin{abstract}
Using dynamical systems theory and 
a detailed numerical analysis, the late-time behaviour of tilting 
perfect fluid Bianchi models of types IV and VII$_h$ are investigated.
In particular, vacuum plane-wave spacetimes are studied and the
important result that the only future
attracting equilibrium points for non-inflationary fluids
are the plane-wave solutions in Bianchi type
VII$_h$ models is discussed. A tiny region of parameter space (the loophole) 
in the  Bianchi type IV model is shown to contain a closed orbit 
which is found to act as an attractor (the Mussel attractor). 
From an extensive numerical analysis it is found 
that at late times the normalised energy-density tends to zero and the 
normalised variables 
'freeze' into their asymptotic values.
A detailed numerical analysis of the type VII$_h$ models
then shows that there
is an open set of parameter space in which solution curves
approach a compact surface that is topologically a torus.

\end{abstract}

\maketitle

\section{Introduction}

In recent years there has been much analysis of  spatially homogeneous  (SH)
perfect fluid cosmological models with equation of state $p = (\gamma -1) \mu$,
where $\gamma$ is constant \cite{DS1}. For the class of tilted (non-orthogonal) SH
models \cite{KingEllis}, the Einstein field equations have been
written as an autonomous system of differential equations in a number of different ways
\cite{rosjan,BN,BS}. Tilted Bianchi models contain additional
degrees of freedom,  and new dynamical features emerge with the
inclusion of tilt. A subclass of models of Bianchi type V
\cite{Shikin,Collins,CollinsEllis,HWV} and models of Bianchi type
II \cite{HBWII} have been studied.

In \cite{BHtilted} a local stability analysis  of tilting Bianchi
class B models and the type VI$_{0}$ model of class A were
studied. The asymptotic late-time behaviour of general tilted
Bianchi type VI$_0$ universes was analysed in detail in
\cite{hervik}, and this was generalized to Bianchi type VI$_0$
cosmological models containing two tilted $\gamma$-law perfect
fluids in \cite{coleyhervik}. Finally, the tilting Bianchi type
VI$_h$ and  VII$_h$ models were recently studied in \cite{CH2} (self-similar
tilting solutions \cite{carrcoley} of Bianchi type VI$_h$ were also studied in
\cite{Apostolopoulos:04}).

In particular, the late-time behaviour of Bianchi models  of type
IV and  VII$_h$ was analysed in \cite{CH2}, with an emphasis on
the stability properties of the vacuum plane-wave spacetimes. It
was shown that for $2/3<\gamma<2$ there will always be future
stable plane-wave solutions in the set of type IV and VII$_h$
tilted Bianchi models and it was proven that the {\em only} future
attracting equilibrium points for non-inflationary fluids
($\gamma>2/3$) are the plane-wave solutions in Bianchi type
VII$_h$ models. A tiny region of the parameter space was
discovered in the type IV model in which a closed orbit occurs.
Moreover, for the type VII$_h$ models a Hopf-bifurcation occurs
and more attracting closed orbits appear.

In this paper we will use a dynamical systems approach  to analyse
the tilting Bianchi type IV and VII$_h$ models  in
more detail, with an emphasis on numerical analysis. Since this current paper
is a companion paper to \cite{CH2}, we shall not reproduce all of the details
of the analysis and we shall frequently refer to equations in the original
paper. 
However, we shall attempt to make the current paper as self-contained as possible.

We note that the Bianchi VII$_h$ models are of special importance in cosmology.
We recall that Bianchi type VII$_h$ models are of maximal generality in the space of
all SH models and are sufficiently general to account for many
interesting phenomena. In particular, these cosmological
models are the most general models that contain the open
Friedmann-Robertson-Walker models and are consequently of special
interest. Therefore, these anisotropic generalizations of the open Friedmann-Robertson-Walker models
have been studied in detail in order to 
examine the dynamical behaviour of exact solutions of the Einstein field equation and, in particular,
to study the evolution of large scale anisotropies in models that contain
the standard low density universe as a limiting case.

There are a number of observational constraints on Bianchi models.  For example, the
structure of the Cosmic Microwave Background radiation anisotropy over intermediate
scales can be used to constrain the open Bianchi type VII$_h$ cosmological models
\cite{Barrow86}.  If the universe is open, then a distorted quadrupole or hotspot effect can
occur \cite{Novikov}.  However, the resulting constraint is of no practical use if the
density lies close to the critical density (as suggested by recent observations).
However, even if the density lies close to the critical density, in Bianchi VII$_h$
models the anisotropic 3-curvature produces a general relativistic geodesic spiralling
effect \cite{CollinsHawking}.  The spiral plus quadrupole are both focussed by the
hotspot effect allowing a possible  discrimination of the open Bianchi VII$_h$ models.

The asymptotic and intermediate evolution of
non-tilting perfect fluid  Bianchi VII$_h$  models has been studied using a
detailed qualitative analysis  \cite{DS1}, 
in conjunction with numerical experimentation \cite{wceh}.
It is known that although Bianchi VII$_h$ universes can isotropize at late times, the set of models that
do so are of measure zero in the the class of all Bianchi models \cite{CollinsHawking}.
The occurrence of intermediate isotropization in this class of non-tilting universes has
been studied in \cite{Doreshkevich1973}
using approximate solutions of the Einstein field equations.
More recently the evolution of the shear in Bianchi VII$_h$ models
was discussed heuristically \cite{Barrow1995};  in particular, it was claimed that if initial
conditions at the Planck time consistent with the Planck equipartition principle are imposed, then
the shear will decay sufficiently by the present epoch in order to be consistent with
the low anisotropy in the Cosmic Microwave Background radiation.
In \cite{wceh} it was found that while intermediate isotropization may occur, 
it does not necessarily do so, and the approximate and heuristic analyses of
\cite{Doreshkevich1973} and \cite{Barrow1995} were critically examined. The linear stability of Bianchi type VII$_h$ vacuum plane-wave solutions has recently been studied in \cite{JohnChristos}. 

The paper is organised as follows. First, we shall write down the
equations of motion using the orthonormal frame formalism in the
`$N$-gauge'. Then, in section \ref{sect:3}, we present all of the equilibrium points and investigate their stability properties. In section \ref{sect:4} we prove the existence of closed loops and provide some criteria for their stability. Then we will study the Bianchi type VII$_h$ (and IV)
models numerically in detail. Some of this numerical analysis is distributed throughout the paper; however, some explicit examples (e.g., the physically interesting case of dust and radiation) are given in Appendix \ref{app:numint}.
In particular, using a detailed numerical analysis of the type VII$_h$ models,
we shall fully justify the claim in \cite{CH2} that there
is an open set of parameter space in which solution curves
approach a compact surface that is topologically a torus.


\section{Equations of motion}

The  companion paper \cite{CH2} contains all the details regarding the  determination of the evolution 
equations for the tilted cosmological models under consideration. 
These equations, written in {\em gauge invariant} form, allow one to choose 
the gauge that is best suited to the application at hand. Here, we shall adopt the $N$-gauge in which the function 
${\bf N}_{\times}$ is purely imaginary; this is ensured by the choice
$\phi'=\sqrt{3}\lambda\Sigma_-$, where $\lambda$ is defined by
$\bar{N}=\lambda\mathrm{Im}({\bf N}_{\times})$. The evolution equation for
$\bar{N}$ can then be replaced with an evolution equation for $\lambda$,
which ensures a closed system of equations. We note that this
choice of gauge is well suited for the study of the dynamics close
to the vacuum plane-wave solutions \cite{CH2}.  For a qualitative analysis of these models, the $N$-gauge
is preferable since the resulting dynamical system is well defined in the Bianchi VI$_h$ case. However, 
care must be exercised when analyzing the Bianchi VII$_h$ case, in which the dynamical system is not
well defined over the entire region of state space, in particular the variable $\lambda$ could diverge.  

In the notation of \cite{CH2}, the expansion-normalised anisotropy and curvature variables used in this paper are
\[ {\mbold\Sigma}_1=\Sigma_{12}+i\Sigma_{13}, \quad {\mbold\Sigma}_{\times}=\Sigma_-+i\Sigma_{23}, \quad {\bf N}_{\times}=iN. \]
We will also adopt the dimensionless time parameter $\tau$, which is related to the cosmological time $t$ via $\mathrm{d} t/\mathrm{d}\tau =(1/H)$, where $H$ is the Hubble scalar. 

Using expansion-normalised variables, the equations of motion in the $N$-gauge
are (see \cite{CH2} for the complete derivation of the equations): 
\beq 
\Sigma_+'&=& (q-2)\Sigma_++{3}(\Sigma_{12}^2+\Sigma^2_{13})-2N^2 +\frac{\gamma\Omega}{2G_+}\left(-2v_1^2+v_2^2+v_3^2\right) \\
\Sigma_-'&=&(q-2-2\sqrt{3}\Sigma_{23}\lambda)\Sigma_-+\sqrt{3}(\Sigma_{12}^2-\Sigma_{13}^2) +2AN +\frac{\sqrt{3}\gamma\Omega}{2G_+}\left(v_2^2-v_3^2\right)\\
\Sigma'_{12}&=& \left(q-2-3\Sigma_+-\sqrt{3}\Sigma_-\right)\Sigma_{12} -\sqrt{3}\left(\Sigma_{23}+\Sigma_-\lambda\right)\Sigma_{13} +\frac{\sqrt{3}\gamma\Omega}{G_+}v_1v_2\\
\Sigma'_{13}&=&\left(q-2-3\Sigma_++\sqrt{3}\Sigma_-\right)\Sigma_{13}-\sqrt{3}\left(\Sigma_{23}-\Sigma_-\lambda\right)\Sigma_{12}+\frac{\sqrt{3}\gamma\Omega}{G_+}v_1v_3\\
\Sigma'_{23}&=&(q-2)\Sigma_{23}-2\sqrt{3}N^2\lambda+2\sqrt{3}\lambda\Sigma_-^2+2\sqrt{3}\Sigma_{12}\Sigma_{13}+ \frac{\sqrt{3}\gamma\Omega}{G_+}v_2v_3\\
N'&=& \left(q+2\Sigma_++2\sqrt{3}\Sigma_{23}\lambda\right){N}\\
\lambda' &=& 2\sqrt{3}\Sigma_{23}\left(1-\lambda^2\right)\\
A'&=& (q+2\Sigma_+)A . 
\eeq 
The equations for the fluids are:
\beq
\Omega'&=& \frac{\Omega}{G_+}\Big\{2q-(3\gamma-2)+2\gamma Av_1
 +\left[2q(\gamma-1)-(2-\gamma)-\gamma\mathcal{S}\right]V^2\Big\}
 \quad \\
 v_1' &=& \left(T+2\Sigma_+\right)v_1-2\sqrt{3}\Sigma_{13}v_3-2\sqrt{3}\Sigma_{12}v_2-A\left(v_2^2+v_3^2\right)-\sqrt{3}N\left(v_2^2-v_3^2\right)\\
 v_2'&=& \left(T-\Sigma_+-\sqrt{3}\Sigma_-\right)v_2-\sqrt{3}\left(\Sigma_{23}+\Sigma_-\lambda\right)v_3+\sqrt{3}\lambda{N}v_1v_3+\left(A+\sqrt{3}N\right)v_1v_2 \\
 v_3'&=& \left(T-\Sigma_++\sqrt{3}\Sigma_-\right)v_3-\sqrt{3}\left(\Sigma_{23}-\Sigma_-\lambda\right)v_2-\sqrt{3}\lambda{N}v_1v_2+\left(A-\sqrt{3}N\right)v_1v_3 \\
 V'&=&\frac{V(1-V^2)}{1-(\gamma-1)V^2}\left[(3\gamma-4)-2(\gamma-1)Av_1-\mathcal{S}\right],
\eeq 
where 
\beq q&=& 2\Sigma^2+\frac
12\frac{(3\gamma-2)+(2-\gamma)V^2}{1+(\gamma-1)V^2}\Omega\nonumber \\
\Sigma^2 &=& \Sigma_+^2+\Sigma_-^2+\Sigma_{12}^2+ \Sigma_{13}^2+\Sigma_{23}^2\nonumber \\
\mathcal{S} &=& \Sigma_{ab}c^ac^b, \quad c^ac_{a}=1, \quad v^a=Vc^a,\quad \nonumber \\
 V^2 &=& v_1^2+v_2^2+v_3^2,\quad  \nonumber \\
 T&=& \frac{\left[(3\gamma-4)-2(\gamma-1)Av_1\right](1-V^2)+(2-\gamma)V^2\mathcal{S}}{1-(\gamma-1)V^2}\nonumber\\
 G_+&=&1+(\gamma-1)V^2\nonumber.
\eeq 
These variables are subject to the constraints 
\beq
1&=& \Sigma^2+A^2+N^2+\Omega \label{const:H}\\
0 &=& 2\Sigma_+A+2\Sigma_-N+\frac{\gamma\Omega v_1}{G_+} \label{const:v1}\\
0 &=&
-\left[\Sigma_{12}(N+\sqrt{3}A)+\Sigma_{13}\lambda{N}\right]+\frac{\gamma\Omega v_2}{G_+} \label{const:v2}\\
0 &=&
\left[\Sigma_{13}(N-\sqrt{3}A)+\Sigma_{12}\lambda{N}\right]+\frac{\gamma\Omega v_3}{G_+} \label{const:v3} \\
0&=& A^2+3h\left(1-\lambda^2\right)N^2.\label{const:group} 
\eeq 
The parameter $\gamma$ will be assumed to be in the interval $\gamma\in (0,2)$. 
The generalized Friedmann equation, (\ref{const:H}), yields an expression which effectively defines the energy density $\Omega$.
Therefore, the state vector can thus be considered 
${\sf X}=[\Sigma_+,\Sigma_-,\Sigma_{12},\Sigma_{13},\Sigma_{23},N,\lambda,A,v_1,v_2,v_3]$ 
modulo the constraint equations (equations (\ref{const:v1}-\ref{const:group})). 
Thus the dimension of the physical state space is seven (for a given value of the parameter $h$).  
Additional details are presented in \cite{CH2}.

The dynamical system is invariant under the following discrete symmetries :
$$\begin{tabular}{l}
$\phi_1:~[\Sigma_+,\Sigma_-,\Sigma_{12},\Sigma_{13},\Sigma_{23},N,\lambda,A,v_1,v_2,v_3] 
\mapsto  [\Sigma_+,\Sigma_-,\Sigma_{12},\Sigma_{13},\Sigma_{23},-N,\lambda,-A,-v_1,-v_2,-v_3] $ \\
$\phi_2:~[\Sigma_+,\Sigma_-,\Sigma_{12},\Sigma_{13},\Sigma_{23},N,\lambda,A,v_1,v_2,v_3] 
\mapsto [\Sigma_+,-\Sigma_-,\Sigma_{13},\Sigma_{12},\Sigma_{23},-N,\lambda,A,v_1,v_3,v_2] $ \\
$\phi_3:~[\Sigma_+,\Sigma_-,\Sigma_{12},\Sigma_{13},\Sigma_{23},N,\lambda,A,v_1,v_2,v_3]
\mapsto [\Sigma_+,\Sigma_-,\mp \Sigma_{12},\pm \Sigma_{13},-\Sigma_{23},N,-\lambda,A,v_1,\mp v_2,\pm v_3]$\\
$\phi_4:~[\Sigma_+,\Sigma_-,\Sigma_{12},\Sigma_{13},\Sigma_{23},N,\lambda,A,v_1,v_2,v_3]
\mapsto[\Sigma_+,\Sigma_-,-\Sigma_{12},-\Sigma_{13},\Sigma_{23},N,\lambda,A,v_1,-v_2,-v_3]$
\end{tabular}$$
These discrete symmetries imply that without loss of  generality we can restrict the variables $\lambda\geq 0$, $A\geq0$, and $N\geq0$,  since the dynamics in the other regions can be obtained by simply applying one or more of  the maps above.  The fourth symmetry listed implies that one can add one additional constraint on one of the variables $\Sigma_{12},\Sigma_{13},v_2$ or $v_3$; however, in general there is no natural way to restrict any one of the variables, and hence we will not do so here.  Note that any equilibrium point in the region $v_2>0$ has a matching equilibrium point in the region $v_2<0$. 

\subsection{Invariant sets}
In this analysis we will be concerned with the following invariant sets: 
\begin{enumerate}
\item{} $T(VII_h)$: The general tilted type VII$_h$  model. Given
by $ 1<|\lambda|$. 
\item{}$T(IV)$: The general tilted type IV model. Given by $\lambda=\pm
1$, $A>0$. 
\item{} $T(V)$: The general tilted type V model. Given
by $N=0$, $A>0$, $\lambda=\pm 1$. 
\item{} $T(II)$: The general type II model. Given
by $\lambda=\pm 1$, $A=0$. 
\item{$B(I)$} Type I: ${N}=A=V=0$.
\item{} $T_1(\mathcal{A})$: One-component tilted model. Given by
$v_2=v_3=0$. 
\item{$\partial T(I)$} ``Tilted'' vacuum type I: $\Omega=N=A=0$.
\end{enumerate}

We note that the closure of the sets $T(IV)$ and $T(VII_h)$ are given by
\beq
\overline{T(VII_h)}&=&T(VII_h)\cup T(II)\cup B(I) \cup \partial T(I), \nonumber\\ 
\overline{T(IV)}&=&T(IV)\cup T(V)\cup T(II)\cup B(I) \cup \partial T(I).
\label{eq:decomp}\eeq
Since the boundaries may play an important role in the evolution of the dynamical system  we must consider all of the sets in the decomposition (\ref{eq:decomp}). 


\section{Future Asymptotic Behaviour}
\label{sect:3}
\subsection {Qualitative Analysis}
In order to study the future asymptotic behaviour of tilted type IV and VII$_h$ models, as a first step we will consider all the equilibrium points of the closed sets $\overline{T(IV)}$ and $\overline{T(VII_h)}$.  
Fortunately, much work has already been completed in this regard.  For instance, the Bianchi tilted type II models were studied in some detail by Hewitt, Bridson and Wainwright \cite{HBWII}.  They found that the future asymptotic state is either a flat Friedman-Lema\^itre model ($0<\gamma\leq2/3)$ , a non-tilted Bianchi type II model ($2/3<\gamma\leq 10/7$), an intermediately tilted Bianchi type II model ($10/7<\gamma<14/9)$, a line of tilted points for $\gamma=14/9$ or an extremely tilted Bianchi type II model when $14/9<\gamma<2$.  Similarly, in a paper by Hewitt and Wainwright \cite{HWV} in which the irrotational Bianchi type V models were analysed, it was found again that the flat Friedman-Lema\^itre model is stable for $0<\gamma<2/3$, a non-tilted Milne model is stable for $2/3<\gamma<4/3$ and an intermediately tilted Milne model is stable for $4/3<\gamma<2$.  Therefore, if we are interested in determining the future asymptotic state of the general tilting Bianchi type IV and VII$_h$ models, we need only  determine whether the corresponding equilibrium points are stable in the full Bianchi IV or VII$_h$ phase space. In particular, all equilibrium points in $\partial T(I)$ are Kasner circles and related equilibrium points. None of these equilibrium points are stable into the future and consequently we will not list these below. On the other hand, the set $\partial T(I)$ is essential in the analysis of the past asymptotic behaviour. We should also point out that for the equilibrium points with $N=0$ a different gauge has been used in order to determine their stability. 

\subsubsection{$B(I)$: Equilibrium points of Bianchi type I} 
\begin{enumerate} 
\item{}$\mathcal{I}(I)$: $[0,0,0,0,0,0,\lambda,0,0,0,0]$ where $\lambda$ is an unphysical parameter.  This represents the flat Friedman-Lema{\^i}tre model. 
\paragraph{Eigenvalues:} 
\[ -\frac{3(2-\gamma)}{2} [\times 5],~ \frac{3\gamma-2}{2} [\times 2].\] 
\end{enumerate}
The remaining equilibrium points are all in $\partial T(I)$.

\subsubsection{$T(II)$: Equilibrium points of Bianchi type II} 
All the tilted equilibrium points come in pairs. These represent identical solutions (they differ by a frame rotation); however, since their embeddings in the full state space are inequivalent, two of their eigenvalues are different. All equilibrium points have an unstable direction with eigenvalue $-2\sqrt{3}\Sigma_{23}$ corresponding to the variable $A$. 
\begin{enumerate}
\item{}{$\mathcal{C}\mathcal{S}(II)$}: $[\frac{2-3\gamma}{16}, 0,0,0,\frac{(2-3\gamma)\sqrt{3}}{16},\frac{\sqrt{3(3\gamma-2)(2-\gamma)}}{8},1,0,0,0,0]$, $2/3<\gamma<2$. This is the Collins-Stewart perfect fluid Bianchi II solution. 
\paragraph{Eigenvalues:} 
\[ \frac{3(3\gamma-2)}{8},~ -\frac{3}{4}(2-\gamma)\left(1\pm\sqrt{1-\frac{(6-\gamma)(3\gamma-2)}{(2-\gamma)}}\right), ~ 
-\frac{3(2-\gamma)}{2} [\times 2], ~\frac{3(7\gamma-10)}{8} [\times 2]\]  

\item{} {$\mathcal{H}_1(II)$}: $[\frac{4-3\gamma}{2}, -\frac{\epsilon\sqrt {3}}{4}\sqrt {\frac {(11\gamma-10)(2-\gamma)(7\gamma-10)}{17\gamma-18}}, 0, 0, -\frac{\sqrt {3}}{4}( 2-\gamma),\frac{\sqrt {3}}{2} \sqrt {\frac{(2-\gamma)(5\gamma-4)(3\gamma-4)}{17\gamma-18}}, 1, 0, \epsilon V, 0, 0]$, $10/7<\gamma<2$. This is Hewitt's tilted Bianchi II model with $\Sigma^2=\frac{(3\gamma-4)(9\gamma^2-20\gamma+12)}{17\gamma-18},\Omega=\frac{3(2-\gamma)( 21\,{\gamma}^{2}-24\gamma+4)}{4(17\gamma-18)},V^2=\frac{(3\gamma-4)(7\gamma-10)}{(11\gamma-10)(5\gamma-4)}$ and $\epsilon^2=1$.\\
 Three of the eigenvalues are $ \frac{3(9\gamma-14)}{4},~\frac{3(2-\gamma)}{2},~\frac{3(7\gamma-10)}{4}$,  with the remaining four eigenvalues being the roots of a nasty equation. 

\item{} {$\mathcal{H}_2(II)$}: $[\frac{9\gamma-14}{8},0 , \frac \epsilon 2\alpha, \frac \epsilon 2\alpha, -\frac{\sqrt {3}}{8}(5\gamma-6),\frac{\sqrt {3}}{2} \sqrt {\frac{(2-\gamma)(5\gamma-4)(3\gamma-4)}{17\gamma-18}}, 1, 0,0, \epsilon \frac{V}{\sqrt{2}},-\epsilon\frac{V}{\sqrt{2}}]$, $10/7<\gamma<2$. This is Hewitt's tilted Bianchi II model where $\Sigma^2$,  $\Omega$ and  $V$ are the same as for $\mathcal{H}_2(II)$, and $\alpha=-\frac{\sqrt {6}}{4}\sqrt {\frac {(11\gamma-10)(2-\gamma)(7\gamma-10)}{17\gamma-18}}$ and $\epsilon^2=1$.\\
Three of the eigenvalues are $ \frac{3(9\gamma-14)}{4},~\frac{3(5\gamma-6)}{4},~-\frac{3(7\gamma-10)}{4}$, with the remaining four being the roots of the same nasty equation as for $\mathcal{H}_1(II)$. 

\item{} {$\mathcal{L}_1(II)$}: $[-\frac 13,-\frac{\epsilon\sqrt{57(27\beta^2+2)(64-81\beta^2)}}{342},\frac 12{\beta},-\frac 12{\beta},-\frac{\sqrt{3}}{9},\frac{\sqrt{19(27\beta^2+4)(17-27\beta^2)}}{114},1,0,\epsilon V,0,0]$, $\gamma=14/9$. This is an intermediately tilted line bifurcation where  $0<\beta^2<\frac{8}{27}$, $V^2=\frac{3(27\beta^2+4)(27\beta^2+2)}{(17-27\beta^2)(64-81\beta^2)}$ and $\Omega=\frac{1}{684}(1458\beta^4-1215\beta^2+472)$. \\
Three of the Eigenvalues are $ 0, ~ \frac 23, ~ \frac 23$.

\item{} {$\mathcal{L}_2(II)$}: $[0,-\frac{\beta\sqrt{19(64-81\beta^2)}}{38},\frac \epsilon 2(\alpha+{\beta}),\frac \epsilon 2(\alpha-{\beta}),-\frac{2\sqrt{3}}{9},\frac{\sqrt{19(27\beta^2+4)(17-27\beta^2)}}{114},1,0,\frac{3\sqrt{3}\beta V}{\rho},\frac{\epsilon V}{\rho},-\frac{\epsilon V}{\rho}]$, $\gamma=14/9$. This is an intermediately tilted line bifurcation where  $0<\beta^2<\frac{8}{27}$, $\alpha=\frac{\sqrt{57(64-81\beta^2)}}{171}$,$\rho=\sqrt{27\beta^2+2}$, $V^2=\frac{3(27\beta^2+4)(27\beta^2+2)}{(17-27\beta^2)(64-81\beta^2)}$, $\Omega=\frac{1}{684}(1458\beta^4-1215\beta^2+472)$ and $\epsilon^2=1$. \\
 Three of the Eigenvalues are $ 0, ~ \frac 43, ~ -\frac 23$.

\item{} {$\mathcal{E}_1(II)$}: $[-\frac 13, -\frac{2\epsilon\sqrt{190}}{57},\frac{\sqrt{6}}{9}, -\frac{\sqrt{6}}{9}, -\frac{\sqrt{3}}{9}, \frac{\sqrt{57}}{19}, 1, 0, \epsilon ,0,0]$, $0<\gamma<2$. This is an extremely tilted Bianchi II, with $\Omega= 20/57$, $\Sigma^2=28/57$ and $\epsilon^2=\delta^2=1$. 
\paragraph{Eigenvalues:} 
\[  \frac{-2(9\gamma-14)}{3(2-\gamma)},~2/3,~ 2/3,~\frac{-1}{3}\pm\frac{1}{57}\sqrt{-20615\pm 456\sqrt{311}}.\]    

\item{} {$\mathcal{E}_2(II)$}: $[ 0, -\frac{4\epsilon\sqrt{285}}{171}, -\frac{\delta\sqrt{6}}{9}\left(1+\frac{\epsilon\sqrt{95}}{19}\right),-\frac{\delta\sqrt{6}}{9}\left(1-\frac{\epsilon\sqrt{95}}{19}\right), -\frac{2\sqrt{3}}{9}, \frac{\sqrt{57}}{19}, 1, 0, \frac{2\epsilon\sqrt{5}}{5}, -\epsilon\delta\frac{\sqrt{10}}{10}, \epsilon\delta\frac{\sqrt{10}}{10}]$, $0<\gamma<2$. This is an extremely tilted Bianchi II, with $\Omega= 20/57$, $\Sigma^2=28/57$ and $\epsilon^2=\delta^2=1$.   \paragraph{Eigenvalues:} 
\[  \frac{-2(9\gamma-14)}{3(2-\gamma)},~4/3,~ -2/3,~\frac{-1}{3}\pm\frac{1}{57}\sqrt{-20615\pm 456\sqrt{311}}.\]    
\end{enumerate}

\subsubsection{$T(V)$: Equilibrium points of Bianchi V}
\begin{enumerate} 
\item{} {$\mathcal{I}(V)$}: $[0,0,0,0,0,0,1,A,0,0,0]$, $0<A<1$, $\gamma=2/3$. Represents a FRW model with a $\gamma=2/3$ fluid. 
\paragraph{Eigenvalues:}
\[ 0[\times 2], ~-2[\times 5]. \]

\item{} {$\mathcal{M}(V)$}: $[0,0,0,0,0,0,1,1,0,0,0]$, $0<\gamma<2$.  This represents the Milne universe. ($\Sigma^2=0, \Omega=0, V^2=0$)  
\paragraph{Eigenvalues:}
\[ 0,~-2 [\times 2]~,-(3\gamma-2),~ (3\gamma-4) [\times 3].\]   
The zero eigenvalue arises due to the fact that this is part of the one-parameter family of plane-wave solutions. 

\item{} { $\widetilde{\mathcal{M}}(V)$}: $[0,0,0,0,0,0,1,1,\frac{3\gamma-4}{2(\gamma-1)},0,0]$, $6/5<\gamma<2$. This represents an ``intermediately tilted'' Milne model ($\Sigma^2=0, \Omega=0, V^2=\frac{(3\gamma-4)^2}{4(\gamma-1)^2}$).  
\paragraph{Eigenvalues:} 
\[ 0, ~-2 [\times 2], \frac{(3\gamma-4)}{2(\gamma-1)} [\times 2], -\frac{(3\gamma-4)(5\gamma-6)}{(9\gamma-10)(\gamma-1)}.\]   
Again the zero eigenvalue is associated with there being a non-isolated line of equilibria. 

\item{} { ${\mathcal{M}_{\pm}}(V)$}: $[0,0,0,0,0,0,1,1,\pm1,0,0]$, $0<\gamma<2$. This represents  ``extremely tilted'' Milne models ($\Sigma^2=0, \Omega=0, V^2=1$).  
\paragraph{Eigenvalues:} 
\[\mathcal{M}_+:~~ 0 [\times 2], ~-2 [\times 2],~ 1 [\times 2],~ 2, \qquad 
\mathcal{M}_-:~~ 0,~-2[\times 2], ~-4, ~\frac{2(5\gamma-6)}{2-\gamma}, ~-1[\times 2]. \]  

\item{} {${\mathcal{R}(V)}$}: $[\Sigma_+,0,0,0,0,0,1,(1+\Sigma_+),1,0,0]$, $-1<\Sigma_+<0$, $0<\gamma<2$. This represents a type V model with a null fluid. 
\paragraph{Eigenvalues:}
\[ 0[\times 2],~-2(1+\Sigma_+)[\times 2], (1-2\Sigma_+)[\times 2], ~2(1-2\Sigma_+).\]
The two zero eigenvalues are associated with this being part of a two-parameter family of equilibrium points ($\mathcal{R}(IV)$ see below).  
\end{enumerate}

\subsubsection{$T(IV)$: Equilibrium points of Bianchi type IV} 
\begin{enumerate}
\item{} $\mathcal{L}(IV):~ [\Sigma_+,\sqrt{-\Sigma_+(1+\Sigma_+)}, 0,0,0,\sqrt{-\Sigma_+(1+\Sigma_+)},1,(1+\Sigma_+),0,0,0]$, $-1<\Sigma_+<0$, $0<\gamma<2$. These represent 'non-tilted' vacuum plane waves. 
\paragraph{Eigenvalues:}
\beq 0,~ -2\left[(1+\Sigma_+)\pm 2i\sqrt{3}N\right], ~ -4\Sigma_+-(3\gamma-2),  ~(3\gamma-4)+2\Sigma_+,~ (3\gamma-4)-\Sigma_+\nonumber
\eeq
\item{} $\widetilde{\mathcal{L}}_{\pm}(IV)$:$[\Sigma_+,\sqrt{-\Sigma_+(1+\Sigma_+)}, 0,0,0,\sqrt{-\Sigma_+(1+\Sigma_+)},1,(1+\Sigma_+),\pm 1,0,0]$, $-1<\Sigma_+<0$, $0<\gamma<2$. These represent 'extremely tilted' vacuum plane waves. 
\paragraph{Eigenvalues:}
\beq &&\widetilde{\mathcal{L}}_+:~
0, ~-2\left[(1+\Sigma_+)\pm 2i\sqrt{3}N\right],~ 0, ~2(1-2\Sigma_+), ~ (1-2\Sigma_+)[\times 2]. \nonumber \\
&&\widetilde{\mathcal{L}}_-:~
0, ~-2\left[(1+\Sigma_+)\pm 2i\sqrt{3}N\right], ~-4(1+\Sigma_+), ~-\frac{2(5\gamma-6+2\gamma\Sigma_+)}{2-\gamma}, -(1+4\Sigma_+)[\times 2].\nonumber
\eeq
\item{} $\widetilde{\mathcal{L}}(IV):~ [\Sigma_+,\sqrt{-\Sigma_+(1+\Sigma_+)}, 0,0,0,\sqrt{-\Sigma_+(1+\Sigma_+)},1,(1+\Sigma_+),\frac{3\gamma-4+2\Sigma_+}{2(\gamma-1)(\Sigma_++1)},0,0]$, $-1<\Sigma_+<0$, $6/(5+2\Sigma_+)<\gamma<2$. These represent 'intermediately tilted' vacuum plane waves.   
\paragraph{Eigenvalues:} 
\beq &&0, ~-2\left[(1+\Sigma_+)\pm 2i\sqrt{3}N\right], ~-\frac{2-\gamma}{\gamma-1}(1-2\Sigma_+),\nonumber \\ && -\frac{(1-2\Sigma_+)(5\gamma-6+2\gamma\Sigma_+)(3\gamma-4+2\Sigma_+)}{(\gamma-1)(9\gamma-10+4\Sigma_+(1-\Sigma_+))}, ~\frac{(1-2\Sigma_+)(3\gamma-4)}{2(\gamma-1)}[\times 2].\nonumber
\eeq
\item{}$\widetilde{\mathcal{F}}_{\pm}(IV):~ [\Sigma_+,\sqrt{-\Sigma_+(1+\Sigma_+)}, 0,0,0,\sqrt{-\Sigma_+(1+\Sigma_+)},1,(1+\Sigma_+),-\frac{4-3\gamma-\Sigma_+}{(3-2\gamma-1)(\Sigma_++1)},v_2,-v_2]$ \\  where $v_2=\pm\frac{\sqrt{2(3\gamma-4)(-1+2\Sigma_+)(3\gamma-4-\Sigma_+)}}{2(3-2\gamma)(1+\Sigma_+)}$, $-1<\Sigma_+<0$, $(4+\Sigma_+)/3<\gamma<\min[4/3,3/(2-\Sigma_+)]$. These represent 'intermediately tilted' vacuum plane waves. 
\paragraph{Eigenvalues:}\beq
0, ~-2\left[(1+\Sigma_+)\pm 2i\sqrt{3}N\right], ~-\frac{(1-2\Sigma_+)(5\gamma-6)}{3-2\gamma},~ 0,~ \lambda_6, ~\lambda_7, \nonumber 
\eeq
where 
\beq
&& {\lambda_{6}\lambda_{7}}=-\frac{(1-2\Sigma_+)^2(4-3\gamma)(3\gamma-4-\Sigma_+)(3-2\gamma+\gamma\Sigma_+)}{(3-2\gamma)G(\gamma,\Sigma_+)}\nonumber  \\
&& \lambda_6+\lambda_7=\frac{(1-2\Sigma_+)F(\gamma,\Sigma_+)} {4(3-2\gamma)(17\gamma^2-40\gamma+24)G(\gamma,\Sigma_+)}
\nonumber \eeq
and
\beq
G(\gamma,\Sigma_+)&\equiv & (5\gamma-6)\Sigma_+^2-(18-25\gamma+9\gamma^2)\Sigma_+-3+2\gamma, \nonumber \\
F(\gamma,\Sigma_+)&\equiv & \left[2(17\gamma^2-40\gamma+24)\Sigma_+-33\gamma^3+121\gamma^2-152\gamma+66\right]^2\nonumber \\&&
- 9(\gamma-1)^2\left(121\gamma^4-736\gamma^3+1664\gamma^2-1656\gamma+612\right)
\label{def:Fdef}\eeq
Here is $G(\gamma,\Sigma_+)<0$ in the whole region under consideration. $F(\gamma,\Sigma_+)=0$, defines a line from $(4/3,0)$ to $(4/3,-1/4)$.

\item{} $\widetilde{\mathcal{E}}_{\pm}(IV):~ [\Sigma_+,\sqrt{-\Sigma_+(1+\Sigma_+)}, 0,0,0,\sqrt{-\Sigma_+(1+\Sigma_+)},1,(1+\Sigma_+),\frac{1+\Sigma_+}{3\Sigma_+},v_2,-v_2]$,\\   where $v_2=\pm\frac{\sqrt{2(4\Sigma_++1)(2\Sigma_+-1)}}{6\Sigma_+}$, $-1<\Sigma_+<-1/4$, $0<\gamma<2$. These represent 'extremely tilted' vacuum plane waves.  
\paragraph{Eigenvalues:}
\beq && 0,~-2\left[(1+\Sigma_+)\pm 2i\sqrt{3}N\right], ~\frac{(1-2\Sigma_+)(1+2\Sigma_+)}{\Sigma_+}, ~0, ~-\frac{(1-2\Sigma_+)(1+4\Sigma_+)}{3\Sigma_+}, ~\frac{2(1-2\Sigma_+)(2\gamma-3-\gamma\Sigma_+)}{3\Sigma_+(2-\gamma)}, \nonumber
\eeq

\item{}$\mathcal{R}(IV)$: $[\Sigma_+,(1-\ell)\sqrt{-\Sigma_+(1+\Sigma_+)}, 0,0,0,(1-\ell)\sqrt{-\Sigma_+(1+\Sigma_+)},1,(1+\Sigma_+),1,0,0]$, $-1<\Sigma_+<0$, $0<\ell<1$, $0<\gamma<2$. These represent non-vacuum plane-waves with a null fluid.  
\paragraph{Eigenvalues:} Same as for $\widetilde{\mathcal{L}}_+(IV)$. 
\end{enumerate}

\subsubsection{$T(VII_h)$: Equilibrium points of Bianchi type VII$_h$} 
For all of the vacuum equilibrium points the group parameter, $h$, is given by 
\[ 3h\Sigma_+(1-\lambda^2)=(1+\Sigma_+),\]
and we have defined $\tilde{h}\equiv 1/\sqrt{h}$. 
\begin{enumerate}
\item{} $\mathcal{L}(VII_h):~ [\Sigma_+,\sqrt{-\Sigma_+(1+\Sigma_+)}, 0,0,0,\sqrt{-\Sigma_+(1+\Sigma_+)},\lambda,(1+\Sigma_+),0,0,0]$, $-1<\Sigma_+<0$, $1<\lambda$, $0<\gamma<2$. These represent 'non-tilted' vacuum plane waves. 
\paragraph{Eigenvalues:}
\beq
0,~-2\left[(1+\Sigma_+)\pm 2i\sqrt{3}N\right], ~-4\Sigma_+-(3\gamma-2), ~(3\gamma-4)+2\Sigma_+, ~(3\gamma-4)-\Sigma_+\pm i\tilde{h}(1+\Sigma_+) .\nonumber
\eeq
\item{} $\widetilde{\mathcal{L}}_{\pm}(VII_h):~  [\Sigma_+,\sqrt{-\Sigma_+(1+\Sigma_+)}, 0,0,0,\sqrt{-\Sigma_+(1+\Sigma_+)},\lambda,(1+\Sigma_+),\pm 1,0,0]$, $-1<\Sigma_+<0$, $1<\lambda$, $0<\gamma<2$. These represent 'extremely tilted' vacuum plane waves. \paragraph{Eigenvalues:}
\beq &&\widetilde{\mathcal{L}}_+: ~
0, ~-2\left[(1+\Sigma_+)\pm 2i\sqrt{3}N\right],~ 0, ~2(1-2\Sigma_+), ~ (1-2\Sigma_+)[\times 2]. \nonumber \\ &&\widetilde{\mathcal{L}}_-: ~
0, ~-2\left[(1+\Sigma_+)\pm 2i\sqrt{3}N\right], ~-4(1+\Sigma_+), ~-\frac{2(5\gamma-6+2\gamma\Sigma_+)}{2-\gamma}, -(1+4\Sigma_+)\pm 2i\tilde{h}(1+\Sigma_+).\nonumber
\eeq
\item{} $\widetilde{\mathcal{L}}(VII_h):~ [\Sigma_+,\sqrt{-\Sigma_+(1+\Sigma_+)}, 0,0,0,\sqrt{-\Sigma_+(1+\Sigma_+)},\lambda,(1+\Sigma_+),\frac{3\gamma-4+2\Sigma_+}{2(\gamma-1)(\Sigma_++1)},0,0]$, $-1<\Sigma_+<0$, $1<\lambda$, $6/(5+2\Sigma_+)<\gamma<2$. These represent 'intermediately tilted' vacuum plane waves.  
\paragraph{Eigenvalues:} 
\beq &&0, ~-2\left[(1+\Sigma_+)\pm 2i\sqrt{3}N\right], ~-\frac{2-\gamma}{\gamma-1}(1-2\Sigma_+),\nonumber \\ && -\frac{(1-2\Sigma_+)(5\gamma-6+2\gamma\Sigma_+)(3\gamma-4+2\Sigma_+)}{(\gamma-1)(9\gamma-10+4\Sigma_+(1-\Sigma_+))}, ~\frac{(1-2\Sigma_+)(3\gamma-4)}{2(\gamma-1)}\pm i\tilde{h}(1-v_1)(1+\Sigma_+).\nonumber
\eeq
\item{}$\mathcal{R}(VII_h)$: $[\Sigma_+,(1-\ell)\sqrt{-\Sigma_+(1+\Sigma_+)}, 0,0,0,(1-\ell)\sqrt{-\Sigma_+(1+\Sigma_+)},\lambda,(1+\Sigma_+),1,0,0]$, $-1<\Sigma_+<0$, $0<\ell<1$, $1<\lambda$, $0<\gamma<2$. Non-vacuum plane-wave with a null fluid.  The group parameter is given by $3h\Sigma_+(1-\lambda^2)(1-\ell)=(1+\Sigma_+)$. 
\paragraph{Eigenvalues:} Same as for $\widetilde{\mathcal{L}}_+(VII_h)$. 
\end{enumerate}

\subsection{Local stability of equilibrium points} 
We quickly observe that none of the equilibria in the Bianchi type I and Bianchi type II  invariant sets are late-time attractors for the Bianchi IV or VII$_h$ models when $2/3<\gamma<2$. For the Bianchi type V model, only the isotropic Milne universes can act as future attractors. However, as can be seen, these equilibrium points are the isotropic limits of the plane-wave equilibrium points of Bianchi type IV and can be extracted directly from this analysis. Furthermore, the isotropic limit of the type VII$_h$ model can also be directly extracted from the type VII$_h$ analysis (even though $\lambda$ diverges here). 

\begin{figure}
\caption{Stability of the Plane-Wave Vacuum Solutions.  The regions where the different plane-wave equilibrium points are attractors for the Bianchi type IV with $v_2+v_3\geq 0$.  In the left figure, all the boundaries along the left edge marks the instability of the energy density, $\Omega$. The right figure is a magnified region of the left one showing the loophole (figure taken from \cite{CH2}). }\label{Fig:MapIV}\vspace{.5cm}
\includegraphics[width=6.5cm]{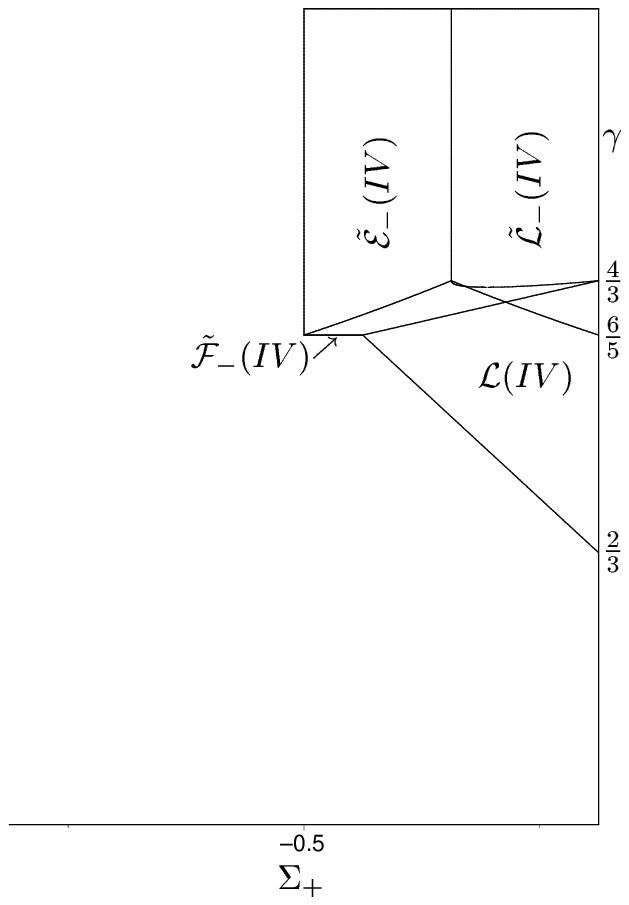}\hspace{1cm}
\includegraphics[width=6.5cm]{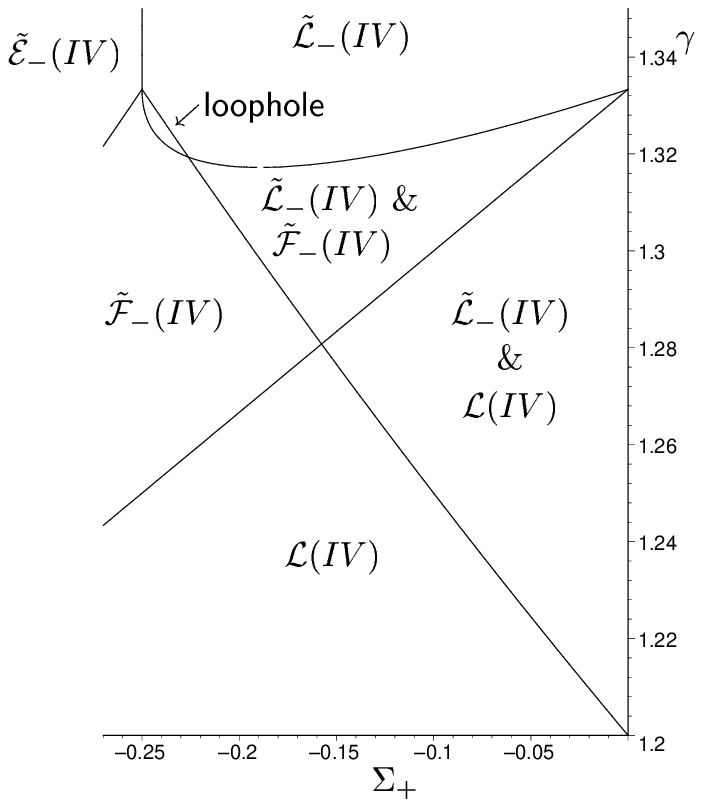}\\

\end{figure}

The stability analysis of the type IV model can be summarized as follows: 
{\it  For $2/3<\gamma<2$, the only equilibrium points that are future attractors in $\overline{T(IV)}$ are:} 
\begin{enumerate}
\item{}{$\mathcal{L}(IV)$:} stable for $\frac{2-4\Sigma_+}{3}<\gamma<\frac{4+\Sigma_+}3$.

\item{}{$\tilde{\mathcal{L}}_-(IV):$} stable $\frac{6}{5+2\Sigma_+}<\gamma<2$, $-\frac 14<\Sigma_+<0$.

\item{}{$\tilde{\mathcal{F}}_{\pm}(IV):$} $\tilde{\mathcal{F}}_{-}(IV)$ stable in $(v_2+v_3)\geq 0$~ \footnote{$\tilde{\mathcal{F}}_{+}(IV)$ is stable in the same region of $\gamma$ for the half $(v_2+v_3)\leq 0$.} for $\max\left(\frac 65,\frac{4+\Sigma_+}{3}\right)<\gamma<\min\left( \gamma_0,\frac{3}{2-\Sigma_+}\right)$

\item{}{$\tilde{\mathcal{E}}_{\pm}(IV):$} $\tilde{\mathcal{E}}_{-}(IV)$ stable for  $-\frac 12<\Sigma_+<-\frac14$, $ \frac{3}{2-\Sigma_+}<\gamma<2$.
\end{enumerate}
Here, we have defined $\gamma_0$ for any given value of $\Sigma_+$
as $F(\gamma_0,\Sigma_+)=0$,  where $F(\gamma,\Sigma_+)$ is
defined in eq.(\ref{def:Fdef}). Note that there is
a region where there are two co-existing future attractors; namely, the region where $\frac{6}{5+2\Sigma_+}<\gamma<\gamma_0$. Also,
there is a tiny region
$\gamma_0<\gamma<\frac{6}{5+2\Sigma_+}$ (from now on called the
'loophole') which does not contain any stable equilibrium points (see
Figure \ref{Fig:MapIV}).

The stability analysis of the equilibrium points of type VII$_h$ can be summarized as follows: 
{\it For $2/3<\gamma<2$, the only equilibrium points that are future attractors in $\overline{T(VII_h)}$ are:} 
\begin{enumerate}
\item{}{$\mathcal{L}(VII_h)$:} stable for $\frac{2-4\Sigma_+}{3}<\gamma<\frac{4+\Sigma_+}3$.
\item{}{$\tilde{\mathcal{L}}_-(VII_h):$} stable for $\frac{6}{5+2\Sigma_+}<\gamma<2$, $-\frac 14<\Sigma_+<0$.
\end{enumerate}

\subsection {The Bianchi type IV loophole}
\begin{figure}
\caption{The figure below displays the dynamical behaviour for values of the parameters near the loophole for Bianchi type IV non-vacuum tilted models.  The value of $\gamma$ is 1.325 and the initial conditions are chosen to be close to the plane wave solutions for illustrative purposes only.  A cross indicates the locations of $\widetilde{\mathcal{E}}_-(IV)$, diamonds indicate the position of $\widetilde{\mathcal{L}}_-(IV)$ while an asterisk indicates the initial condition.  The equilibrium point $\widetilde{\mathcal{F}}_-(IV)$ is located in the center of the spiral and the center of the closed loop. Note how the dynamical behaviour depends on the final value of $\Sigma_+$. } \label{figure1}\vspace{.5cm}
\includegraphics*[scale=0.8]{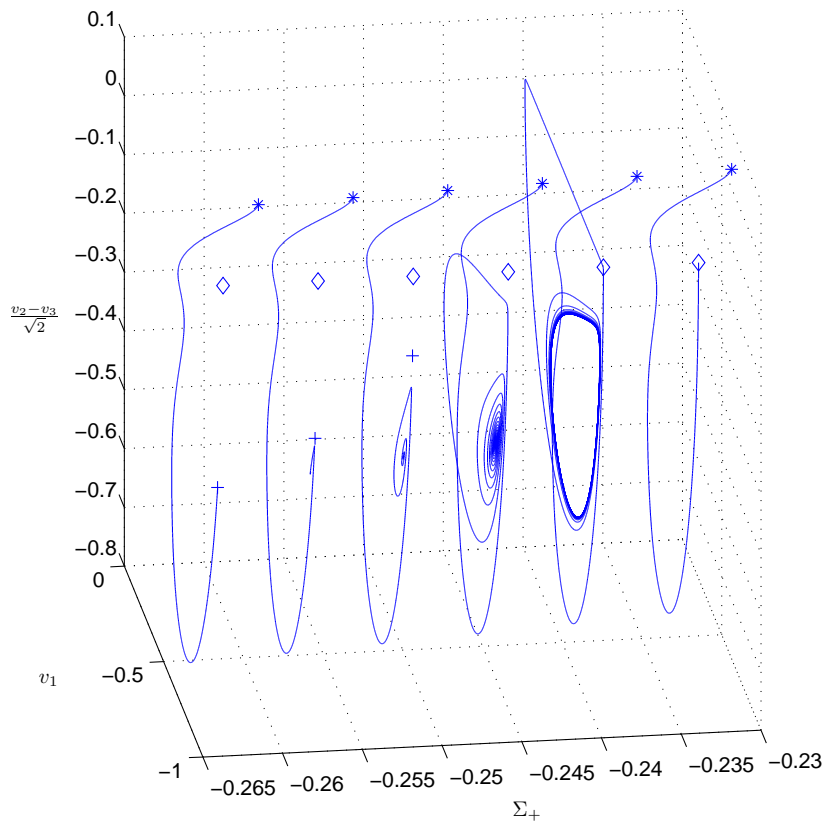}
\end{figure}

\begin{figure}
\caption{The figure below displays the dynamical behaviour for values of the parameters near the loophole for Bianchi type IV non-vacuum tilted models.  A cross indicates the locations of $\widetilde{\mathcal{E}}_-(IV)$, diamonds indicate the position of $\widetilde{\mathcal{L}}_-(IV)$ while an asterisk indicates the initial condition.  The equilibrium point $\widetilde{\mathcal{F}}_-(IV)$ is located in the center of the spiral and the center of the closed loop. Note how the dynamical behaviour depends on the value of $\Sigma_+^*$.  Each trajectory represents the vacuum limit of the orbits observed in the figure above.  The circular arc represents the boundary of the phase space.} \label{figure2}
 \includegraphics*[scale = 0.45]{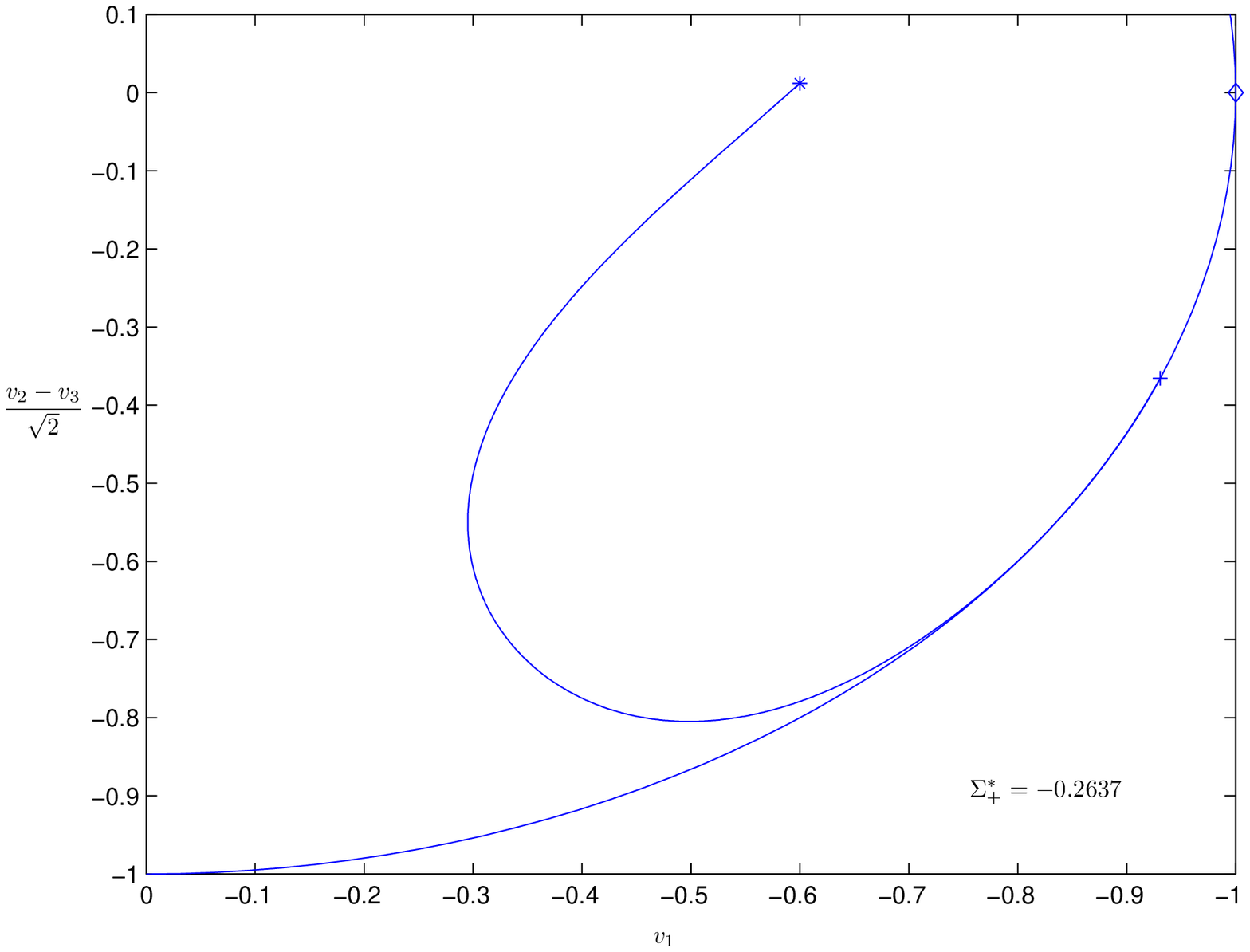}
 \includegraphics*[scale = 0.45]{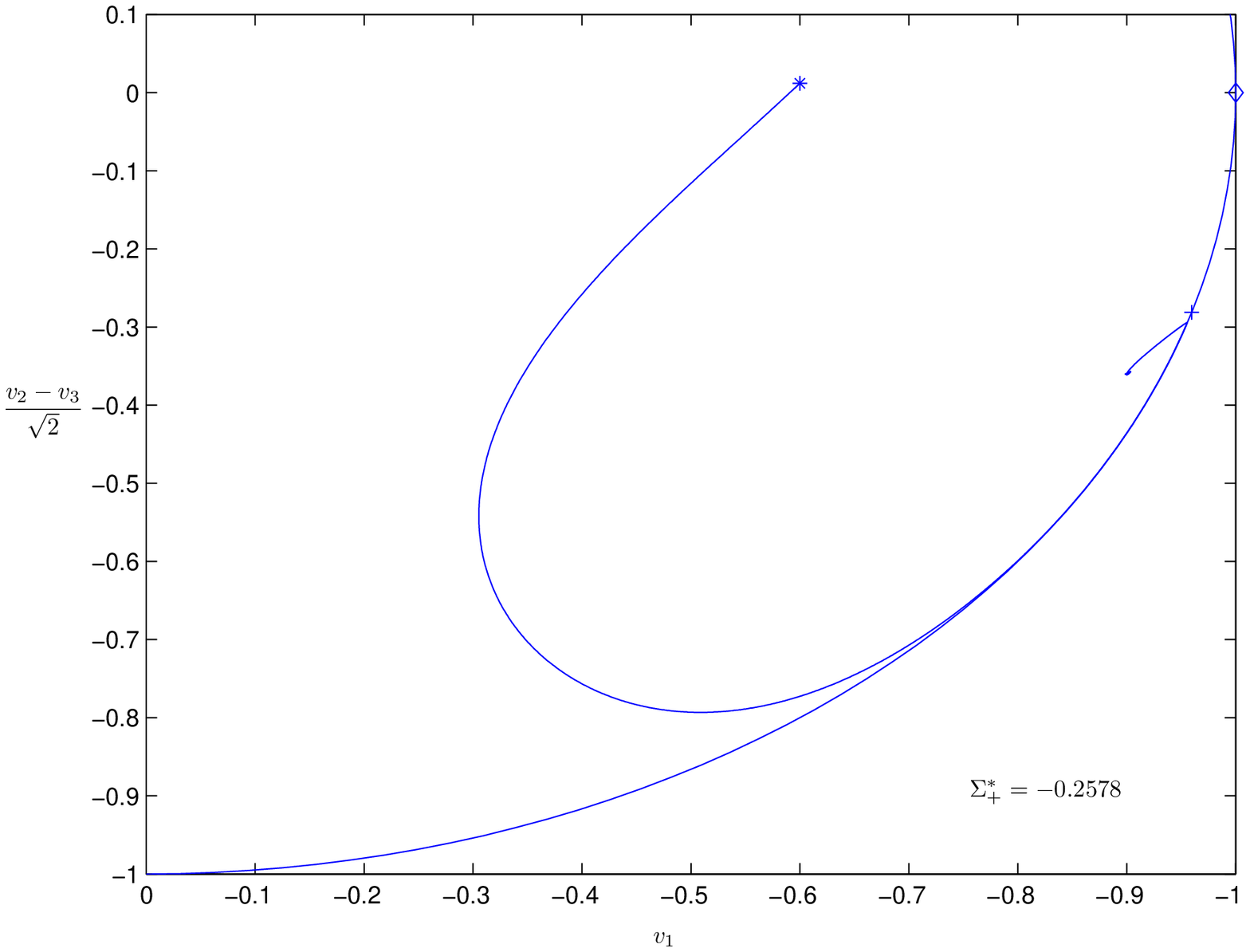}
 \linebreak
 \includegraphics*[scale = 0.45]{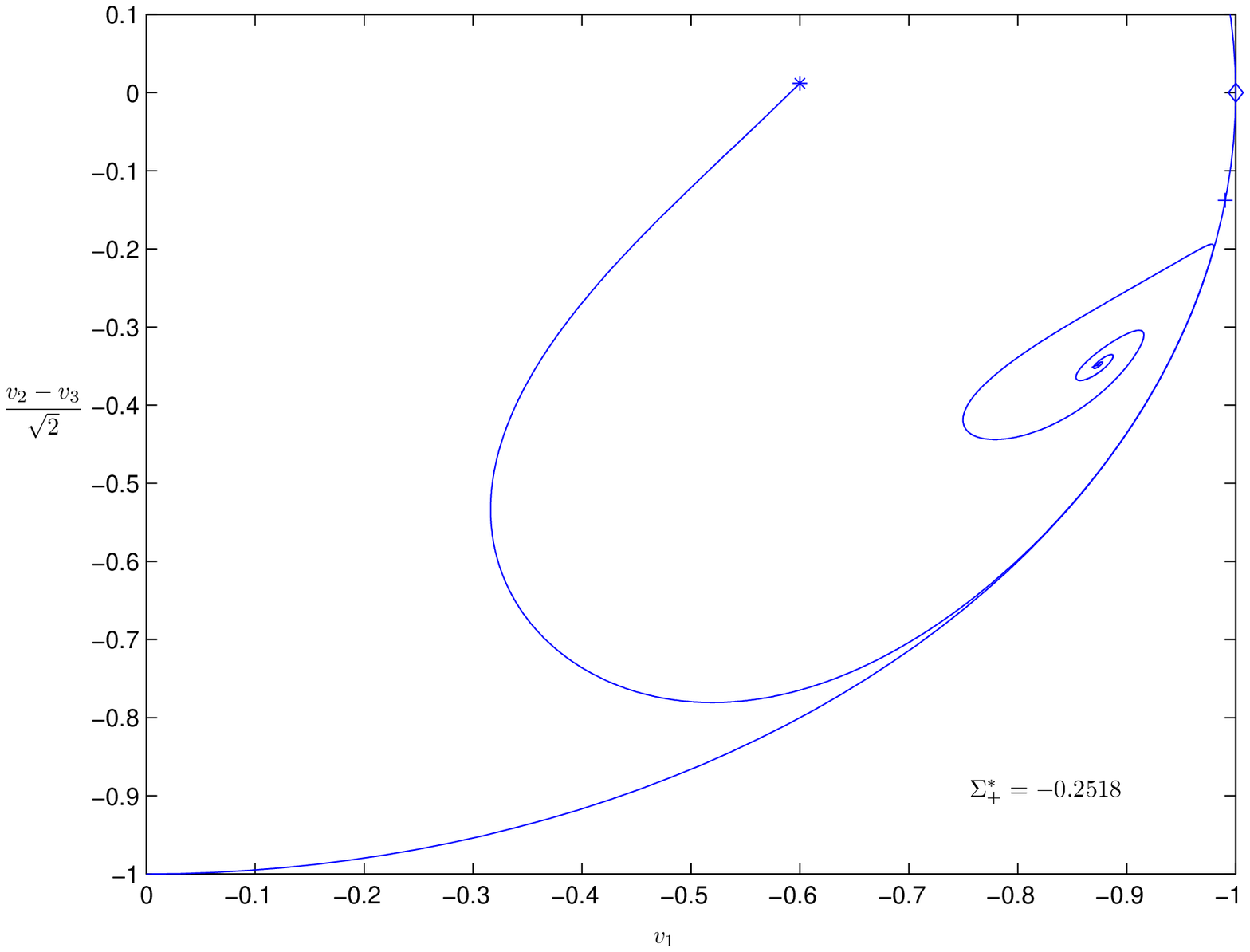}
 \includegraphics*[scale = 0.45]{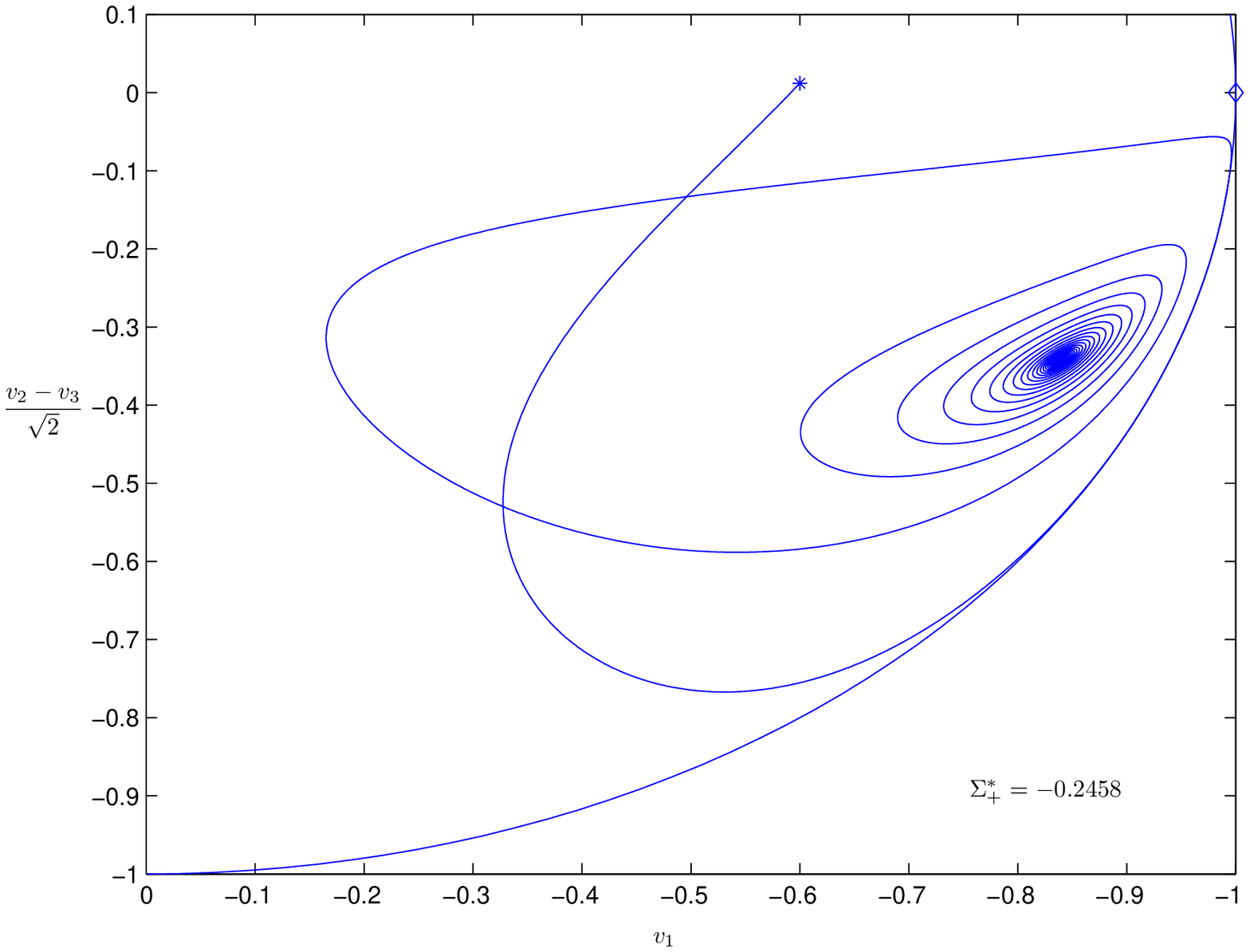}
 \linebreak
 \includegraphics*[scale = 0.45]{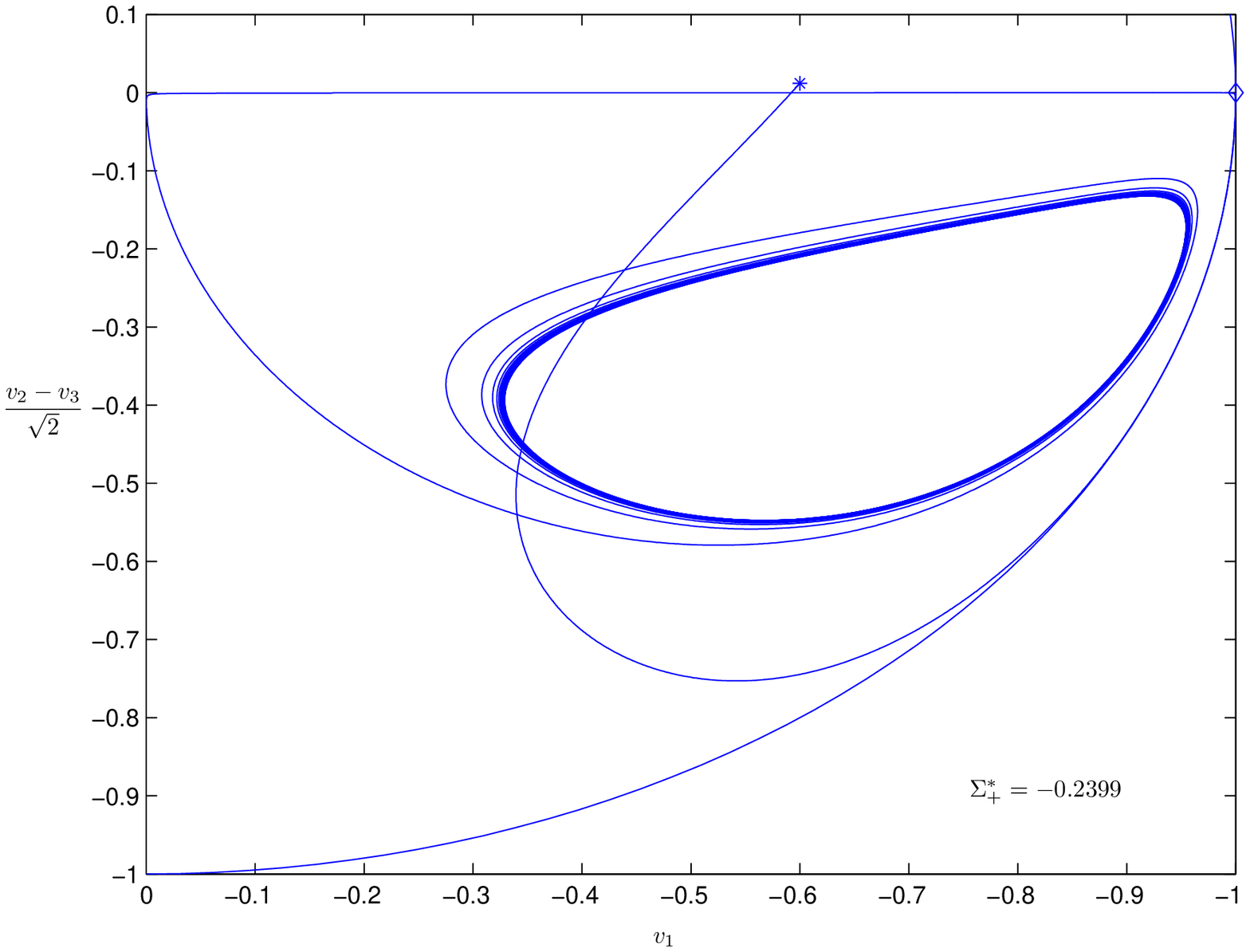}
 \includegraphics*[scale = 0.45]{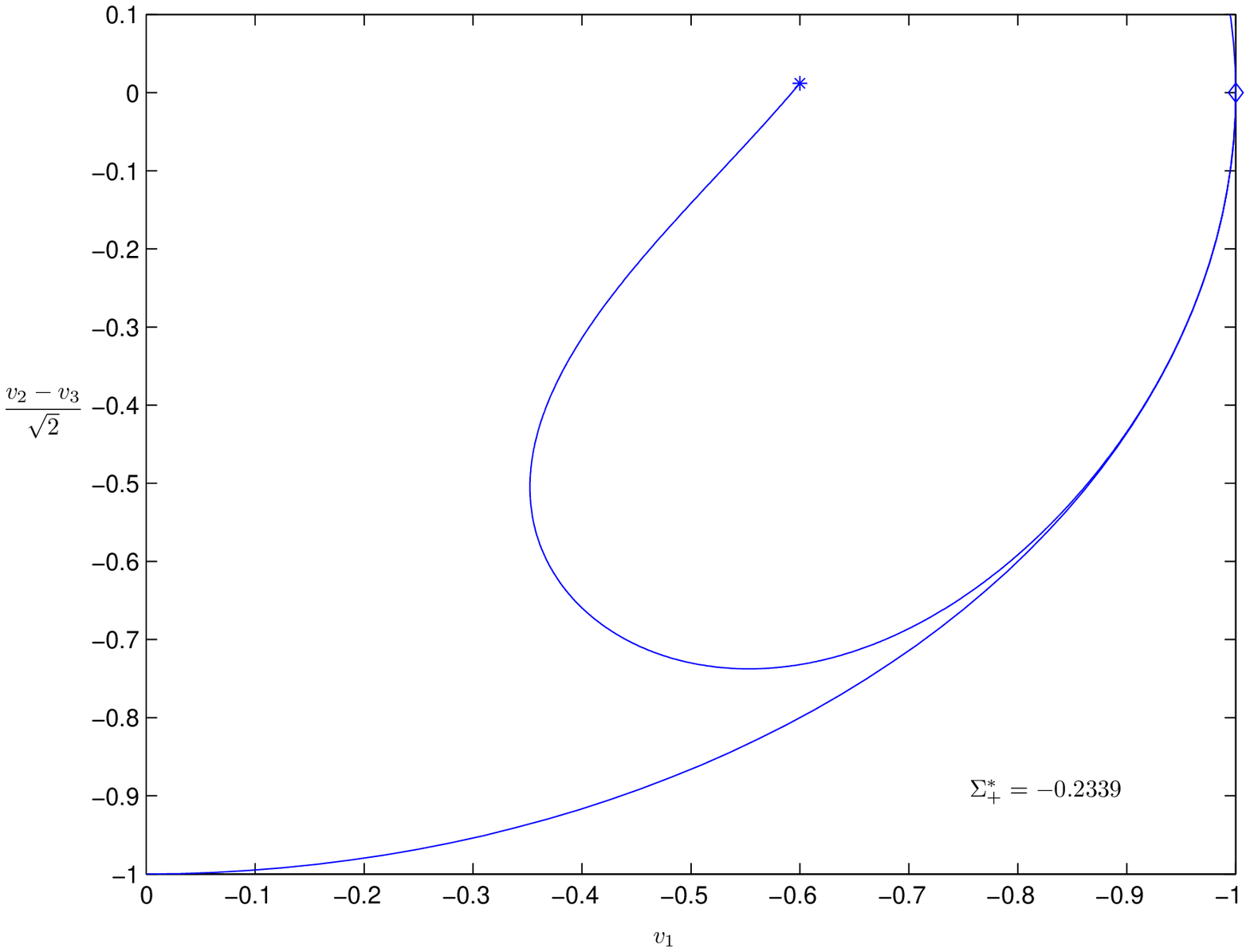}
 \end{figure}

As pointed out in \cite{CH2}, in  the limiting vacuum case (i.e., $\lim_{\Omega\to 0}$), the stability of the plane wave attractors changed as a function of both the parameter $\gamma$ and the terminal value of $\Sigma_+$, henceforth denoted as $\Sigma_+^*$.  As we have seen from the above there also exists a region of parameter space (here we consider the terminal value  $\Sigma_+^*$ as a parameter) which does not contain any stable plane-wave equilibrium points  (the loophole).   With the use of the bifurcation diagram and through subsequent numerical experimentation, a limit cycle was found for values of $\gamma$ and $\Sigma_+^*$ inside the loophole, see Figure \ref{Fig:MapIV}. This limit cycle acts as an attractor and will be referred to as \emph{the Mussel attractor} (see section 5 and Figure 2 in \cite{CH2}). Figures \ref{figure1} and \ref{figure2} indicate how the dynamical behaviour of the Mussel attractor changes for $\gamma=1.325$ and different limiting values of $\Sigma_+$.

We note that the limit cycles observed in Figure \ref{figure1} above are actually a continuum of limit cycles side by side.  The limiting value of $\Sigma_+$ is the determining factor in the final asymptotic state of the system.  Of course one might argue that the likelihood of actually falling into this family of limit cycles is small.  Again through some numerical experimentation it can be shown that there is a set of Bianchi IV initial conditions far from the plane-wave solutions of non-zero measure that have these limit cycles as future attractors for values of $\gamma$ inside the loophole (see Fig. \ref{far-torus} where this fact is illustrated for the Bianchi type VII$_h$ loophole).

\subsection{Behaviour of the System in General}
One of the most important points and argument in this and the companion paper is that at late times $\Omega\to 0$ and the variables $\Sigma_{ab}$, $N$, etc.,  'freeze' into a particular asymptotic value.  We have not been able to prove this rigorously, however, there is much evidence to support the following:
\begin{con}
Except for a set of measure zero, non-inflationary ($2/3<\gamma<2$) tilted perfect fluid Bianchi type IV and VII$_h$ models will at late times approach a vacuum state; i.e. 
\[ \lim_{\tau\rightarrow\infty} \Omega=0.\]
\label{conj}\end{con}
There are several results that support this. First, a local stability analysis shows that there are always stable vacuum solutions. So the conjecture is at least valid locally. Second, we know that there are no non-vacuum equilibrium points acting as local attractors; the preceding analysis excluded this. Third, in our fairly extensive numerical analysis no evidence for any other behaviour has been found. In particular, Figure \ref{figure1} shows how the variable $\Sigma_+$ freezes in to a particular value of $\Sigma_+$. In Appendix \ref{app:numint} more examples of this  behaviour are shown and we can see explicitly how $\Omega\rightarrow 0$ and $\Sigma_+\rightarrow \Sigma_+^*$.   

The local analysis also shows that the variable $\Sigma_{ab}$, $N$, etc. 'freeze in' to particular values $\Sigma_{ab}^*$, $N^*$. After this freezing has occurred, the system of equations effectively decouples and the  equations governing the tilt velocities separate. In effect, the tilt equations reduce to the much simpler form $v_i=F(v_j;\Sigma_+^*,\Sigma_-^*,\cdots)$, and hence, we can, at late times, treat all the variables, except $v_1$, $v_2$ and $v_3$, as constants. This 'freezing in' process makes these models amenable to analytical investigations, which was done in \cite{CH2}; the analysis provides numerical justification for this freezing process (e.g., see Figure \ref{figure1}). In fact, the numerical investigations show that the time scale of this freezing process is typically much shorter than many of the other features present; e.g. the oscillation in the type IV loophole (see Appendix \ref{app:numint}). In the Appendix we have included a sample of numerical integrations. For the plots we have chosen the particularly interesting values $\gamma=1$ (dust), $1.325$ and $4/3$ (radiation). In all of the plots we can see how quickly the energy-density, $\Omega$, tends to zero and the variable $\Sigma_+$ freezes in. On the other hand, the velocities typically take longer to settle into a particular asymptotic state (as explicit examples of this, see Figures \ref{figure(VIIg=1.325a)} and \ref{figure(VIIg=1.325b)}). We should point out that the numerical integrations are done in the \emph{full 7-dimensional state space}; however, for some figures the reduced system is used for illustrative purposes only.

\section{Dealing with Closed Loops}
\label{sect:4}
The dynamical system contains closed periodic orbits. Some of these closed curves act as limit cycles and take the role as attractors for the dynamical system. However, closed loops are difficult to find and describe in full generality. 

In our case, the 'freezing in' process enables us to consider the reduced system of equations describing the behaviour of the tilt velocities in a vacuum plane-wave background,  see \cite{CH2}.  We utilize the identity 
\[ (v_2^2+v_3^2)^2=(2v_2v_3)^2+(v_2^2-v_3^2)^2,  \]
and define $(x,\rho,\theta)$ by 
\beq
x=v_1, \quad  \rho\equiv v_2^2+v_3^2, \quad 2v_2v_3=\rho\sin\theta,\quad (v_2^2-v_3^2)= \rho\cos\theta.
\eeq
Then we have for the reduced system
\beq
x'&=& (T+2\Sigma^*_+)x-(A^*+\sqrt{3}N^*\cos\theta)\rho, \nonumber \\
\rho'&=& 2\left(T-\Sigma^*_++A^*x-\alpha\cos\theta\right)\rho, \nonumber \\
\theta'&=& 2\alpha(\lambda^*+\sin\theta),
\label{eq:redtilt}\eeq
where $\alpha=\sqrt{3}(1-x)N^*$ and, by use of the discrete symmetry $\phi_4$, we can assume that $0\leq \theta<2\pi$. Furthermore, these variables are bounded by
\beq
0\leq \rho, \quad V^2= x^2+\rho\leq 1.
\eeq
An asterisk has been added to the variables to emphasize that these should be thought of as  the limit values for the full system.

We will utilize the following fact: Assume that $c(\tau)$ is a (solution) curve, then if $f$ is a function in the state space,
\beq
\int_{t_1}^{t_2}f'(c(\tau))d \tau=f(c(t_2))-f(c(t_1)). 
\eeq
In particular, if $c(\tau)$ is a closed periodic orbit, 
\beq 
\oint f'(c(\tau))d \tau=0. 
\eeq

For a periodic orbit, $c$, with period $T_n$, we introduce the average of a variable $B$: 
\beq
\av{B}\equiv \frac{1}{T_n}\oint_c Bd\tau.
\eeq
We can also say something about the stability of a closed periodic orbit. For example, consider 
the evolution equation for $\Omega$, which we write as $\Omega'=\Omega\lambda_{\Omega}$. Assume that $c(\tau)$ is 
a closed periodic orbit with period $T_n$. Then for every $\tau_0$ and $\epsilon>0$,  there exists a solution curve, $\tilde{c}(\tau)=c(\tau)+\delta(\tau)$, for which
\beq
|\delta(\tau)|<\epsilon,\quad \tau_0<\tau<\tau_0+T_n.
\eeq
(This follows from Proposition 4.2, page 104, in \cite{DSReza}.) Hence, the curve $\tilde{c}(\tau)$ can be used to approximate the closed curve $c(\tau)$. Using $ \tilde{c}(\tau)$ we get
\beq
\ln\left[\frac{\Omega(\tau_0+T_n)}{\Omega(\tau_0)}\right]=\int_{\tau_0}^{\tau_0+T_n}\frac{\Omega'}{\Omega}d\tau=\int_{\tau_0}^{\tau_0+T_n}\lambda_{\Omega}d\tau=T_n\av{\lambda_{\Omega}}+\mathcal{O}(\epsilon).
\eeq
Since $\epsilon$ can be arbitrarily small, we can approximate
\beq
\Omega(\tau_{0}+T_n)\approx \Omega(\tau_0)\exp(T_n\av{\lambda_{\Omega}}).
\label{eq:stabC}\eeq
Hence, if $\av{\lambda_{\Omega}}<0$, then $\Omega(\tau_{0}+T_n)< \Omega(\tau_0)$ for a sufficiently small perturbation; i.e., the closed curve is \emph{stable} with respect to $\Omega$. If, on the other hand, $\av{\lambda_{\Omega}}>0$ then the closed curve is \emph{unstable} with respect to a perturbation of $\Omega$. This result is the corresponding local stability criterion for periodic orbits. 

By calculating $\av{1-x}$ (in two different ways) we are able to find an expression for $T_n$ for orbits in the type VII$_h$ model:
\beq
T_n=\frac{n\pi \sqrt{h}}{(1+\Sigma_+^*)(1-\av{x})},
\eeq
where $n$ is an integer. This expression involves the group parameter $h$, and we note that a larger  $h$ implies longer orbital period.

We also need,
\begin{lem} \label{lem:alphacos}
Assume that there is a closed properly\footnote{By \emph{properly periodic} we will mean a periodic orbit which is not an equilibrium point. Equilibrium points are trivially periodic in the sense that $c(\tau)=c(\tau+T_n)$ for any $T_n$.}  periodic orbit, $c(\tau)$, for the dynamical system (\ref{eq:redtilt}). Then 
\beq
\av{\frac{(1-x)(v_2^2-v_3^2)}{v_2^2+v_3^2}}=0. 
\eeq
\end{lem} 
\begin{proof} 

For $(\lambda^*)^2>1$, $\theta$ is a strictly monotonic function, so in this case we can use $\theta$ as a variable along the  curve. Furthermore, $\theta(\tau_0+T_n)=\theta_0+2\pi n$, where $n\in \mathbb{Z}$, which gives
\beq
\oint 2\alpha\cos\theta d\tau=\int_{\theta_0}^{\theta_0+2\pi n}\frac{\cos\theta}{\lambda^*+\sin\theta}d\theta=\left.\ln\left(\lambda^*+\sin\theta\right)\right|^{\theta_0+2\pi n}_{\theta_0}=0.
\eeq 
Since $\alpha=\sqrt{3}N^*(1-x)$, the Lemma now follows for $(\lambda^*)^2>1$. For $\lambda^*=1$ the equation for $\theta$ implies that the periodic curve must be in the invariant set $\sin\theta=-1$. Thus the Lemma is trivially satisfied in that case.
\end{proof} 
\begin{thm}\label{thm:xbar}
Assume that there is a closed properly periodic orbit, $c(\tau)$, for the dynamical system (\ref{eq:redtilt}). Then either
\beq\mathcal{C}:\quad 
\av{x}=-\frac{(3\gamma-4)-\Sigma^*_+}{(1+\Sigma^*_+)(3-2\gamma)}, \quad \av{\lambda_{\Omega}}=-\frac{(1-2\Sigma^*_+)(5\gamma-6)}{3-2\gamma}, 
\eeq
or $V=1$ and 
\beq
\mathcal{E}:\quad\av{x}=\frac{1+\Sigma^*_+}{3\Sigma^*_+}, \quad \av{\lambda_{\Omega}}=\frac{(1-2\Sigma^*_+)(1+2\Sigma^*_+)}{\Sigma^*_+}.
\eeq
\end{thm} 
\begin{proof} 
Assume first that there exists a closed periodic orbit with $V<1$ and $\rho>0$. Then, using the evolution equation for $V$, we get 
\beq
(3\gamma-4)\av{f(V)}-2(\gamma-1)A^*\av{xf(V)}-\av{f(V)\mathcal{S}}=0, 
\label{eq:avV}\eeq
for any analytic function $f(V)$ on $(0,1)$. Using Lemma \ref{lem:alphacos} the equation for $\rho'$ yields 
\[ \av{T}-\Sigma^*_++A^*\av{x}=0.\] 
Using eq.
(\ref{eq:avV}) with $f(V)=V^2/G_-$ we obtain $\av{T}=\av{\mathcal{S}}$. Using eq. (\ref{eq:avV}) with $f(V)=1$ we can solve for $\av{x}$, which yields
\[ \av{x}=-\frac{(3\gamma-4)-\Sigma^*_+}{(1+\Sigma^*_+)(3-2\gamma)}.\] 
A similar calculation yields the desired expression for $\av{\lambda_{\Omega}}$. 

For $V=1$ (which is an invariant subspace), we have $\rho=1-x^2$, and we can rewrite the equation for $\rho$:
\beq
\frac{(1-x^2)'}{1-x^2}=2x\left(-3x\Sigma^*_++A^*+2\alpha\cos\theta\right).
\eeq
Dividing by $2x$ and integrating yields $\av{x}=A^*/3\Sigma^*_+$ as desired. The expression for $\av{\lambda_{\Omega}}$ is found analogously. 

Now consider the invariant subspace $\rho=0$. This subspace is homeomorphic to the closed interval $[-1,1]$, and hence, no proper periodic orbit can exist there. 

The Theorem now follows. 
\end{proof}

This theorem states exactly when a closed periodic orbit, if it exists, is stable with respect to the variable $\Omega$. Equation (\ref{eq:stabC}) implies that small deviations from $\Omega=0$ will effectively decay exponentially if $\av{\lambda_\Omega}<0$. 

So far we have only given necessary criteria for closed periodic orbits. However, in some cases we can even prove 
existence of such orbits. 
\begin{thm}
For every $\Sigma^*_+\in (-1,-1/4)$, $(\lambda^*)^2>1$, and $\gamma\in (0,2)$ there exists one, and only one up to winding number, 
closed periodic orbit, $\widetilde{\mathcal{E}}(VII_h)$, with $V=1$ for the dynamical system (\ref{eq:redtilt}).   
\end{thm}
\begin{proof}
\emph{Existence}: Given $\Sigma^*_+\in (-1,-1/4)$, $\lambda^2>1$, and $\gamma\in (0,2)$, we consider the invariant subspace $V=1$. 
This extremely tilted set is topologically a sphere, $S^2$. For $(\lambda^*)^2>1$, the variable $\theta$ is strictly 
monotonic and can thus be considered as a "time" variable. On the sphere $S^2$, there are two, and only two,  equilibrium 
points; the North pole, $x=1$ and the South pole, $x=-1$. For $\Sigma^*_+\in (-1,-1/4)$, both of these 
equilibrium points have two unstable directions, and hence there exist open neighbourhoods $U_N$ and $U_S$ around the
North and South Pole, respectively, such that $S^2-(U_N\cup U_S)$ is a future trapping region. 
The compact set $S^2-(U_N\cup U_S)$ is diffeomorphic to a compact annulus in $\mathbb{R}^2$, which allows us to use 
the Poincar\'e-Bendixson theorem. This says that there must be either a future attracting equilibrium point, 
a future attracting closed orbit, or an arbitrary union of points and curves. However, 
we know there are no other equilibrium points; i.e., there must be a  closed curve. Existence is thus proven. 

\emph{Uniqueness}: Since $\theta$ is strictly monotonic, these periodic curves have to be winding around 
the $x$-axis. Assume there exists two such curves, $c(\theta)$, and $\tilde{c}(\theta)$. Since these two curves
cannot cross each other, $|x(\theta)-\tilde{x}(\theta)|\geq 0$, and equality holds if and only if $c=\tilde{c}$ (up to winding number). By considering the 
$x$-coordinate for these two closed period orbits, we obtain after a straight-forward manipulation, 
\[ \left|\av{x}-\av{\tilde{x}}\right|\geq 0,\]
where equality holds if and only if $c=\tilde{c}$. Using Theorem \ref{thm:xbar} we have $\av{x}=\av{\tilde{x}}$; 
hence, $c=\tilde{c}$ (up to winding number) and uniqueness is proved.  
\end{proof}

\begin{thm}[Mussel attractor]
For $\lambda^*=1$ and every $\Sigma^*_+$ and $\gamma$ taking values in the type IV loophole ( $\gamma_0<\gamma<\frac{6}{5+2\Sigma_+^*}$ )  there exists a
closed periodic orbit, $\mathcal{C}(IV)$,  for the dynamical system \ref{eq:redtilt}.
\end{thm}
\begin{proof}
In this case we can consider the 2-dimensional invariant subspace $\sin\theta=-1$. Again we can use Poincar\'e-Bendixson theory, however, in a slightly more complicated form. By a consideration of the equilibrium points and the separatrices the existence of a closed curve can be shown (see \cite{2Ddynsys} for details and exact formulation of the theory).   
\end{proof}

\subsection{Stability of the  Bianchi type VII$_h$ plane waves}
\begin{figure}
\caption{The figure below displays the dynamical behaviour for values of the parameters near the loophole $\gamma=1.325$ for Bianchi type VII$_h$ non-vacuum tilted models (for the figures $h=1$ is used).   When $\Sigma_+^*=-.2637$, the future asymptotic behaviour is dominated by the stable closed curve $ \widetilde{\mathcal{E}}(VII_h)$.  As $\Sigma_+^*$ increases to a value of $\Sigma_+^*=-.2578$, the closed loop $\widetilde{\mathcal{ E}}(VII_h)$ becomes unstable (still exists), and a new stable closed loop appears in the form of $\widetilde{\mathcal{F}}(VII_h)$.  When $\Sigma_+^*=-.2518$, the future asymptotic behaviour is still dominated by the stable closed loop $\widetilde{\mathcal{F}}(VII_h)$.    As $\Sigma_+^*$ increases to a value of $\Sigma_+^*=-.2458$, the transverse wrapping of the orbits around the closed loop $ \widetilde{\mathcal{F}}(VII_h)$ is increasing.  The orbits wrap transversely around the closed curve $ \widetilde{\mathcal{F}}(VII_h)$ so much so, that they catch up with themselves and create a new stable attractor that is topologically a torus (which we represent as $\mathcal{T}(VII_h)$). The creation of the torus is not especially clear in this view, because the interior radius of the torus is extremely small.  Note the size of the throat of the torus.} \label{figure3A} \vspace{.5cm}
 \includegraphics*[scale = 0.55]{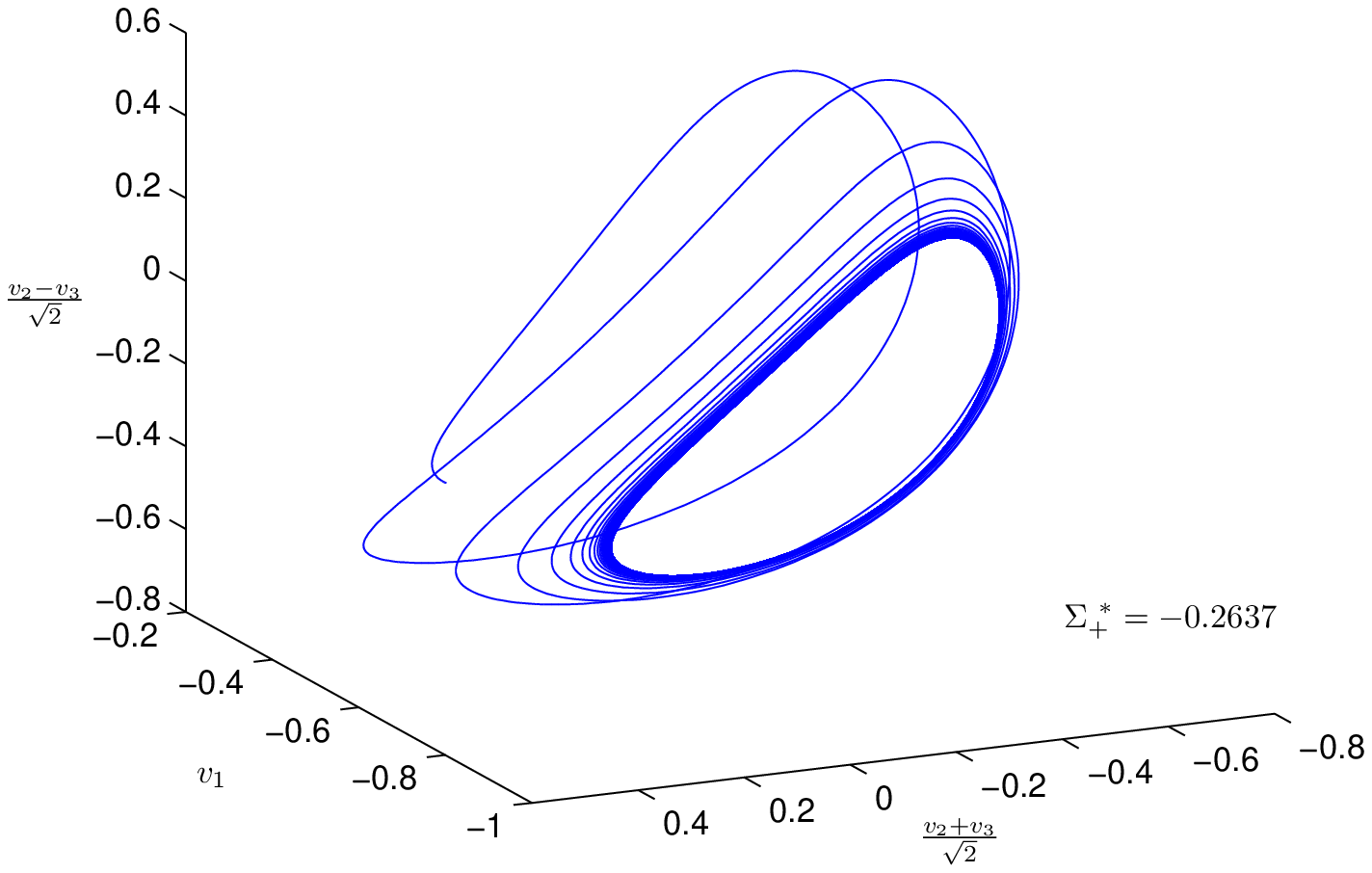}
 \includegraphics*[scale = 0.55]{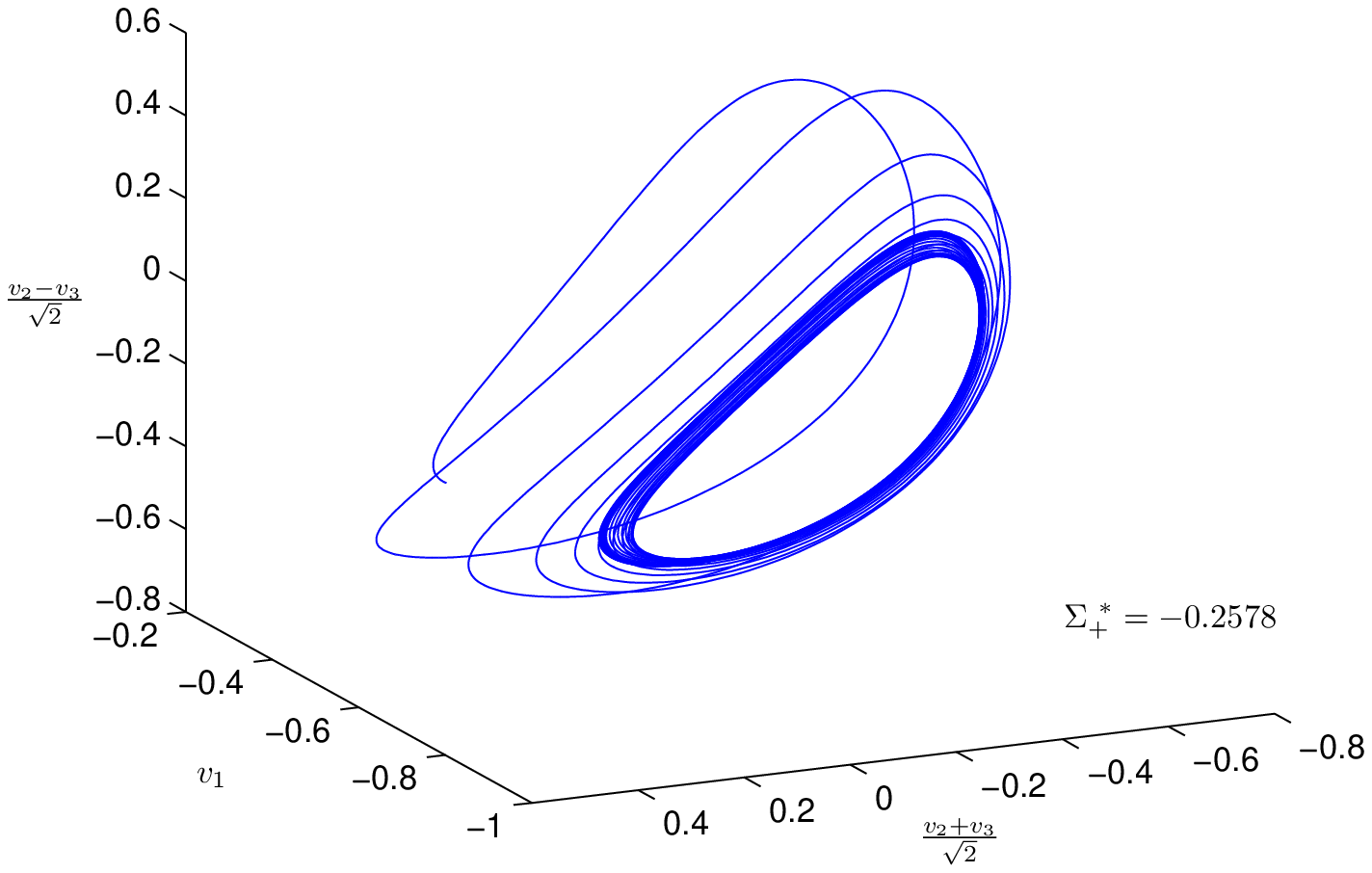}\\ 
\includegraphics*[scale = 0.55]{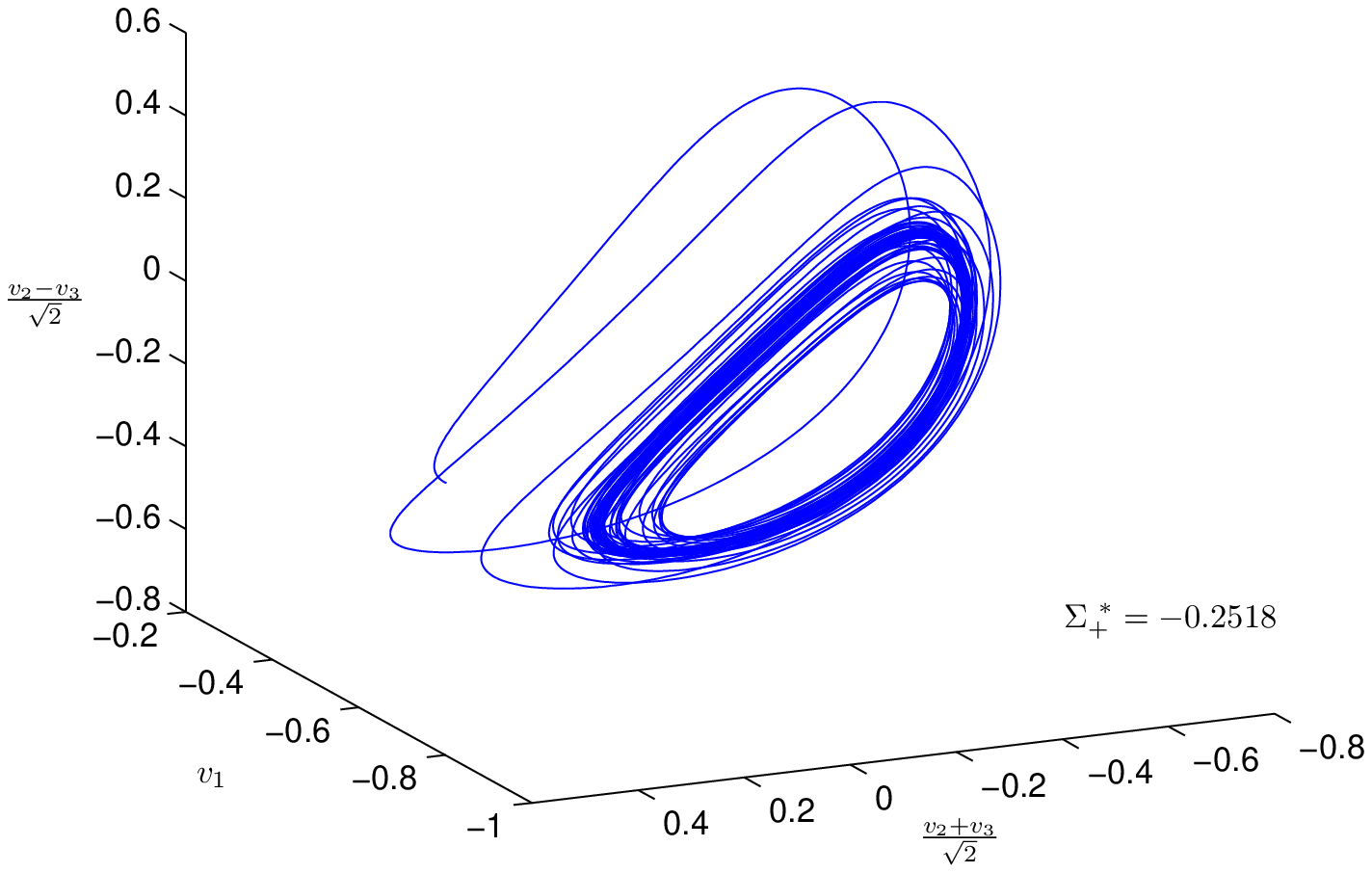}
 \includegraphics*[scale = 0.55]{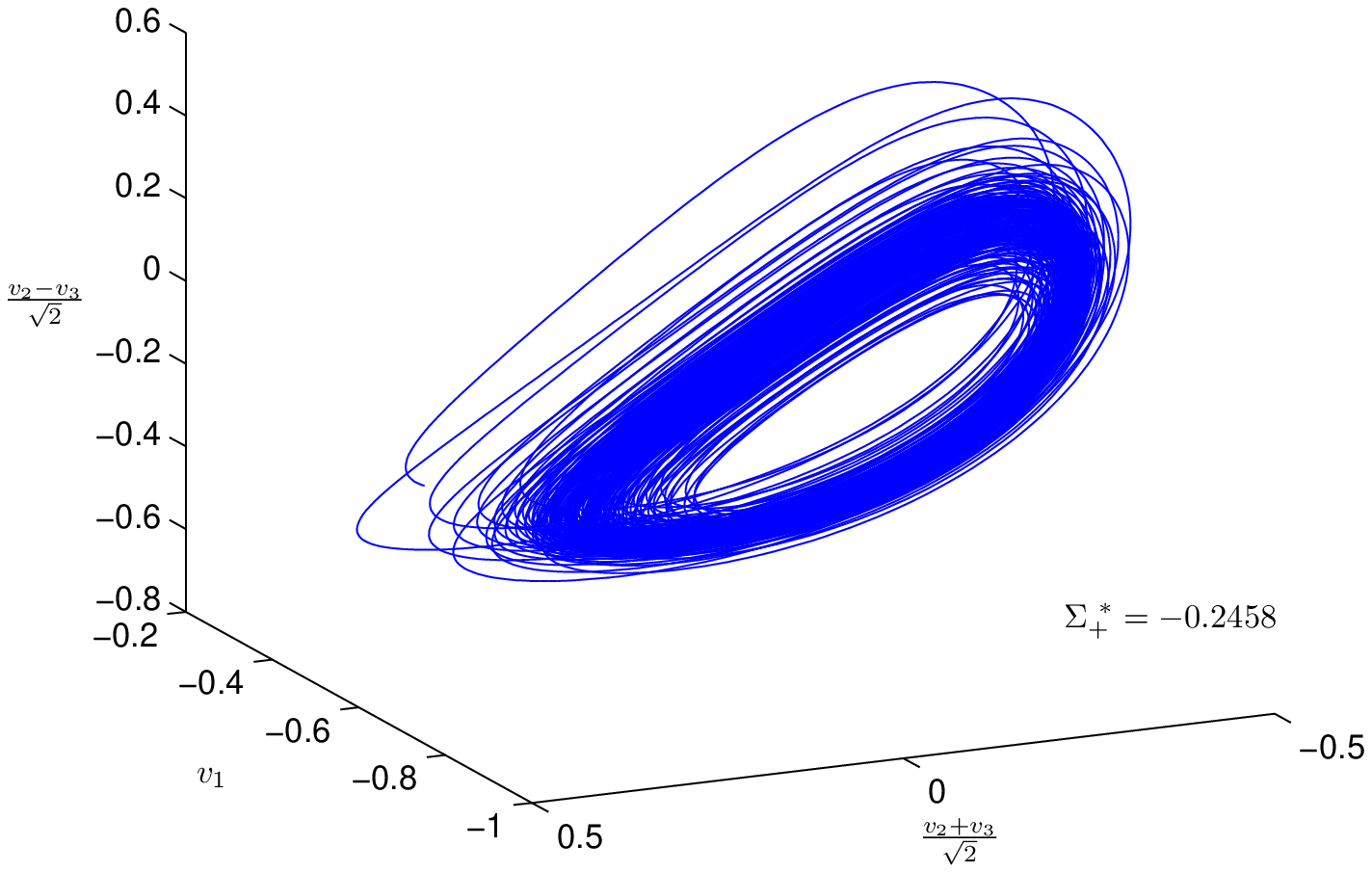}
\end{figure}

\begin{figure}
\caption{The figure below displays the dynamical behaviour for values of the parameters near the loophole $\gamma=1.325$ for Bianchi type VII$_h$ non-vacuum tilted models.  As $\Sigma_+^*$ increases from $\Sigma_+^*-.2458$ to $\Sigma_+^*=-.2399$, the interior radius of the torus increases.  The closed curve $\widetilde{\mathcal{ F}}(VII_h)$ lies in the interior of the torus. The future asymptotic behaviour is dominated by the attracting torus $\mathcal{T}(VII_h)$.   Note how the throat of the torus narrows as $\Sigma_+^*$ increases. As $\Sigma_+^*$ increases from $\Sigma_+^*-.2339$ to $\Sigma_+^*=-.2250$, the throat of the torus collapses, and the stable attractor becomes the equilibrium point $\widetilde{\mathcal{ L}}_-(VII_h)$.} \label{figure3C}\vspace{.5cm}
\includegraphics*[scale = 0.55]{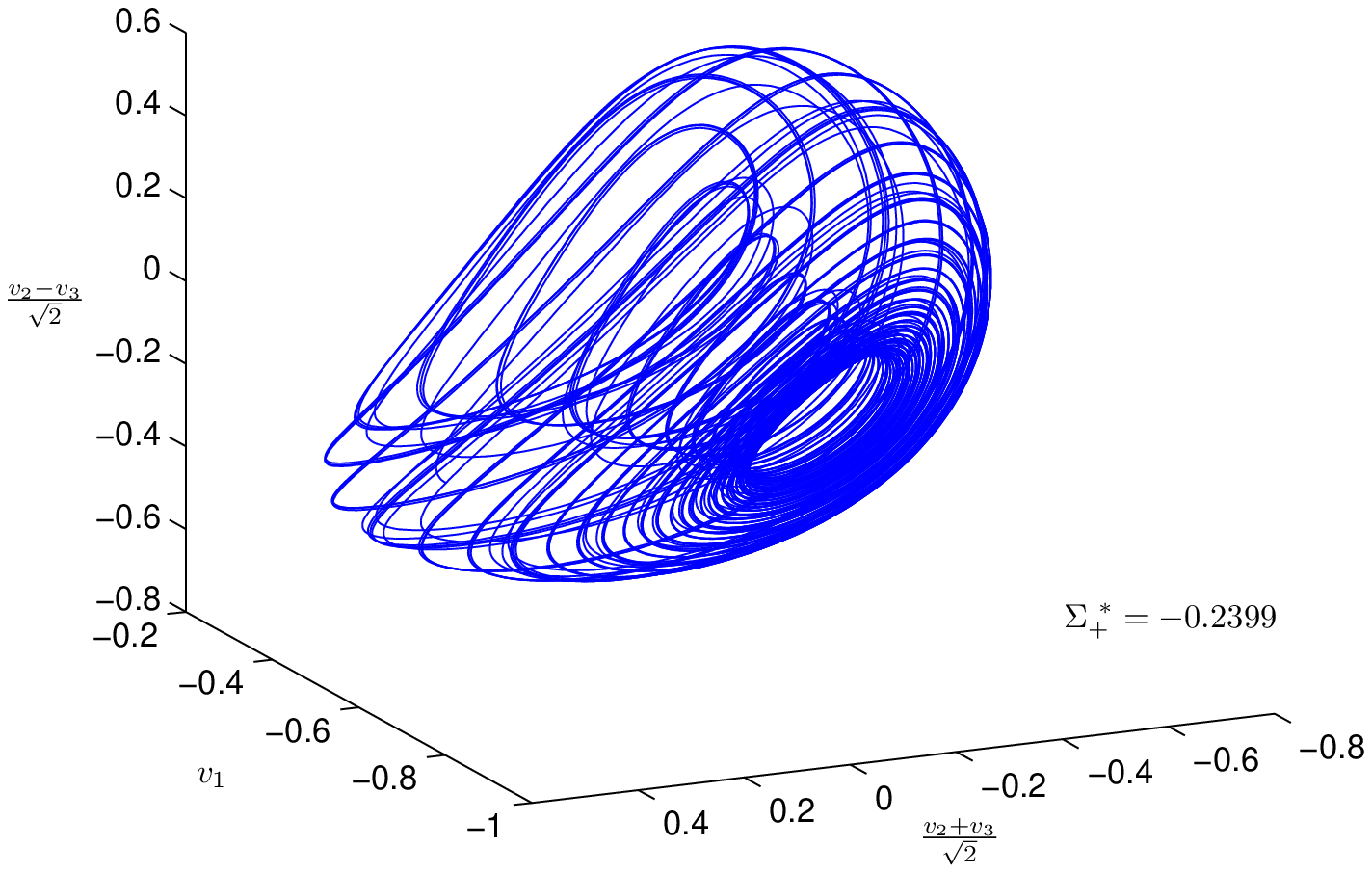}
\includegraphics*[scale = 0.55]{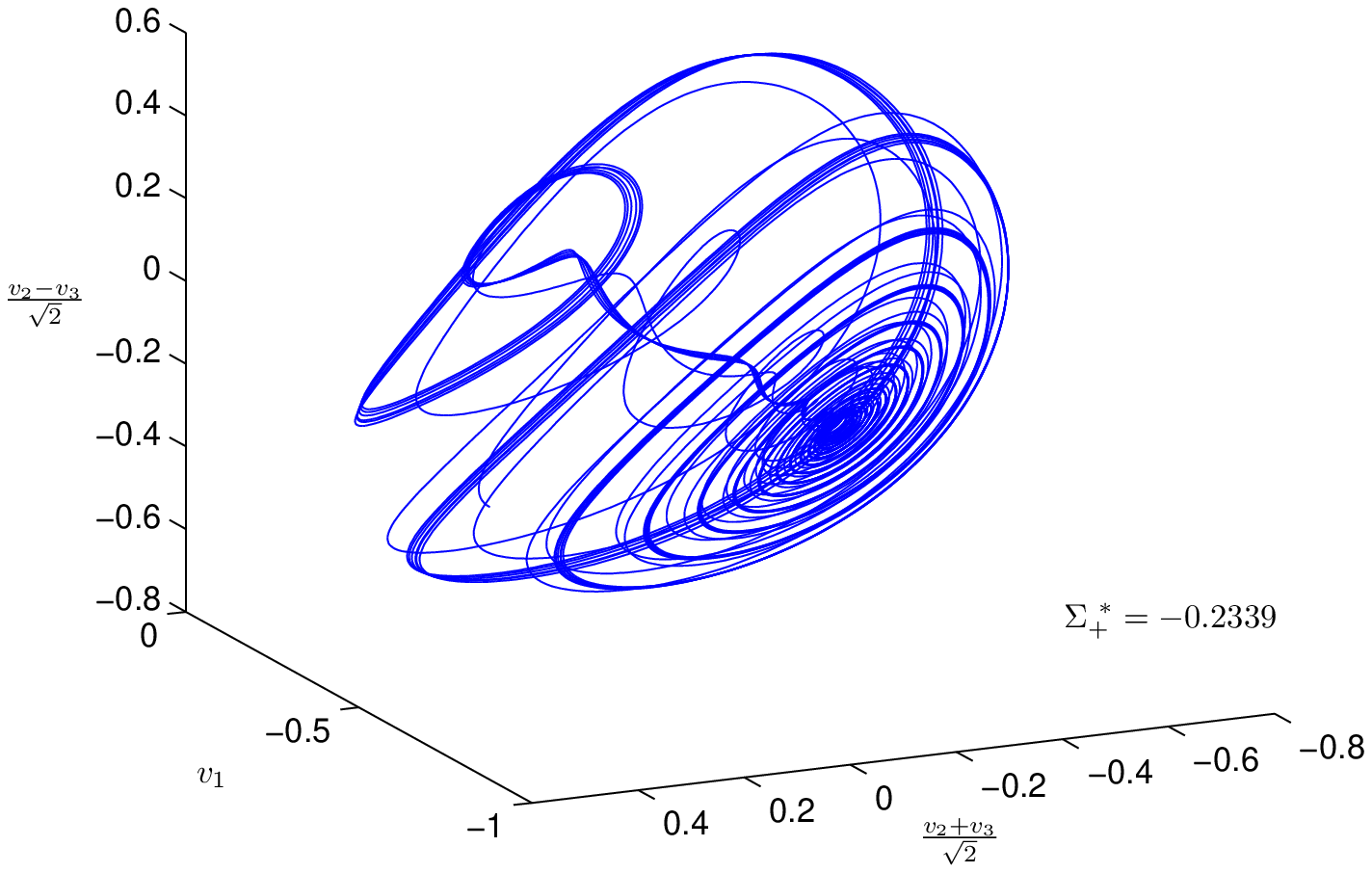}\\
\includegraphics*[scale = 0.55]{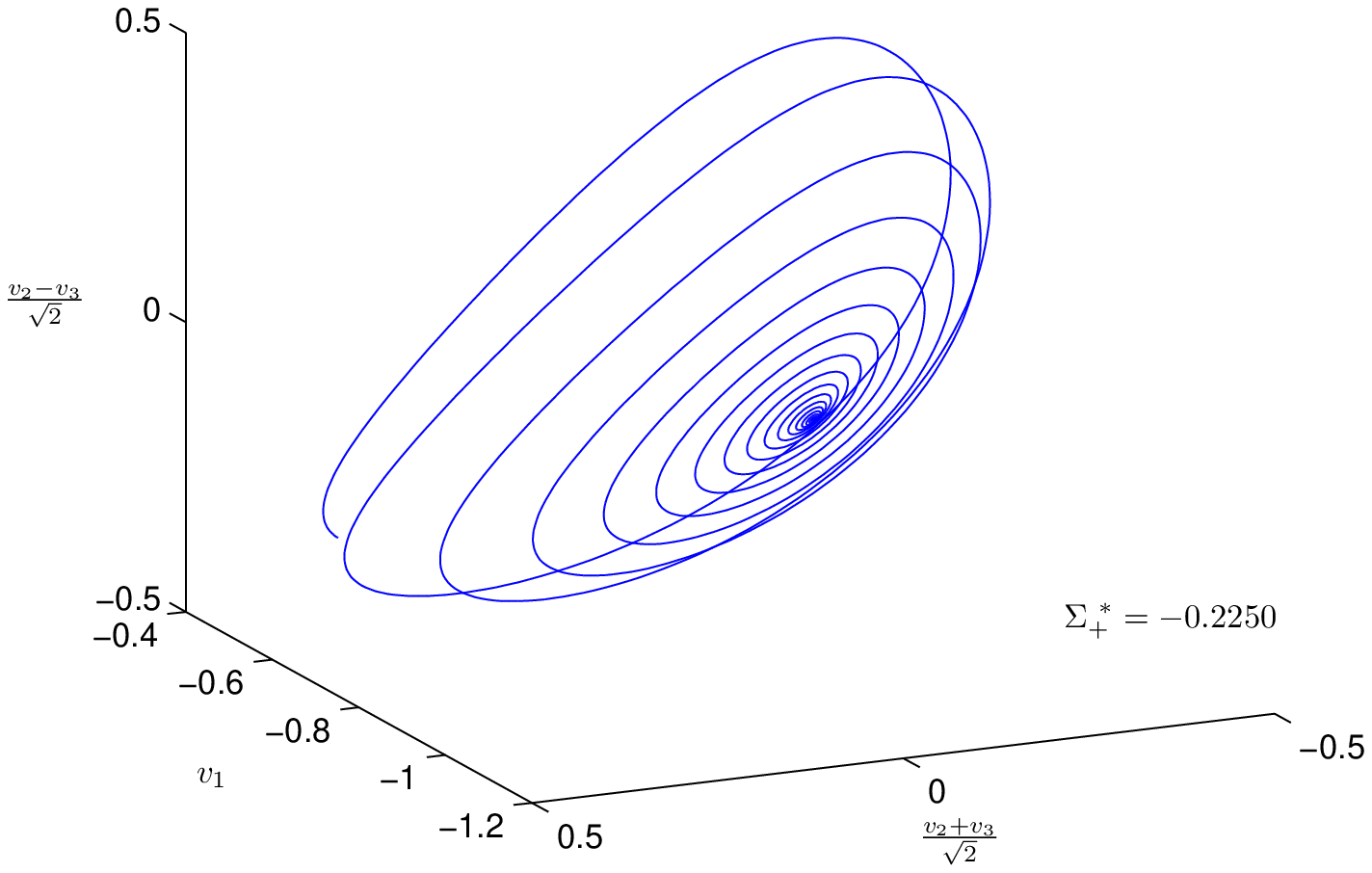}
\end{figure}

We now return to the general Bianchi type VII$_h$ models. In addition to the analytical results stated above, we have done a comprehensive numerical analysis in the full 7-dimensional state space which is summarised in the Figures to follow. 
The above results, and from the comprehensive  numerical analysis, indicate that the nature of the late-time attractors, as we change the parameter $\lambda^*$ from $\lambda^*=1$ to $\lambda^*>1$, change according to 

\beq
\text{IV} & \longmapsto & \text{VII}_h \nonumber \\
\mathcal{L}(IV) & \longmapsto & \mathcal{L}(VII_h)\quad  \text{ (equilibrium point)}\nonumber \\
\widetilde{\mathcal{L}}_-(IV) & \longmapsto & \widetilde{\mathcal{L}}(VII_h) \quad\text{ (equilibrium point)}\nonumber \\
\widetilde{\mathcal{F}}_-(IV) & \longmapsto & \widetilde{\mathcal{F}}(VII_h)\quad\text{ (closed orbit)} \nonumber \\
\widetilde{\mathcal{E}}_-(IV) & \longmapsto & \widetilde{\mathcal{E}}(VII_h)\quad\text{ (closed orbit)} \nonumber \\
\mathcal{C}(IV) & \longmapsto & \mathcal{T}(VII_h) \quad\text{ (torus)}\nonumber
\eeq
These closed curves, and the torus, are illustrated in Figures \ref{figure3A} and \ref{figure3C}. 
Analysis indicate that the stability of these attractors is preserved under the transition $\text{IV}\longmapsto \text{VII}_h$. For the curve $\widetilde{\mathcal{E}}(VII_h)$, this can be shown using similar techniques as above; the region of stabilty in terms of the limiting value $\Sigma_+^*$ and $\gamma$, is exactly the same as for $\widetilde{\mathcal{E}}_-(IV)$ (see Figure \ref{Fig:MapIV}). 

However, the torus case has to be treated with care; the late-time behaviour is described by a nowhere vanishing flow on a torus. The future behaviour of such integral curves come in two classes according to the nature of the future limit sets \cite{2Ddynsys}:
\begin{enumerate} 
\item{} Rational curves: The attractor is a closed periodic curve which can be described by its homotopy class, $H$, which takes values in the fundamental group of the torus, $\pi_1(T^2)=\mathbb{Z}\times\mathbb{Z}$. 
\item{} Irrational curves: The attractor is an everywhere dense curve on the torus. 
\end{enumerate} 
The rational curves immediately lend themselves to the analysis above; if $H=(m,n)$, where $n$ is defined as the winding number of $\theta$, then $T_n=\frac{n\pi\sqrt{h} }{(1-\av{x})(1+\Sigma_+^*)}$, where $\av{x}$ is given by Theorem \ref{thm:xbar}.  
For an irrational curve, $c(\tau)$, we can utilize the following fact. For any $\tau_0$ and  $\delta>0$, there exists a closed curve $\tilde{c}(\tau)$ with period $T_n$ (not an integral curve) such that $|c(\tau)-\tilde{c}(\tau)|<\delta/2$ for $\tau_0<\tau<\tau_0+T_n$. In particular, this means that $|c(\tau_0)-c(\tau_0+T_n)|<\delta$, which implies that the limiting curve $c(\tau)$ can be arbitrary close to a closed curve. This implies that we can get arbitrary close to the values given in Theorem \ref{thm:xbar} and hence, we get similar restrictions on the limit cycle. The stability analysis for the variable $\Omega$ can then  be applied in a similar manner. 
Note that for most parameter values in the type VII$_h$ loophole we expect these irrational curves to be attractors. Since the flow on the torus is described continuously by the parameters $\gamma$ and $\Sigma^*_+$, we would expect that locally the direction of the flow takes all real values in a certain interval. Since the rational numbers only form a set of measure zero we expect that ``most'' curves will be irrational. 

\section{Conclusion}
We have used a dynamical systems approach and 
a detailed numerical analysis to analyse the late-time behaviour of tilting 
Bianchi models of types IV and  VII$_h$. We established the
equations of motion in the `$N$-gauge' and we found all of the equilibrium points 
and 
investigated their stability, with an emphasis on the vacuum plane-wave 
spacetimes.
In particular, it was shown that for $2/3<\gamma<2$ there will always be future
stable plane-wave solutions in the set of type IV and VII$_h$
tilted Bianchi models and it was proven that the {\em only} future
attracting equilibrium points for non-inflationary fluids
($\gamma>2/3$) are the plane-wave solutions in Bianchi type
VII$_h$ models. 

The future asymptotic behaviour of tilted $\gamma$-law perfect fluid models of 
Bianchi type 
IV and VII$_h$ can be summarised as follows\footnote{Models with $0<\gamma<2/3$ 
are 
subject to the 'no-hair' theorem (originally due to Wald for a cosmological 
constant 
\cite{Wald}). The tilted version was proven in \cite{CH2}.}:
\paragraph{Bianchi type IV:}
\begin{enumerate} 
\item{} $0<\gamma\leq 2/3$: Asymptotically FRW.
\item{} $2/3<\gamma<6/5$: Asymptotically a vacuum plane wave. Tilt velocity tends 
to zero. 
\item{} $6/5<\gamma<4/3$: Asymptotically a vacuum plane wave. Tilt velocity is 
asymptotically  zero, intermediately tilted, intermediately tilted and oscillatory (in the 
loophole), or extreme. 
\item{} $4/3\leq\gamma<2$: Asymptotically a vacuum plane wave. Tilt velocity is 
asymptotically extreme. 
\end{enumerate}
\paragraph{Bianchi type VII$_h$:}
\begin{enumerate} 
\item{} $0<\gamma\leq 2/3$: Asymptotically FRW.
\item{} $2/3<\gamma<6/5$: Asymptotically a vacuum plane wave. Tilt velocity tends 
to zero. 
\item{} $6/5<\gamma<4/3$: Asymptotically a vacuum plane wave. Tilt velocity is 
asymptotically  zero, intermediately tilted and oscillatory (either 
$\widetilde{\mathcal{F}}(VII_
h)$ or the torus attractor), extreme, or extreme and oscillatory. 
\item{} $4/3\leq \gamma<2$: Asymptotically a vacuum plane wave. Tilt velocity is 
asymptotically extreme or extreme and oscillatory. 
\end{enumerate}
The stability of the plane-wave solutions is a key result.

A tiny region of the parameter space (the loophole) was
discovered in the  Bianchi type IV model which contains no stable
plane-wave equilibrium points. We proved the existence of a closed orbit 
in the loophole and provided criteria for its stability. With the use 
of theory and extensive numerical experimentation,
a limit cycle was found inside the loophole 
which acts as an attractor (the Mussel attractor).

We then studied the Bianchi type VII$_h$ 
models numerically. In particular, 
from a local stability analysis and extensive numerical analysis we found 
that at late times $\Omega\to 0$ and the variables 
'freeze' into their asymptotic values (in a time scale much shorter than 
the other dynamical features present).
After this freezing has occurred, the system of equations effectively decouples
making these models amenable to analytical investigations. This, in turn, showed the existence  closed curves which interestingly had a characteristic frequency $f=1/T_1=(1-\av{x})(1+\Sigma_+^*)/(\pi\sqrt{h})$. 
From a detailed numerical analysis of the type VII$_h$ models,
we then showed that there
is an open set of parameter space in which solution curves
approach a compact surface that is topologically a torus.
The stability of these late-time attractors seems to be preserved under the 
transition $\text{IV}\longmapsto \text{VII}_h$  (e.g., the closed curve in the loophole in the type IV 
models 
to the torus in the type VII$_h$ models). 
Finally, we would like to emphasise that the comprehensive numerical integrations 
presented in this paper
(in the full Bianchi state space) serve to fully justify all of the claims in 
\cite{CH2}, particularly
the 'freezing in' process, the existence of the mussel attractor
in the loophole and the existence of the torus.

\section{Acknowledgments}

This work was supported by NSERC (AC and RvdH) and the Killam Trust
and AARMS (SH).

\appendix
\section{Numerical Integrations}
\label{app:numint}
\subsection{Bianchi type IV and VII$_h$}
\subsubsection{ Initial Conditions}
The initial conditions are chosen so that the constraint equations are satisfied initially. In both the Bianchi IV and Bianchi VII$_h$ cases the initial conditions are 
chosen so that 
$\Sigma_- =0.1,\Sigma_{12}= 0.01,\Sigma_{13}=0.01,N= 0.1, A=0.1$ with $\lambda 
=1$ for Bianchi type IV, and $\lambda=2/\sqrt{3} $ for Bianchi type VII$_h$. The initial values for  $\Sigma_+$ and $\Sigma_{23}$ are chosen so that the constraint equations are satisfied.  Note how the initial value of $\Sigma_{23}$ changes as a function of $\gamma$. 

$$\begin{tabular}{|c|c|c|c|}
               \hline
               $\Sigma_+$ & \multicolumn{3}{c|}{$\Sigma_{23}$}\\
               \hline
               \ &  $\gamma=1$&$\gamma=1.325$ &$\gamma=4/3$\\
               \hline
              -0.7   & 0.48980 & 0.43387 & 0.43349\\
              -0.6   & 0.55218 & 0.51171 & 0.51144\\
              -0.5   & 0.59992 & 0.57048 & 0.57028\\
              -0.4   & 0.63632 & 0.61566 & 0.61552\\
              -0.3   & 0.66325 & 0.65009 & 0.65000\\
              -0.2   & 0.68184 & 0.67539 & 0.67534\\
              -0.1   & 0.69275 & 0.69159 & 0.69158\\
               0.0   & 0.69635 & 0.69003 & 0.68999\\
               0.1  & 0.69275 & 0.68016 & 0.68008\\
               0.2  & 0.68184 & 0.66260 & 0.66247\\
               0.3  & 0.66325 & 0.63674 & 0.63657\\
               0.4  & 0.63632 & 0.60154 & 0.60131\\
               0.5  & 0.59992 & 0.55520 & 0.55490\\
               0.6  & 0.55218 & 0.49461 & 0.49422\\
               0.7  & 0.48980 & 0.41357 & 0.41304\\ 
                 \hline
\end{tabular}$$

\subsubsection{Type IV: Numerical Integrations}
\begin{figure}[th] 
\caption{Type IV: The figure below displays the future dynamical evolution for $\gamma=1$.  The asterisk indicates the initial condition.  Note how $\Omega\to 0$ and $V^2\to 0$ but $\Sigma_+\not \to 0$.  In this case, the local sink is $ {\mathcal L}(IV)$.} \label{figure(g=1)}
\includegraphics*[scale = 0.50,angle= 90]{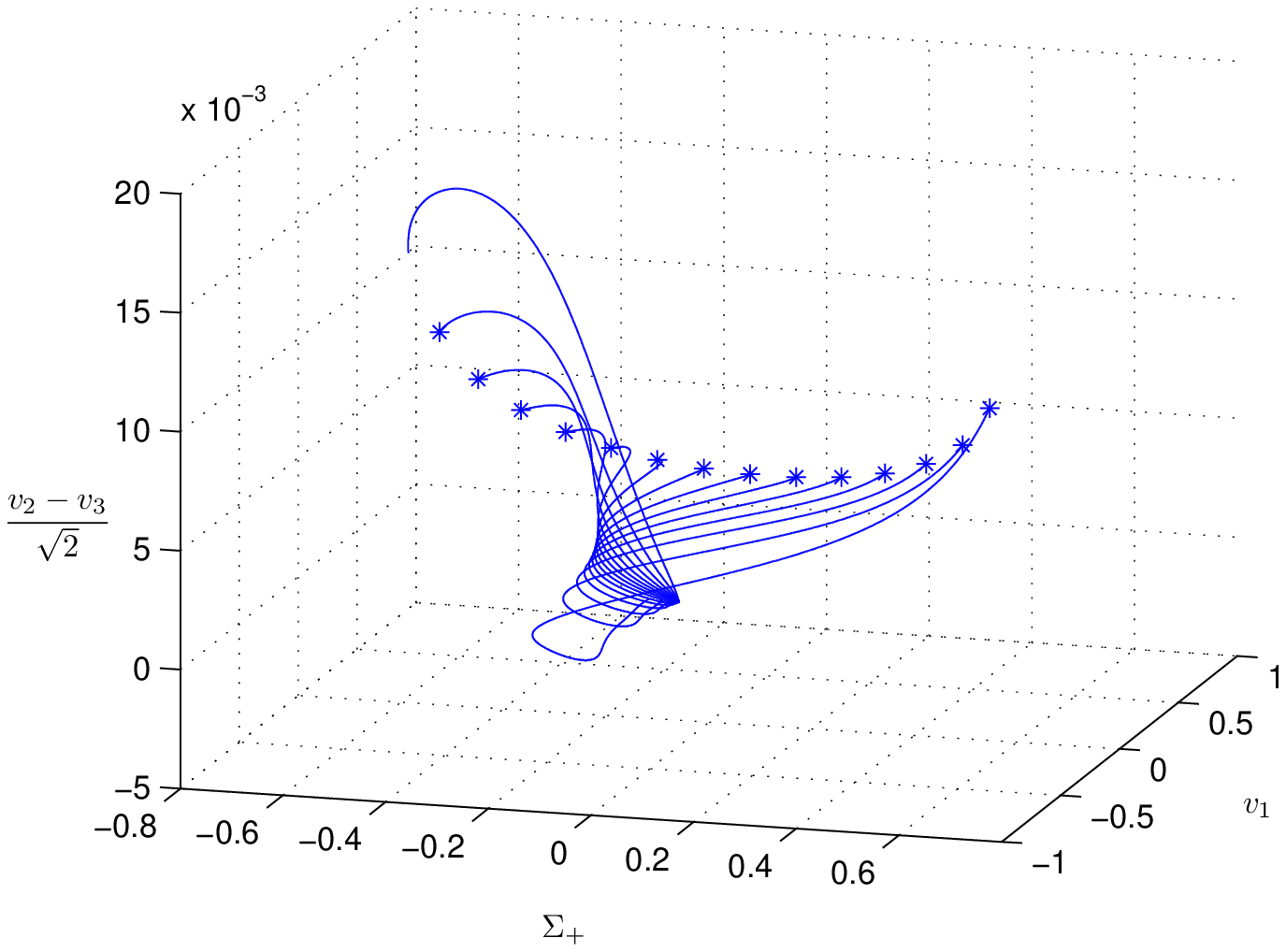}
\includegraphics*[scale = 0.50,angle= 90]{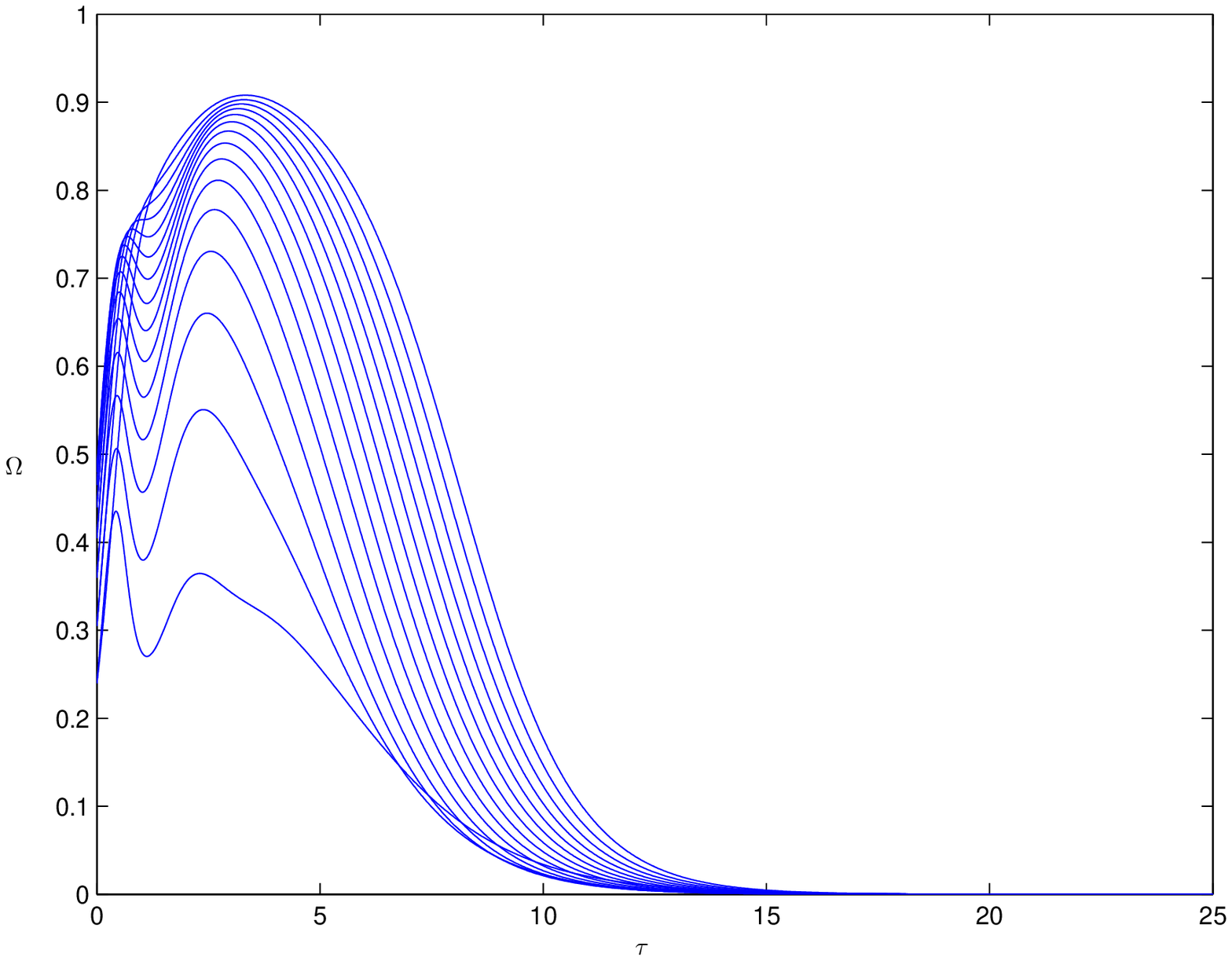}
\linebreak
\linebreak
\includegraphics*[scale = 0.50,angle= 90]{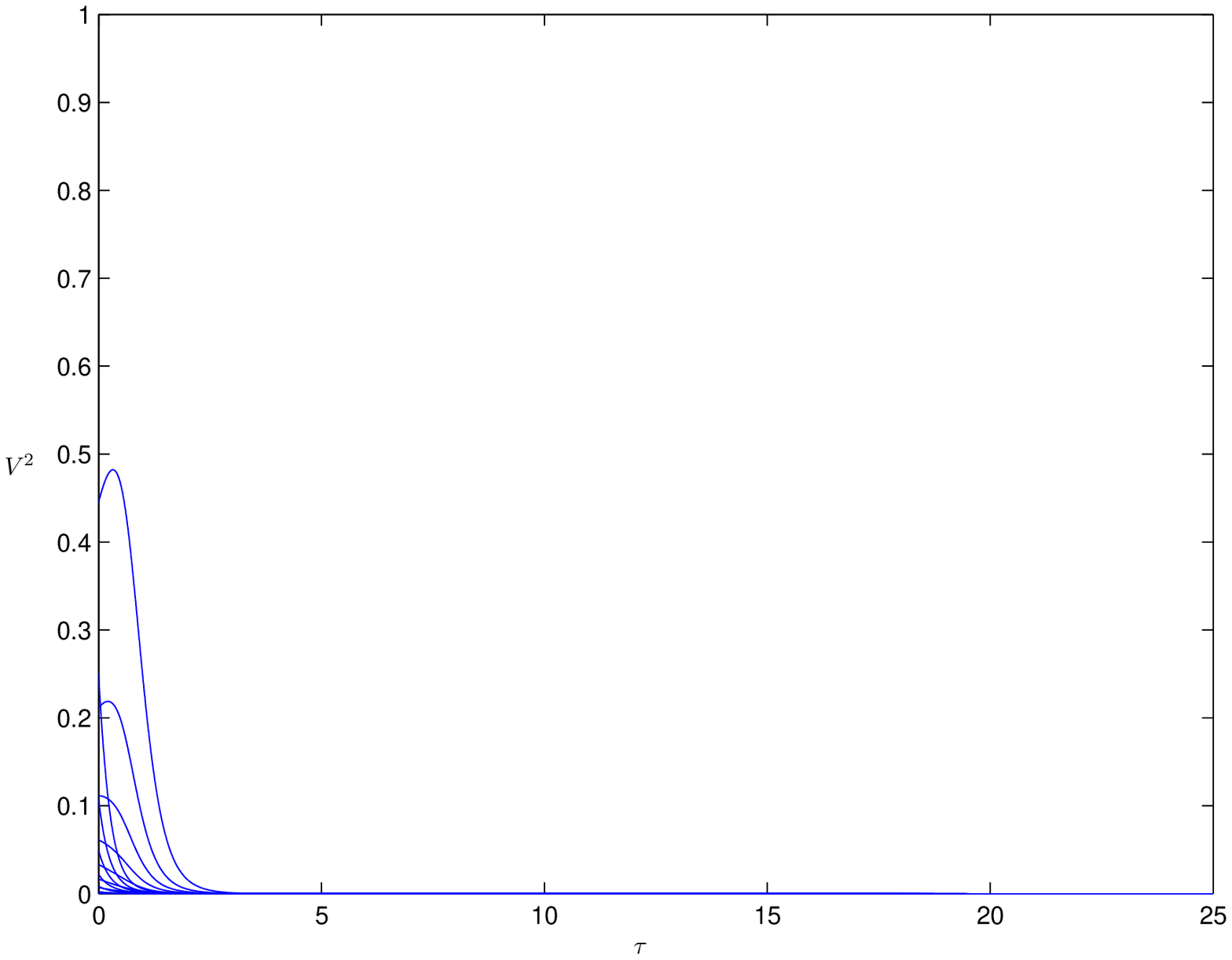}
\includegraphics*[scale = 0.50,angle= 90]{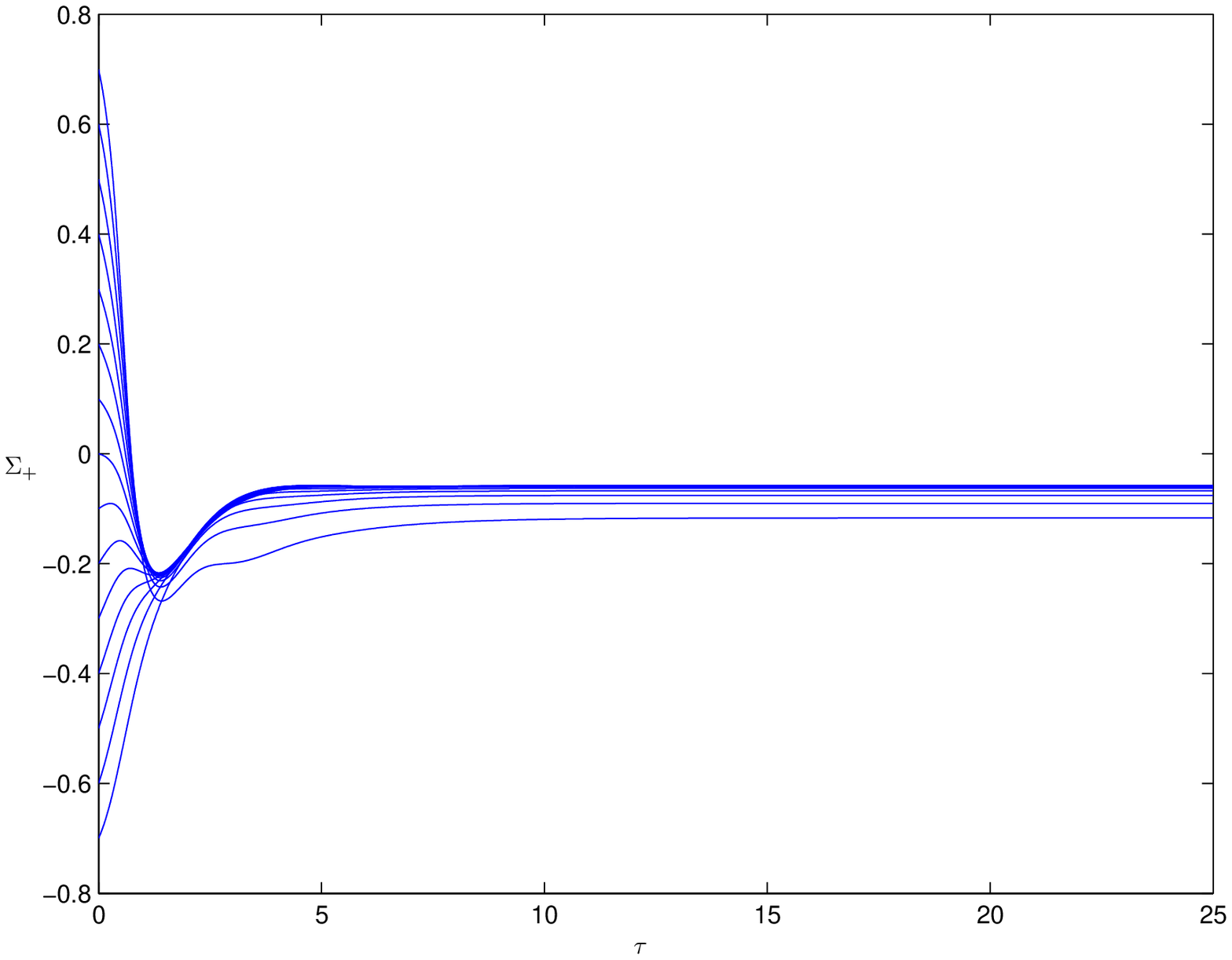}
\end{figure}

\begin{figure} 
\caption{Type IV: The figure below displays the future dynamical evolution for $\gamma=1.325$.  The asterisk indicates the initial condition.  Note how $\Omega\to 0$ and $V^2$ intermediately approaches 1,  but $\Sigma_+\not \to 0$. Also note the timescale.} \label{figure(g=1.325a)}
\includegraphics*[scale = 0.50,angle= 90]{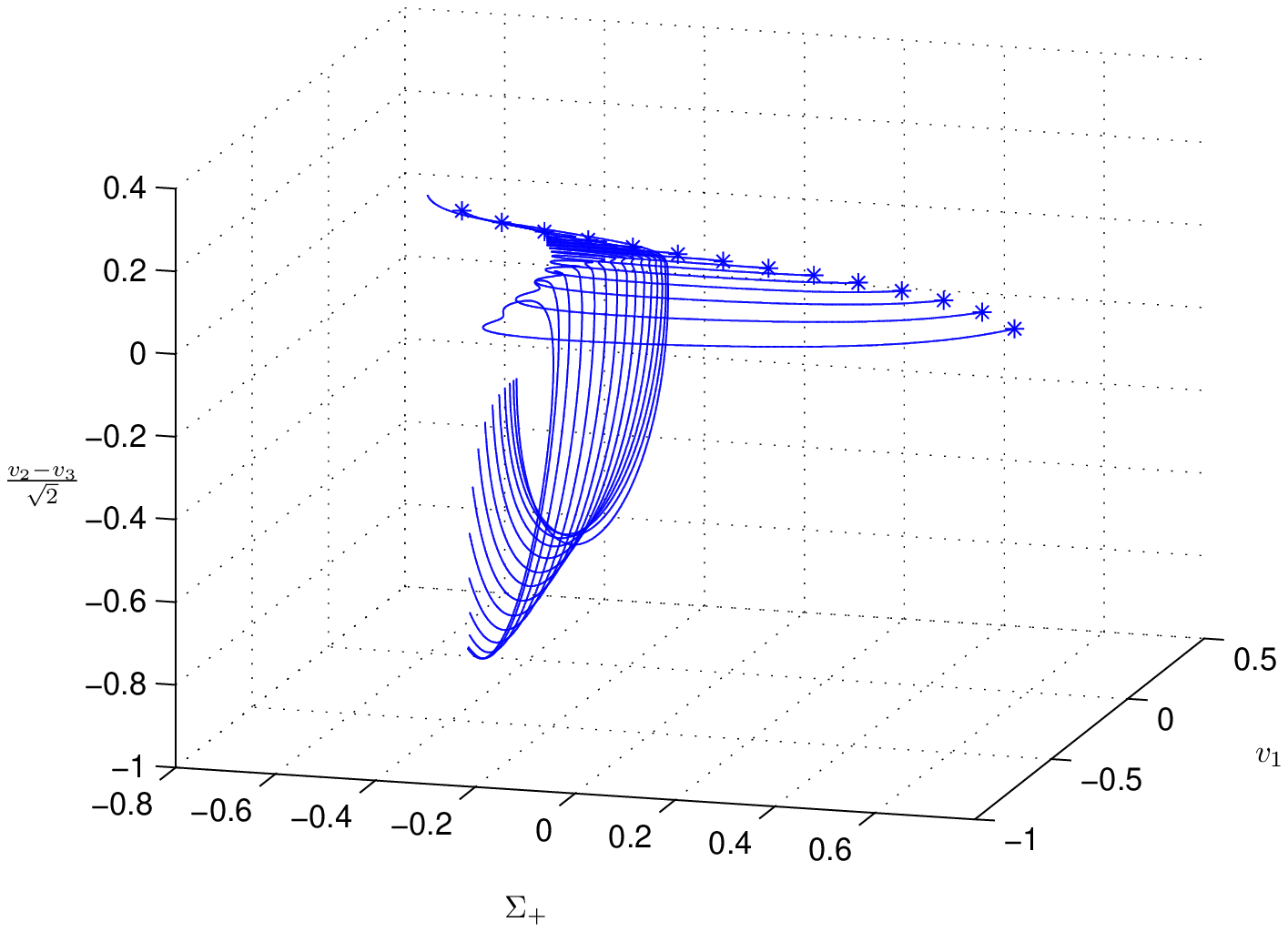}
\includegraphics*[scale = 0.50,angle= 90]{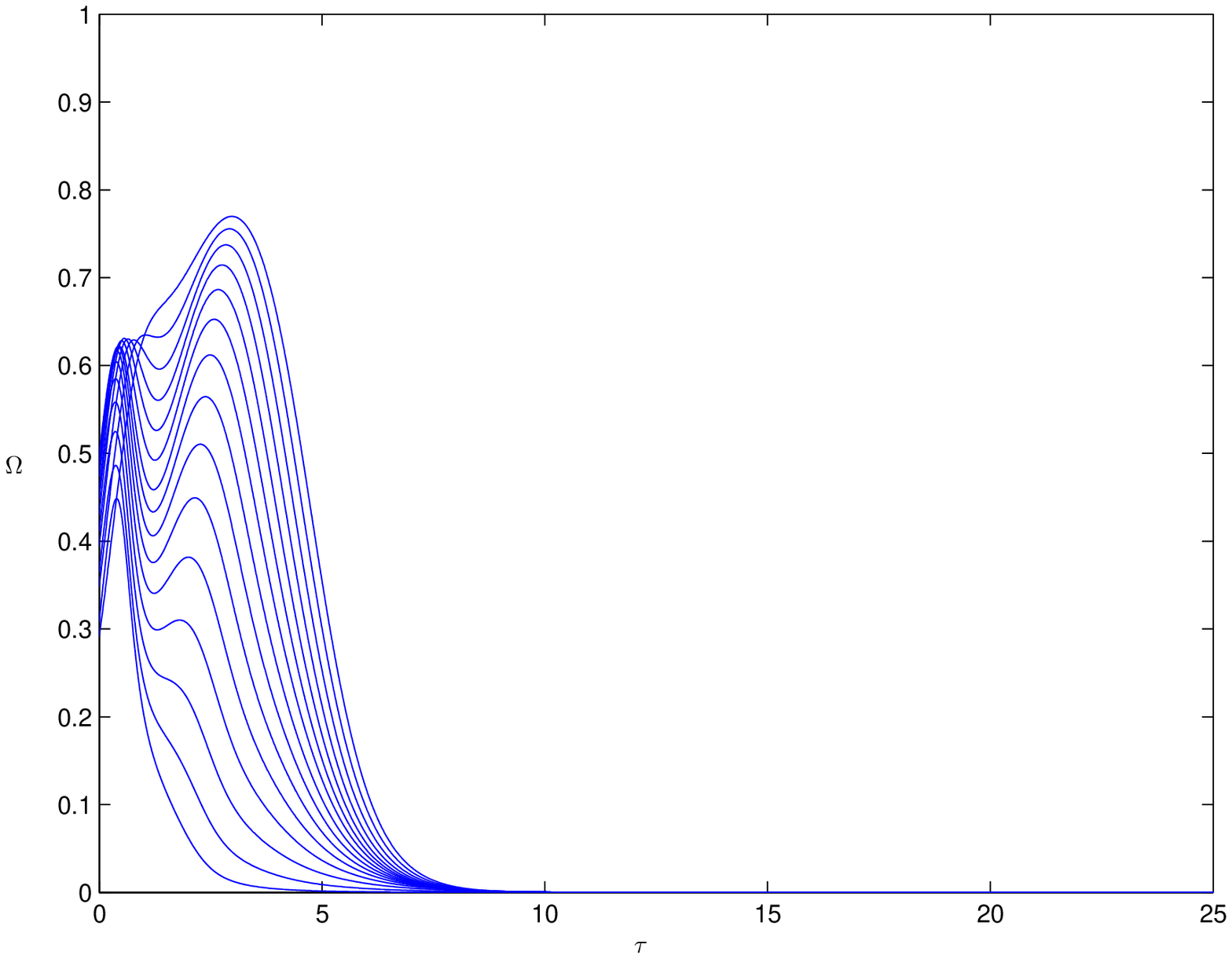}
\linebreak
\linebreak
\includegraphics*[scale = 0.50,angle= 90]{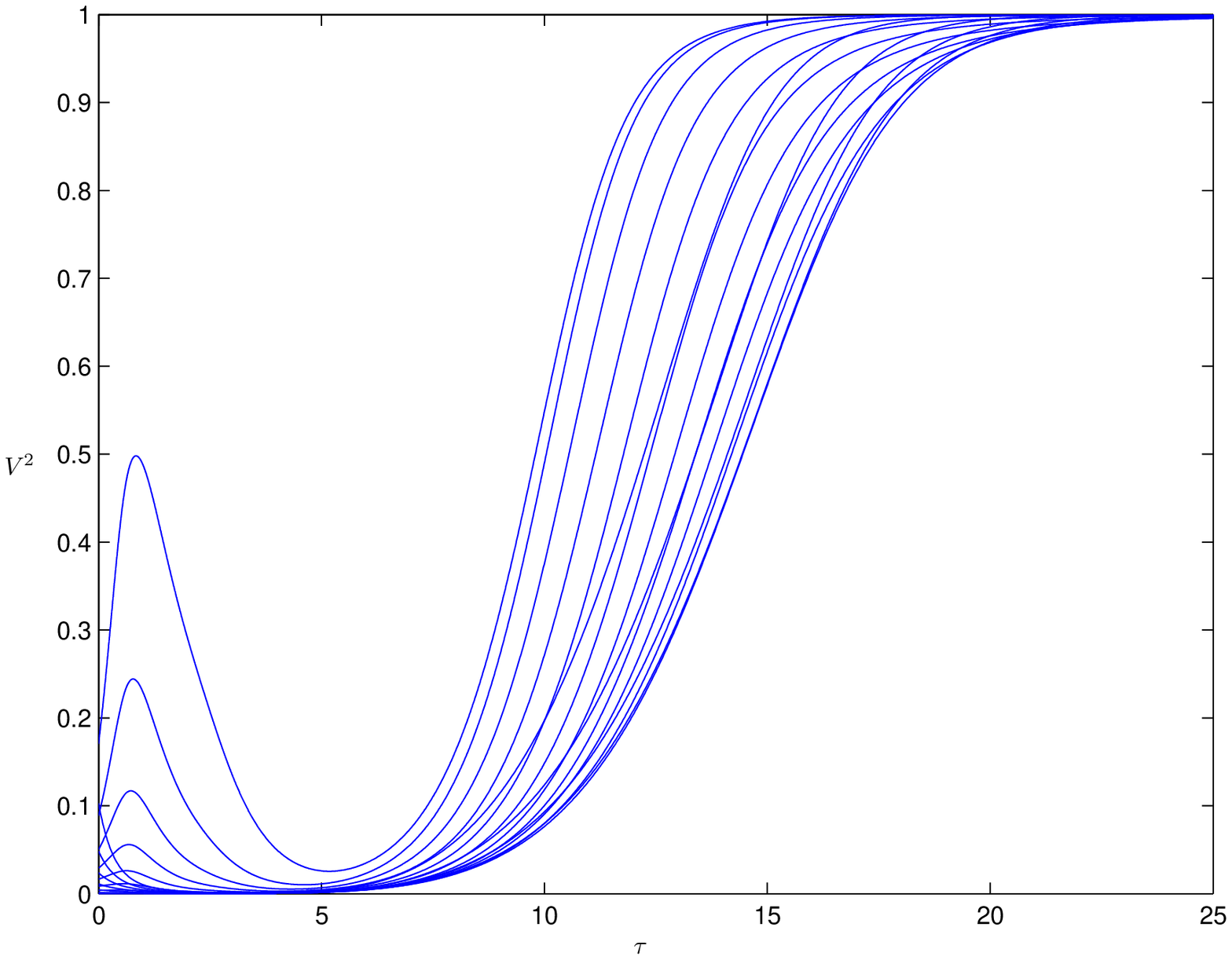}
\includegraphics*[scale = 0.50,angle= 90]{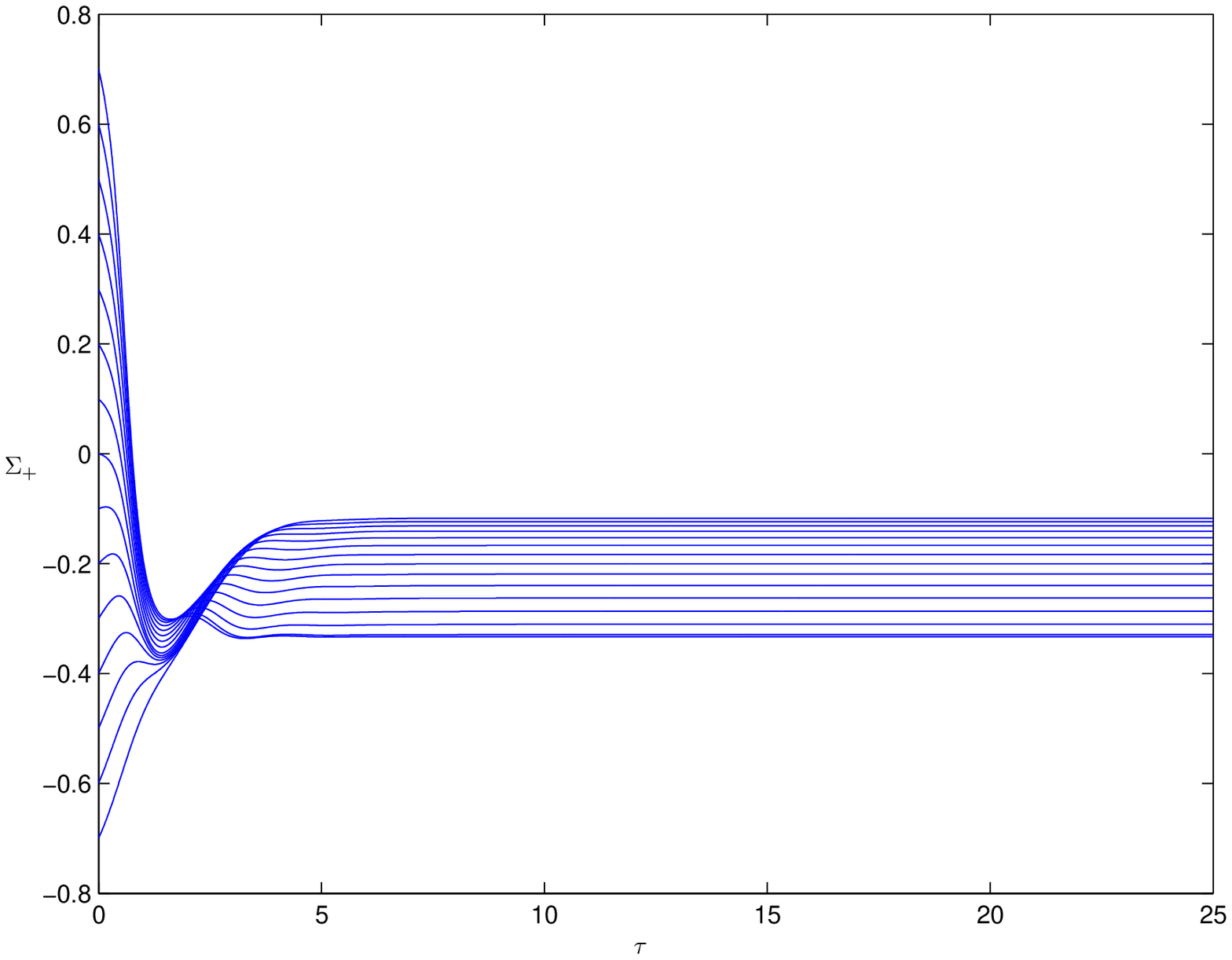}
\end{figure}

\begin{figure} 
\caption{Type IV: The figure below displays the future dynamical evolution for $\gamma=1.325$.  The timescale has been increased.  The asterisk indicates the initial condition.  Note how $\Omega\to 0$ and $V^2\to  1$ or $V^2$ oscillates, but $\Sigma_+\not \to 0$. Further analysis of this "exotic" behavior is described in the text.  In this case, the local sinks are $\widetilde{\mathcal E}(IV)$, $\widetilde{\mathcal F}(IV)$, $\widetilde{\mathcal C}(IV)$ and $\widetilde{\mathcal L}_-(IV)$.  If we look closely, one closed loop $ {\mathcal C}(IV)$ can be seen.} \label{figure(g=1.325b)}
\includegraphics*[scale = 0.50,angle= 90]{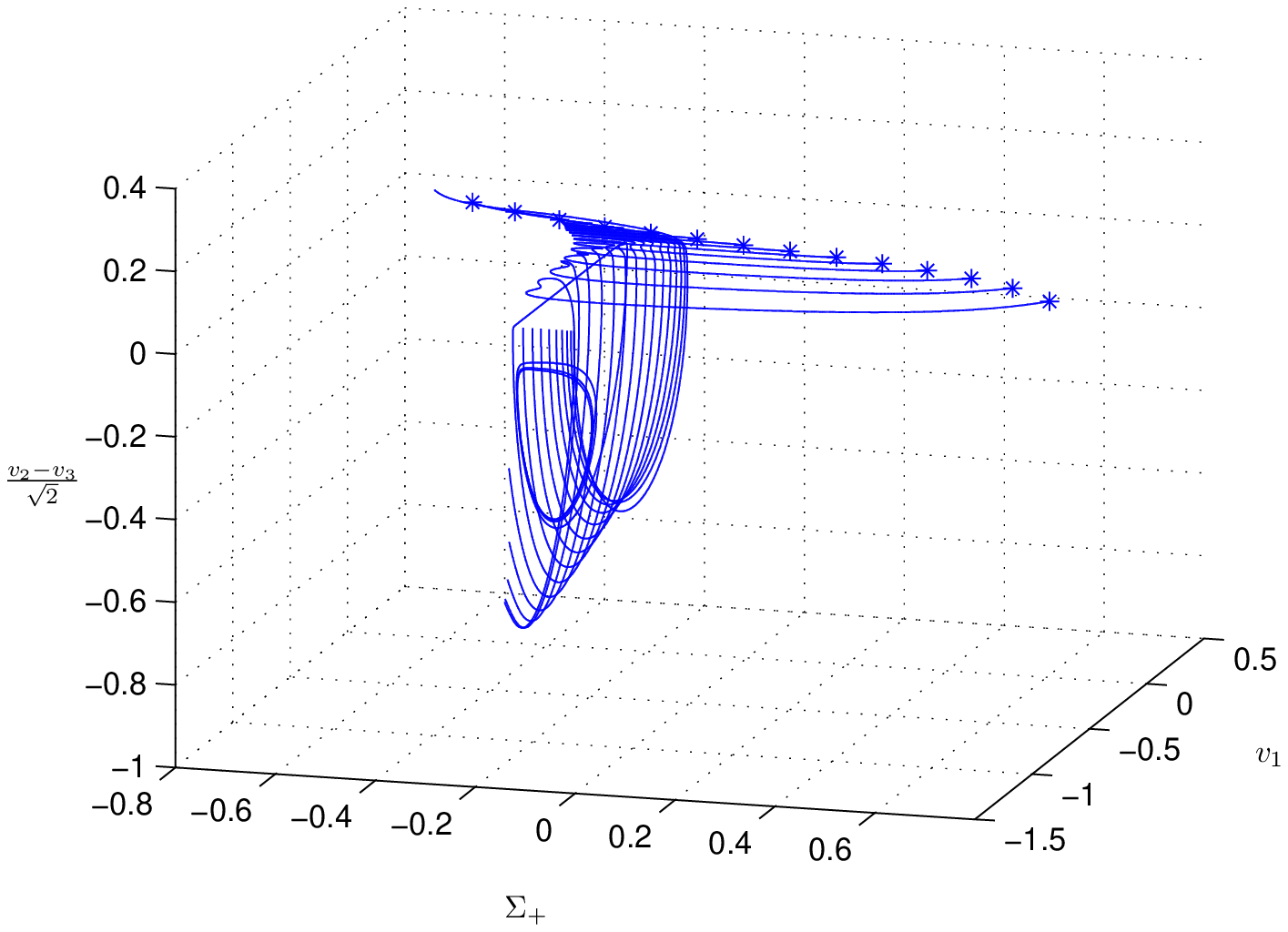}
\includegraphics*[scale = 0.50,angle= 90]{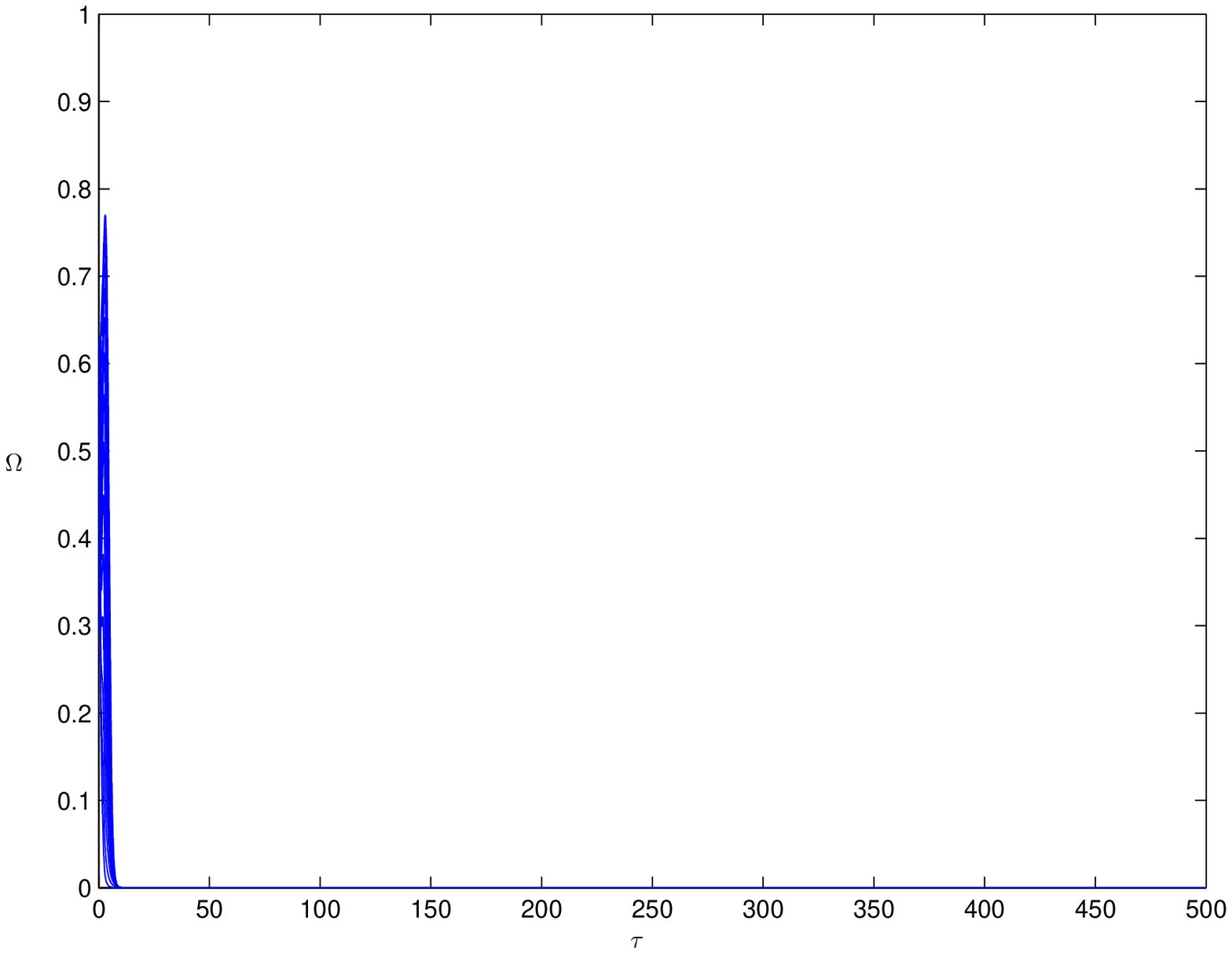}
\linebreak
\linebreak
\includegraphics*[scale = 0.50,angle= 90]{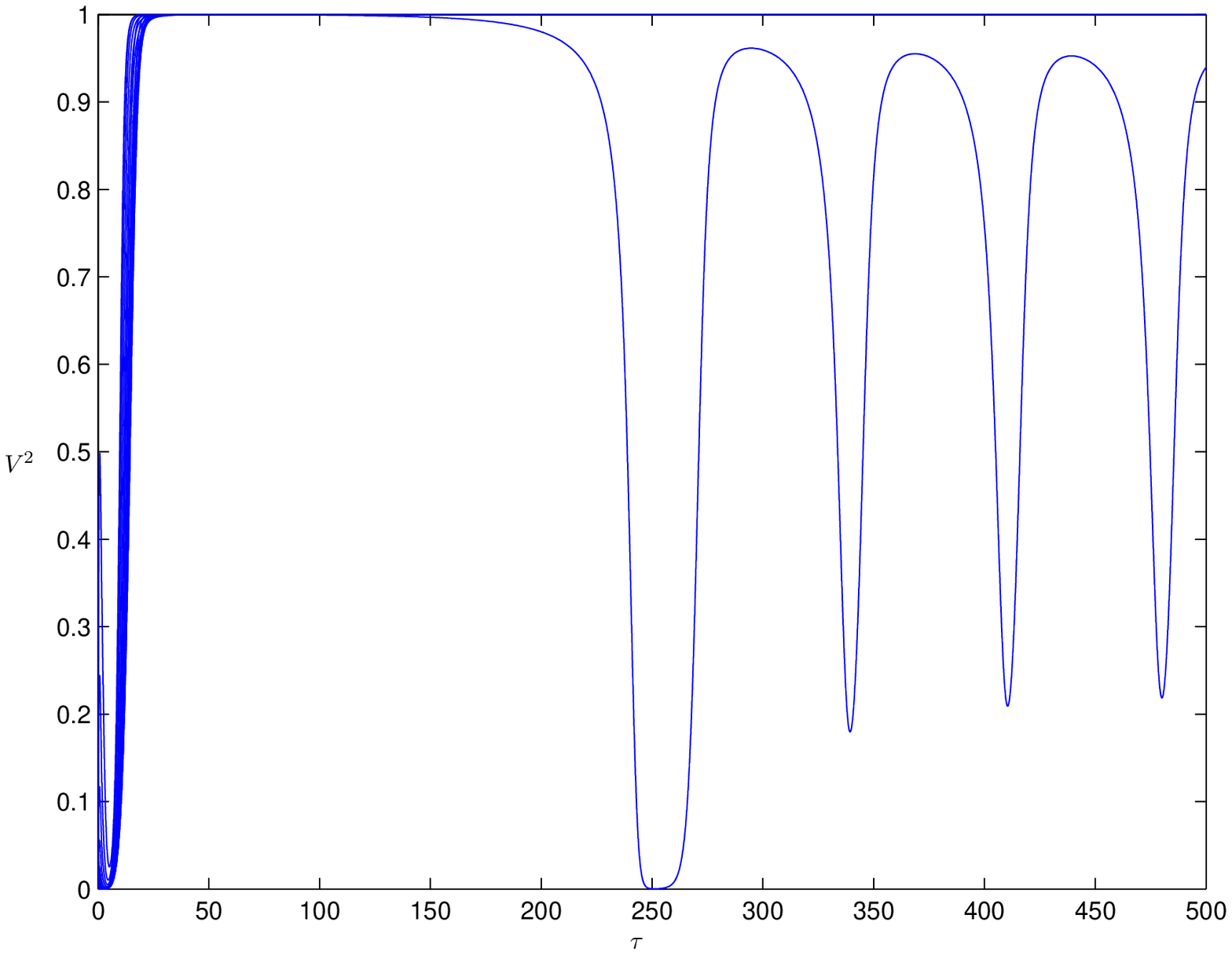}
\includegraphics*[scale = 0.50,angle= 90]{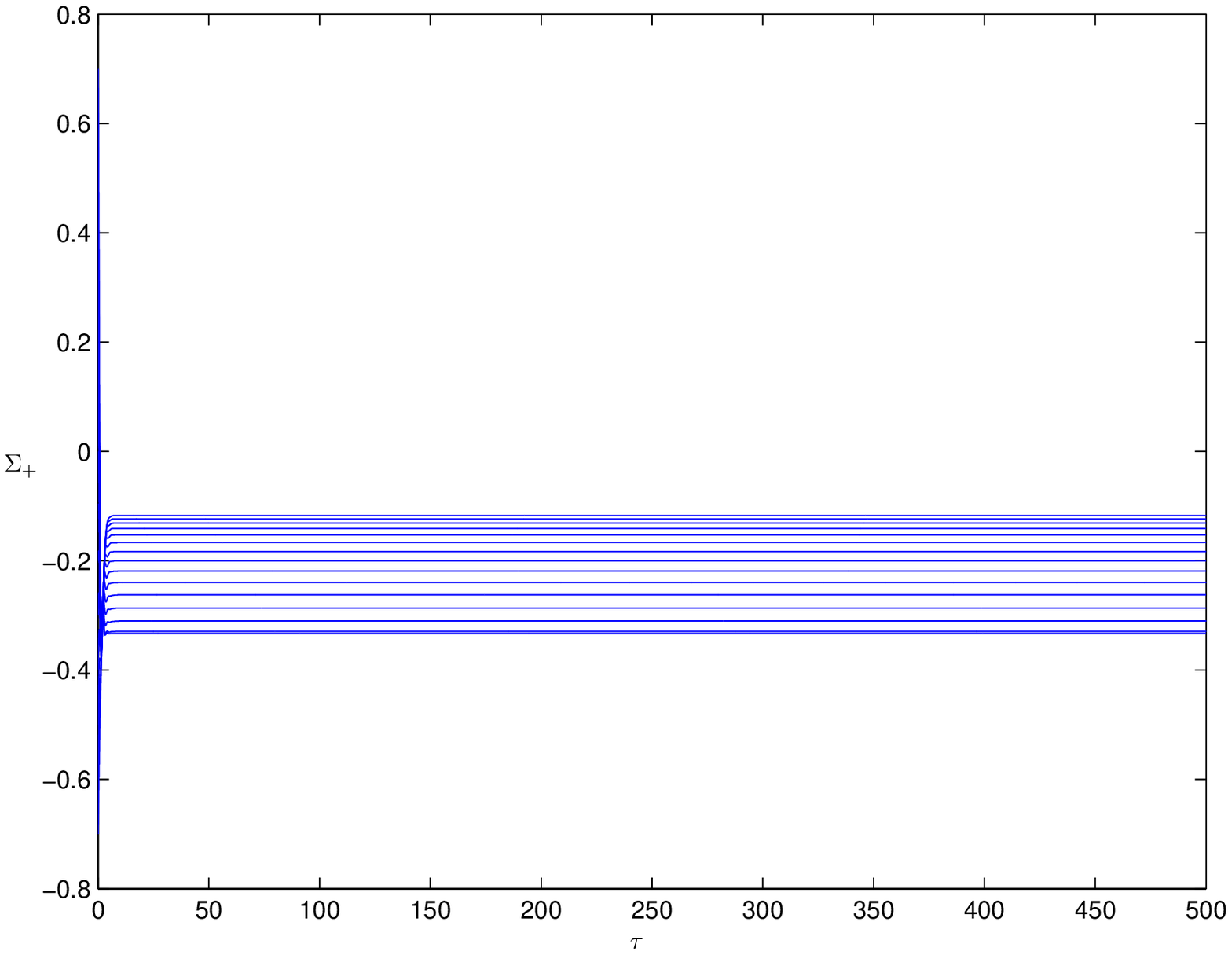}
\end{figure}

\begin{figure} 
\caption{Type IV: The figure below displays the future dynamical evolution for $\gamma=4/3$.  The asterisk indicates the initial condition.  Note how $\Omega\to 0$ and $V^2\to 1$, but $\Sigma_+\not \to 0$.  In this case, the local sinks are $\widetilde{\mathcal E}(IV)$ and $\widetilde{\mathcal L}_-(IV)$.} \label{figure(g=4/3)}
\includegraphics*[scale = 0.50,angle= 90]{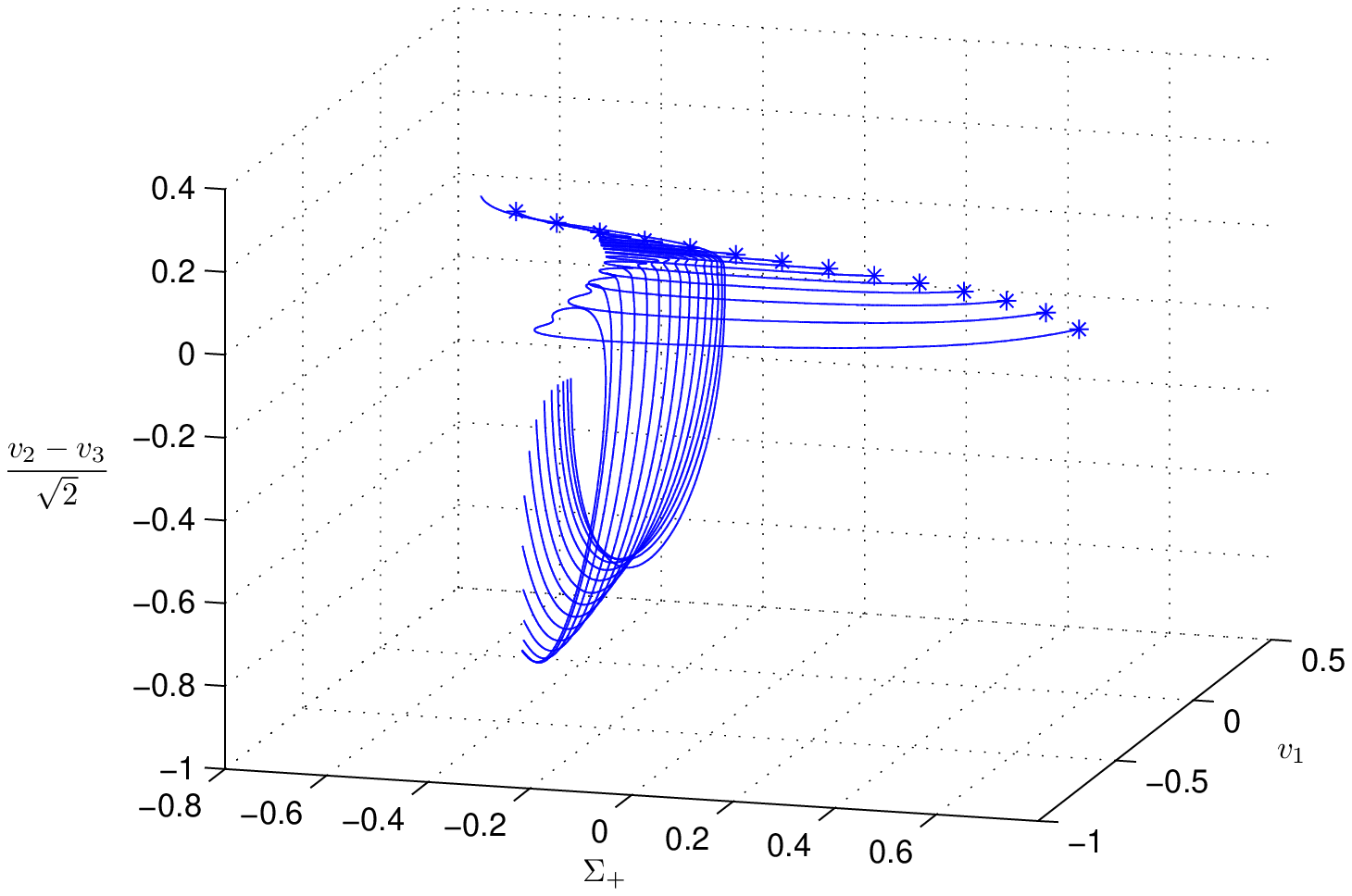}
\includegraphics*[scale = 0.50,angle= 90]{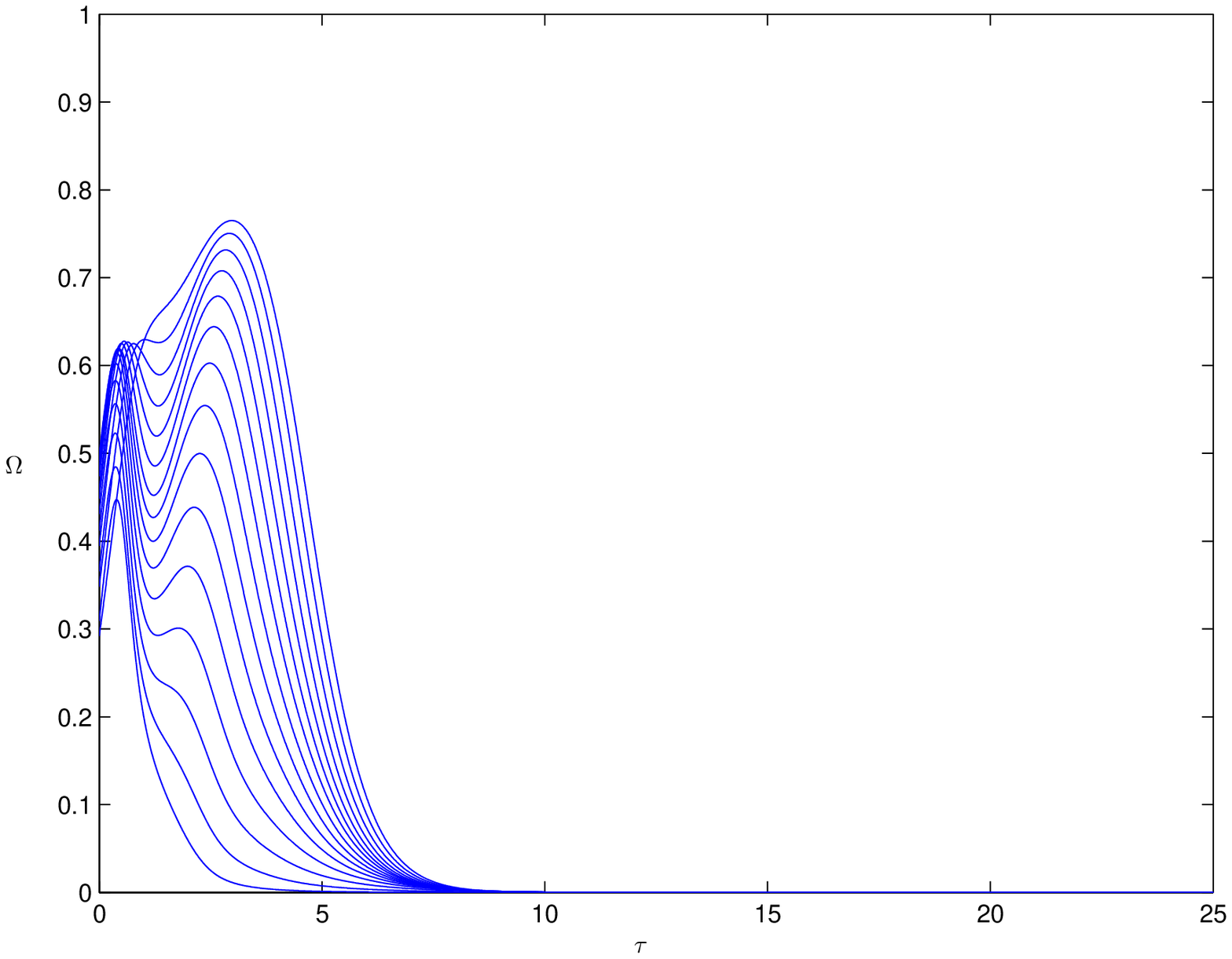}
\linebreak
\linebreak
\includegraphics*[scale = 0.50,angle= 90]{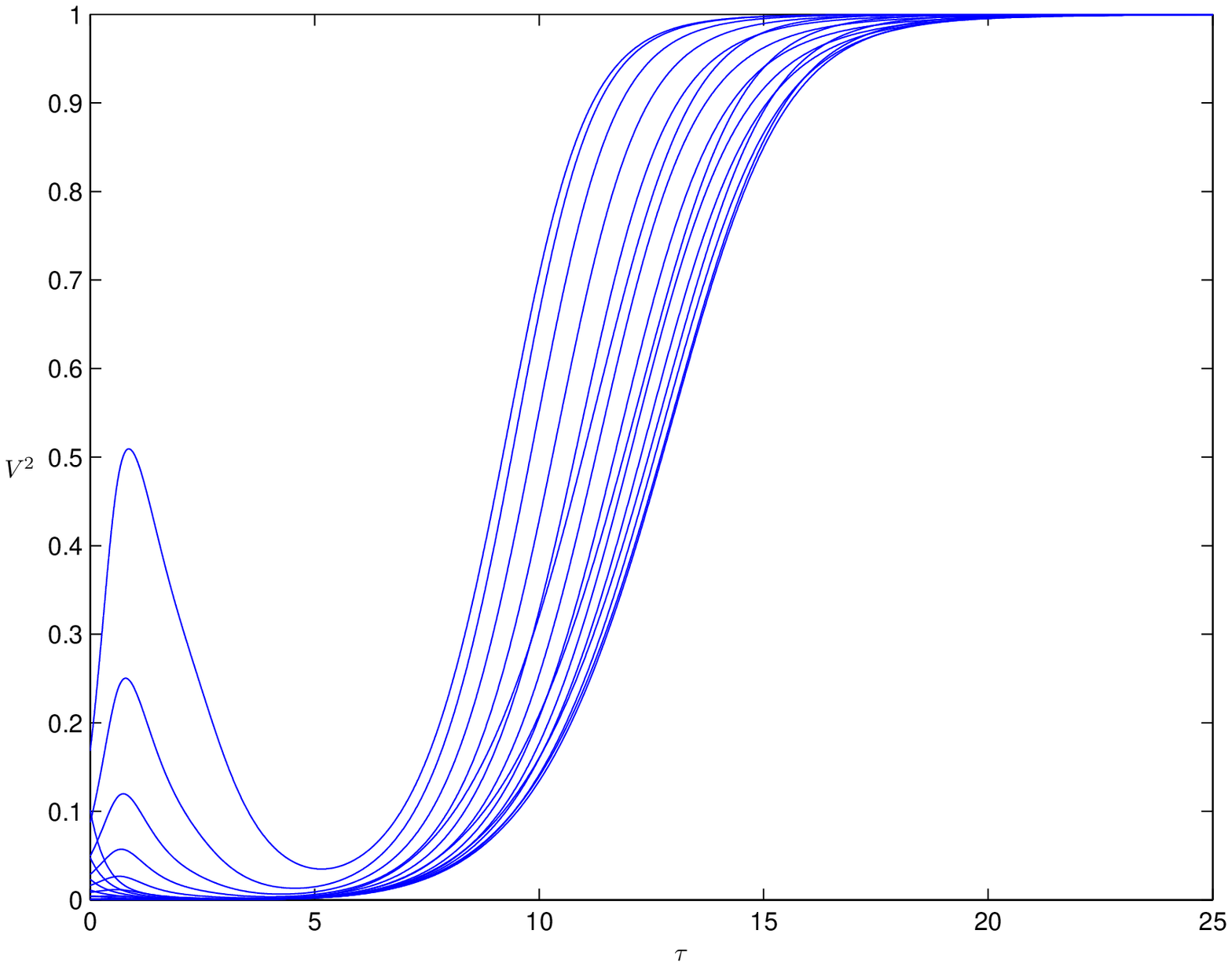}
\includegraphics*[scale = 0.50,angle= 90]{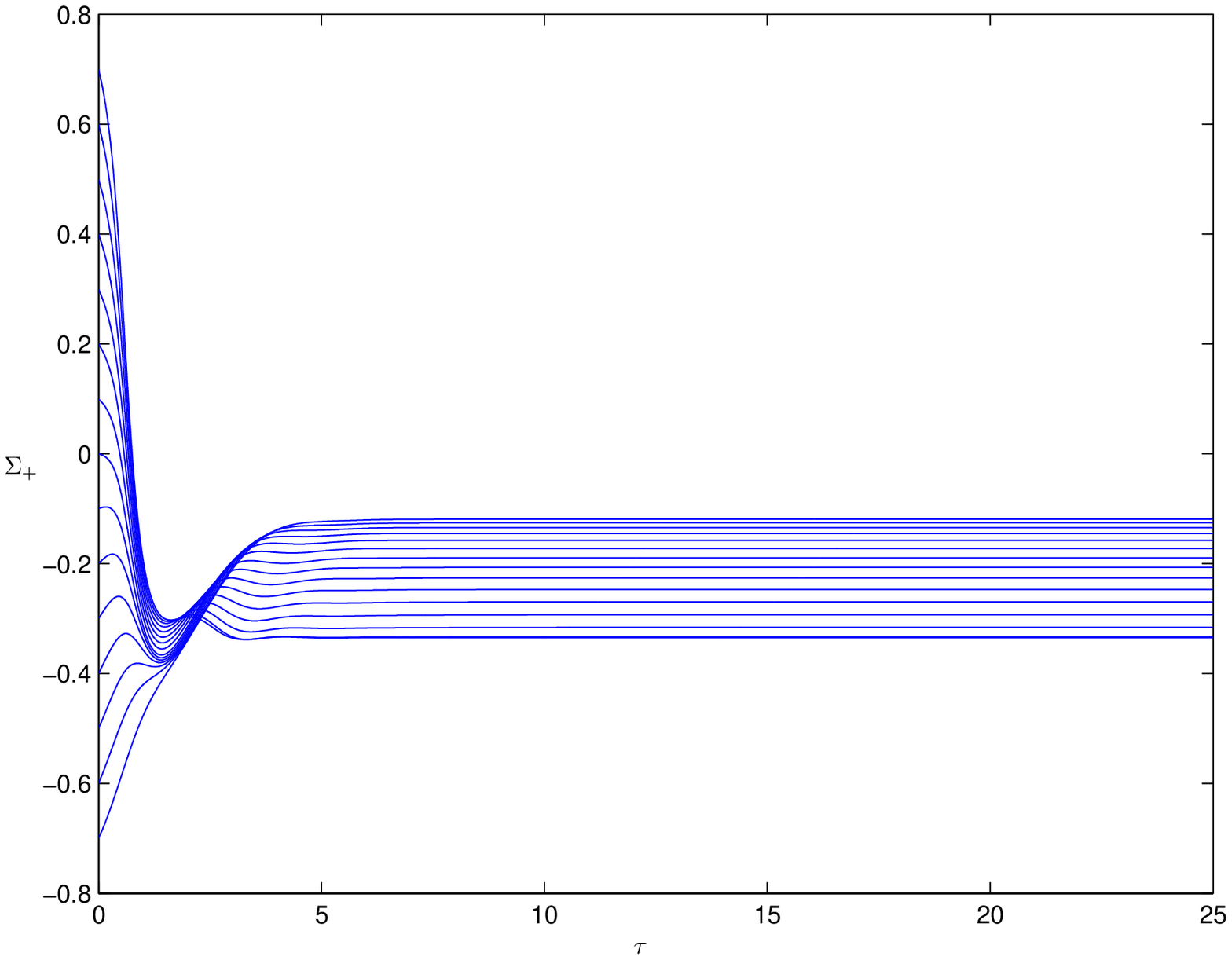}
\end{figure}

For completeness we present numerical integrations of the dynamical system for the cosmologically interesting values of $\gamma$ equal to $1$ ({\em dust}), $1.325$ ({\em a dust/radiation mixture}) and $4/3$ ({\em radiation}).  Figures \ref{figure(g=1)} to \ref{figure(g=4/3)} depict some of the results of this integration. The integrations were done over sufficiently long time intervals $[0,800]$, but the plots given here are for time intervals $[0,25]$.

\subsubsection{Type VII$_h$: Numerical Integrations}

\begin{figure} 
\caption{Type VII$_h$:  The figure below displays the dynamical behavior for 
$\gamma=1$.  Note how $\Omega\to 0$ and $V^2\to 0$ but $\Sigma_+\not \to 0$.  
In this case, the local sink is $ {\mathcal L}(VII_h)$.} \label{figure(VIIg=1)}
\includegraphics*[scale = 0.65,angle= 90]{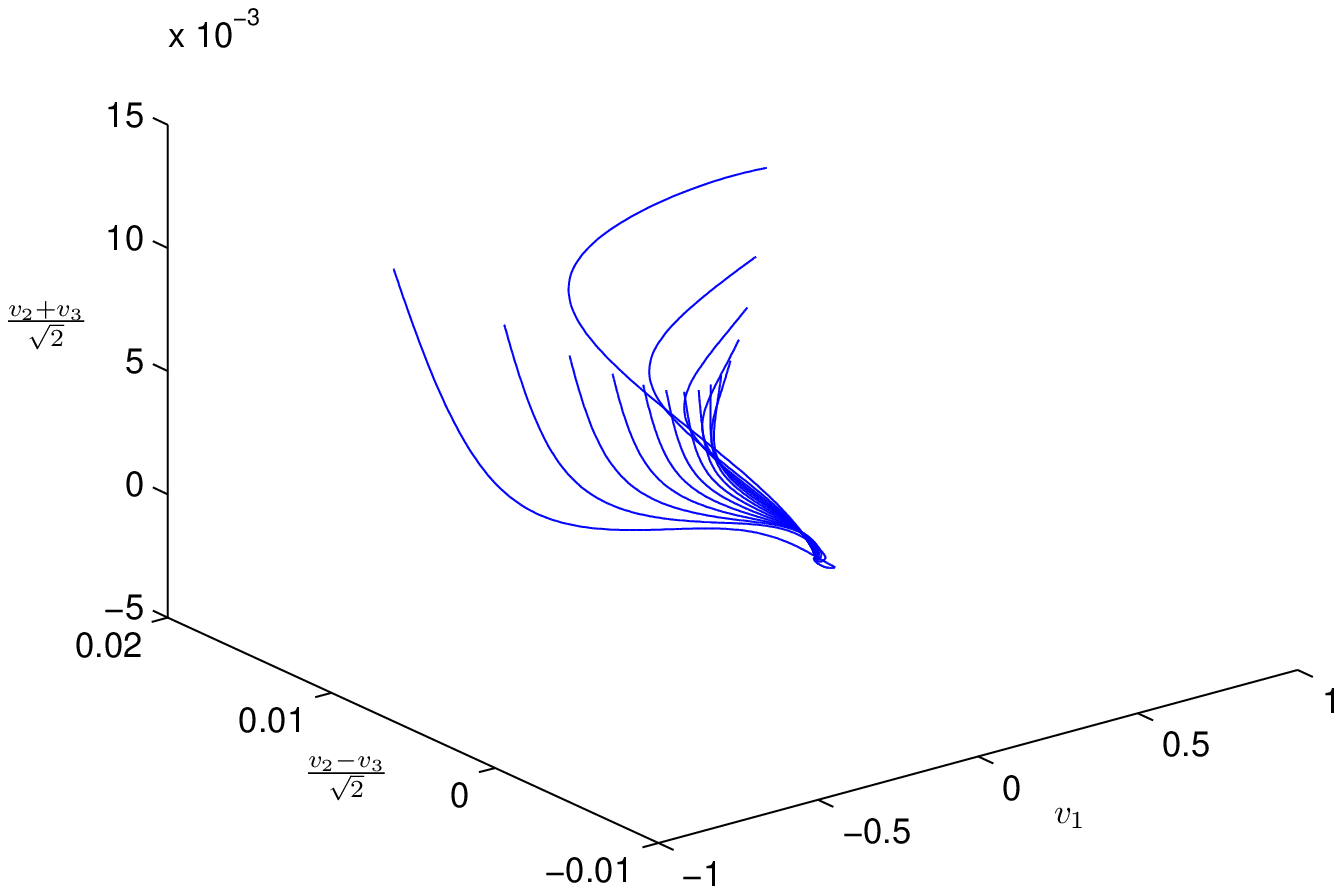}
\includegraphics*[scale = 0.65,angle= 90]{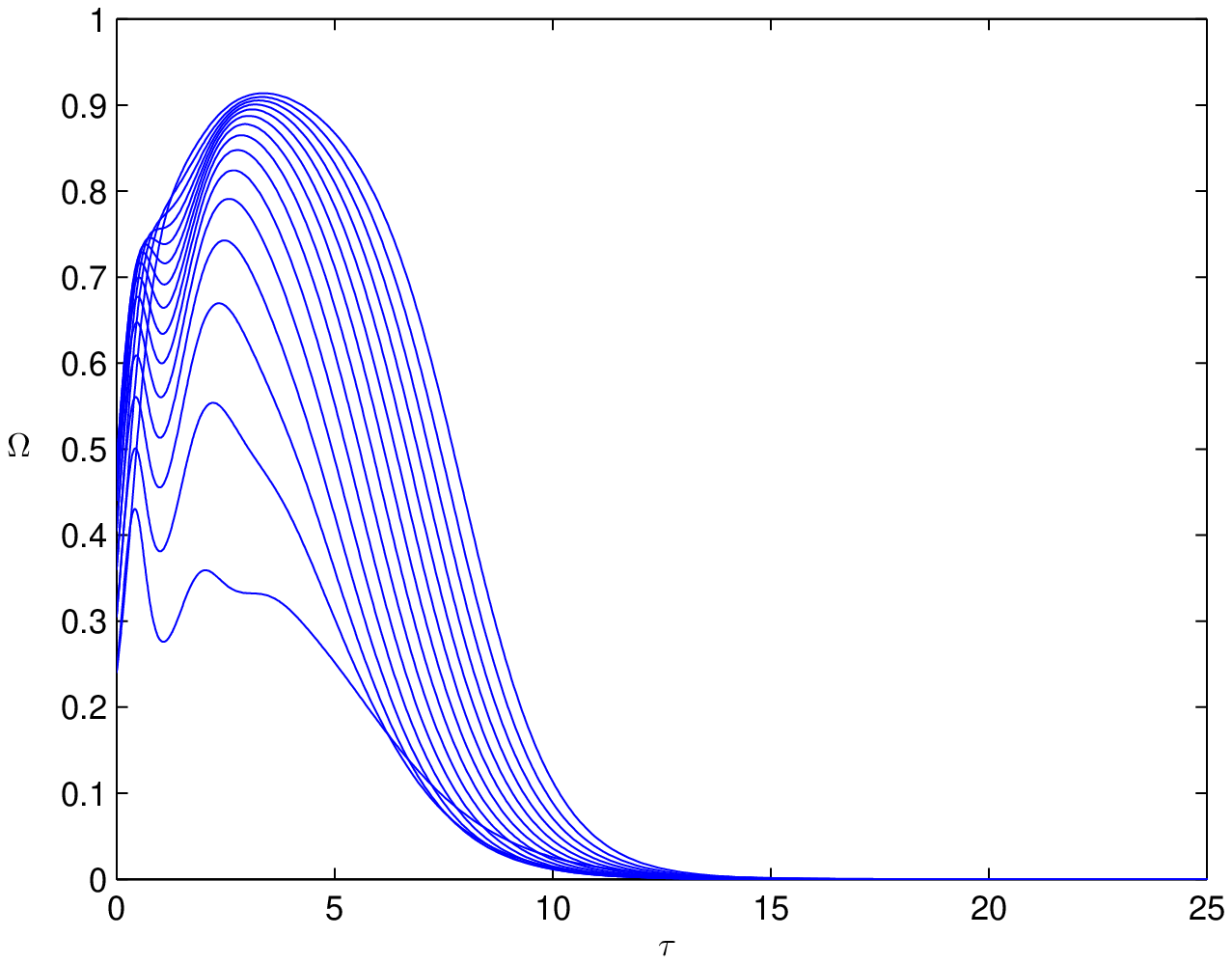}
\linebreak
\linebreak
\includegraphics*[scale = 0.65,angle= 90]{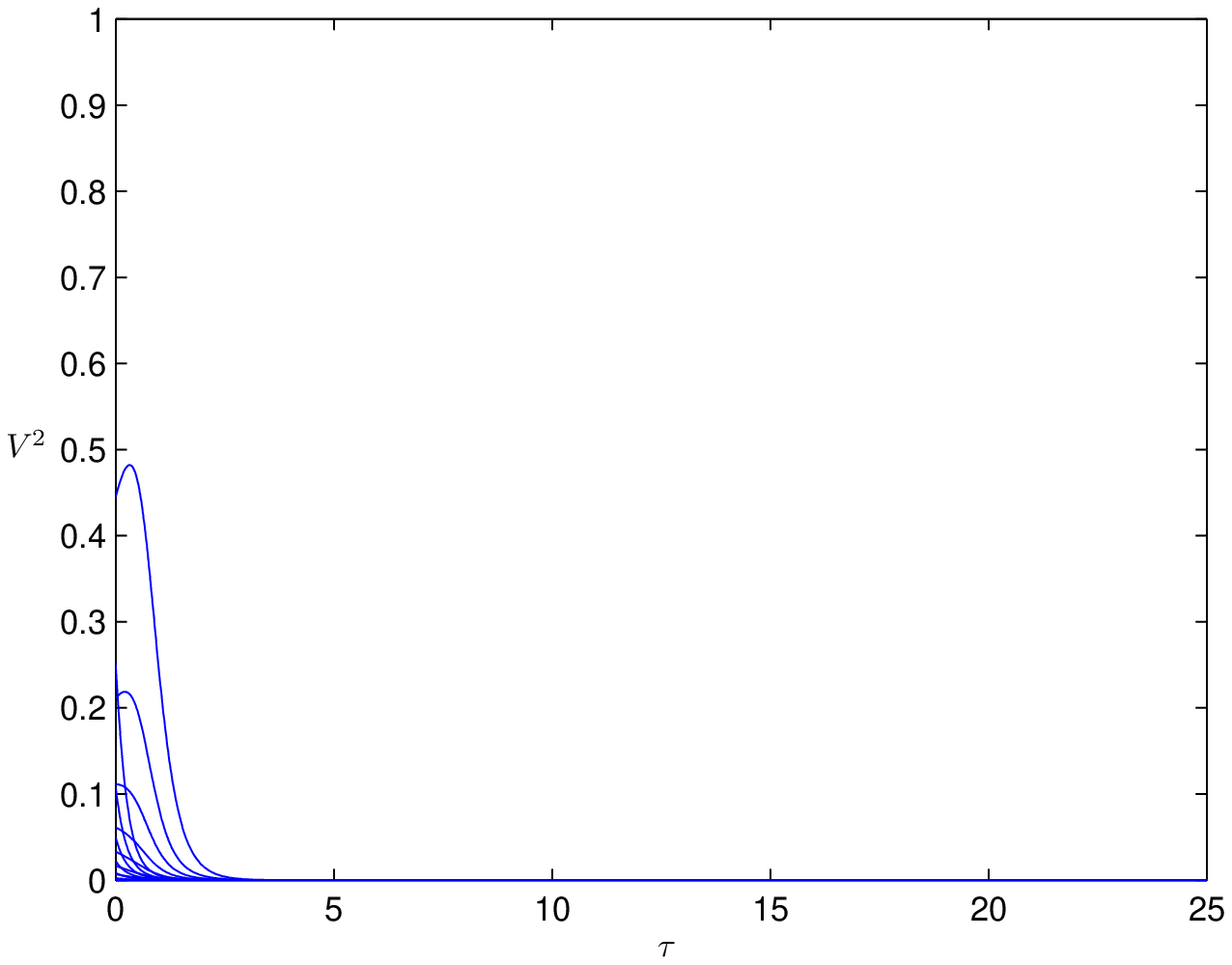}
\includegraphics*[scale = 0.65,angle= 90]{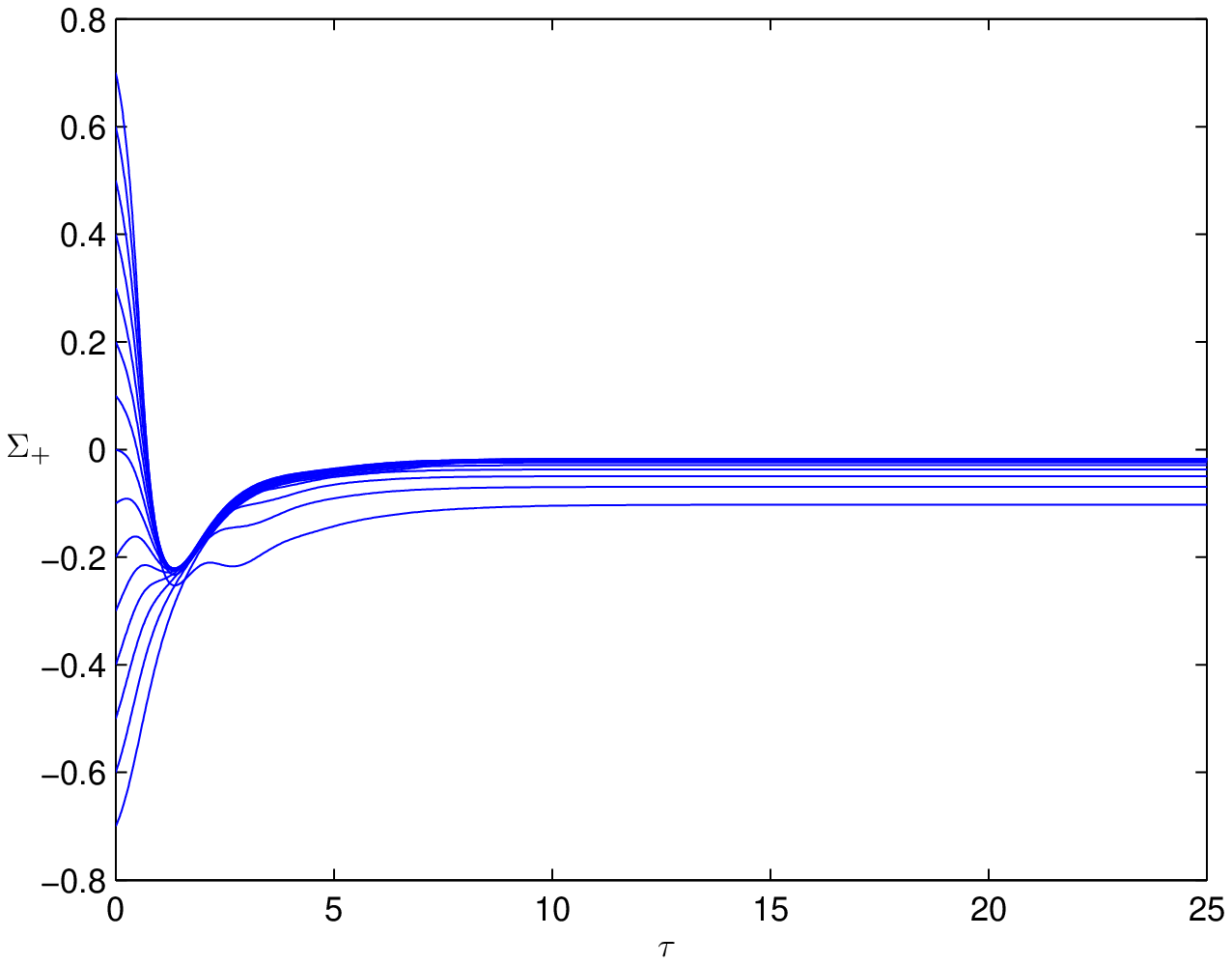}
\end{figure}

\begin{figure} 
\caption{Type VII$_h$:  The figure below displays the dynamical behavior for 
$\gamma=1.325$.  Note how $\Omega\to 0$  but $\Sigma_+\not \to 0$.  Note the 
increased timescale in the graph of $\tau$ vs $V^2$.  Only two trajectories 
were sketched in the first figure to show some of the structure of the future 
asymptotic attractor.  Note how both of the trajectories in the first figure 
spiral into what appears to be a point near $v_1=-1$, but then one of the 
trajectories starts to spiral out again.  This is evidence of the torus 
attractor $\mathcal{T}(VII_h)$.} \label{figure(VIIg=1.325a)}
\includegraphics*[scale = 0.65,angle= 90]{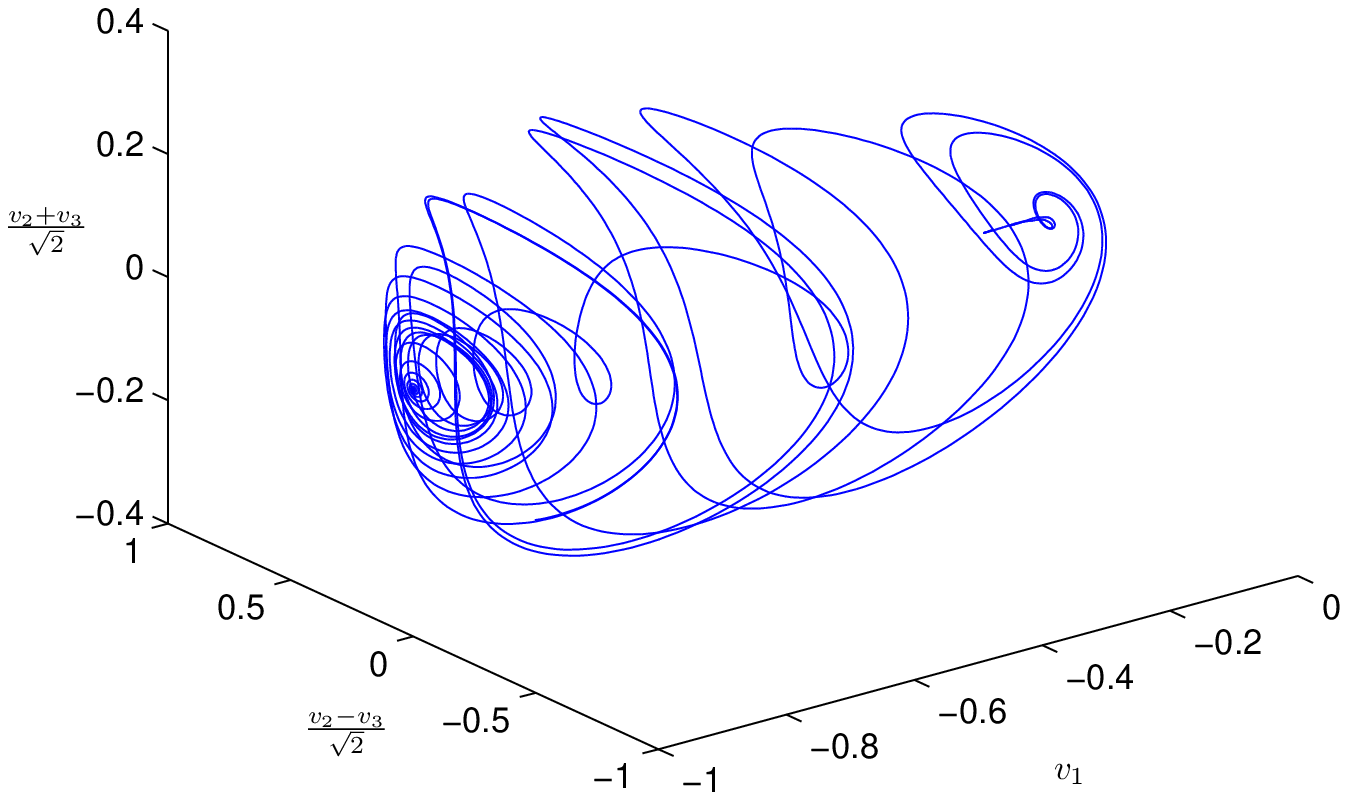}
\includegraphics*[scale = 0.65,angle= 90]{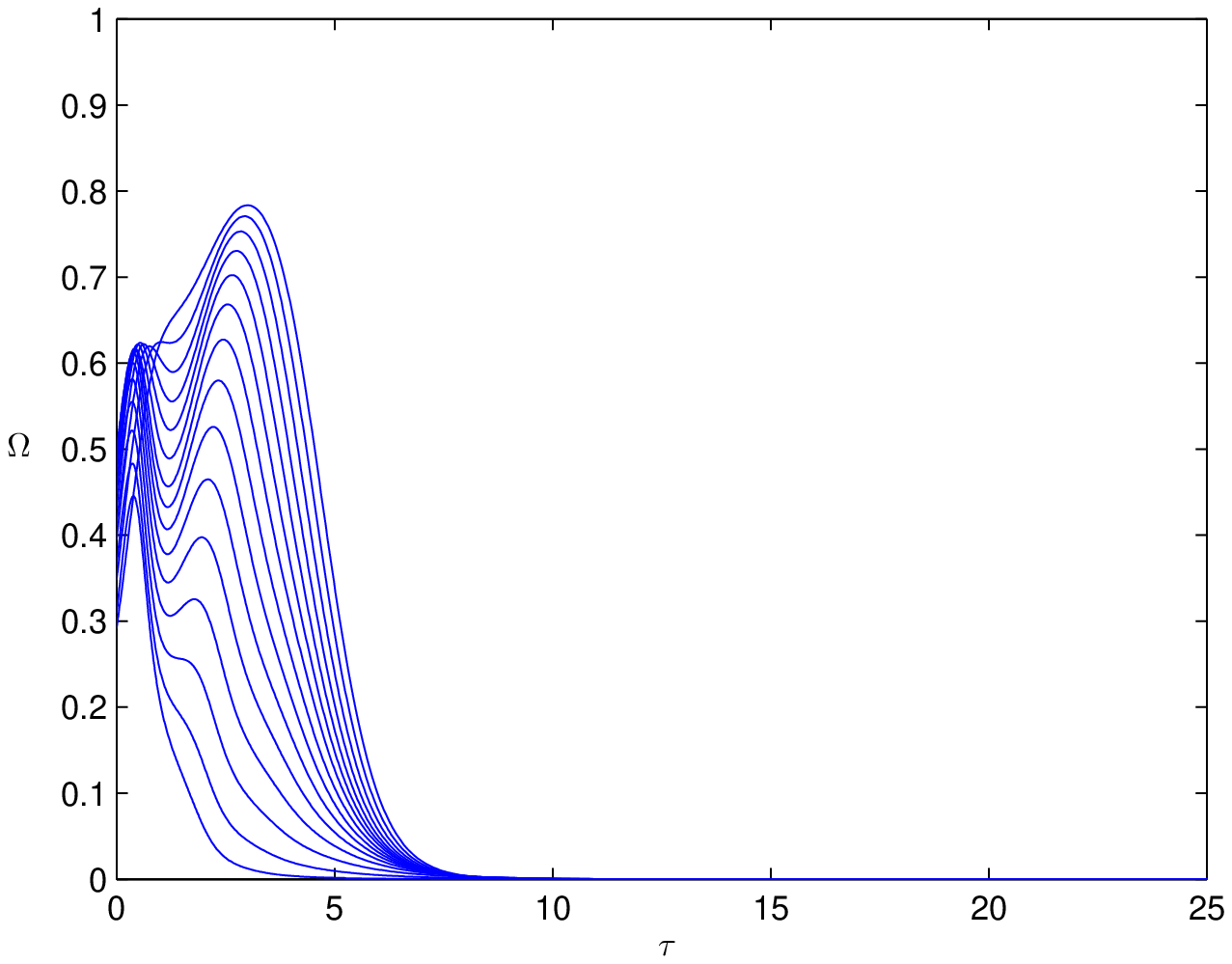}
\linebreak
\linebreak
\includegraphics*[scale = 0.65,angle= 90]{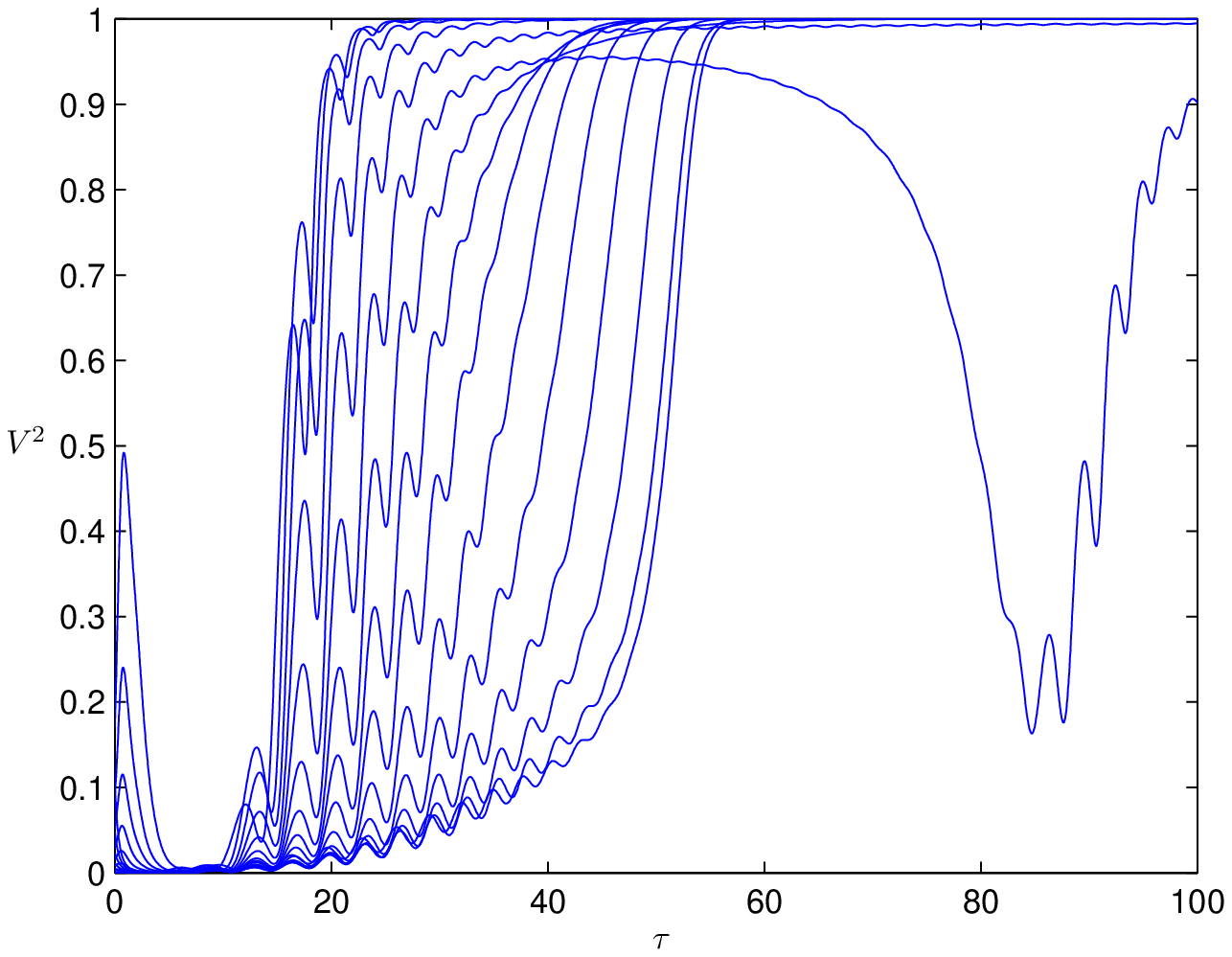}
\includegraphics*[scale = 0.65,angle= 90]{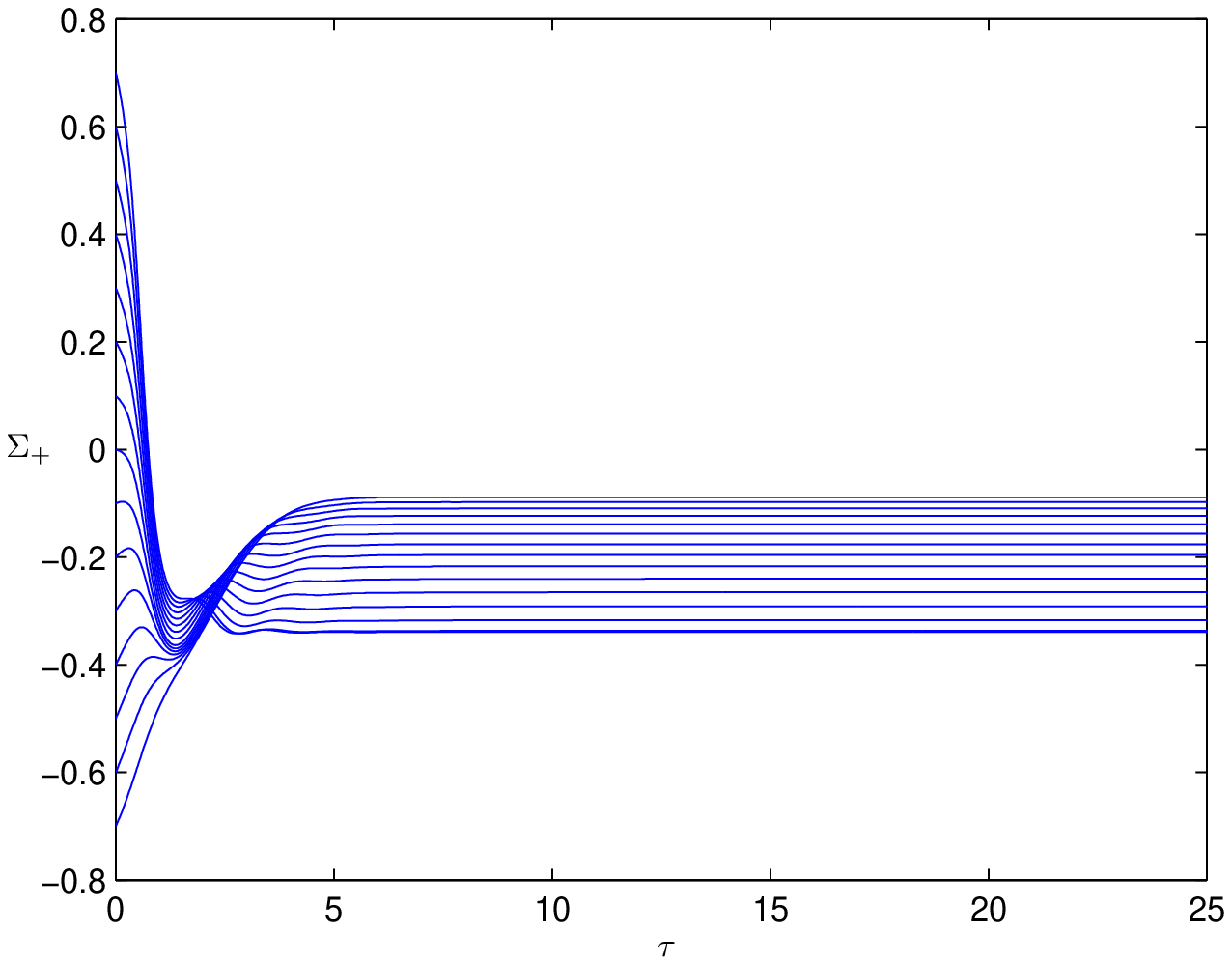}
\end{figure}

\begin{figure} 
\caption{Type VII$_h$:  The figure below displays the dynamical behavior for 
$\gamma=1.325$.  The timescale has been increased in the graph of $\tau$ vs 
$V^2$.    Note how $V^2\to 1$ or $V^2$ oscillates.  Further analysis of this 
"exotic" behavior takes place in the text.  In this case, the local sinks are 
$\widetilde{\mathcal E}(VII_h)$, $\widetilde{\mathcal F}(VII_h)$, ${\mathcal T}(VII_h)$ and 
$\widetilde{\mathcal L}_-(VII_h)$.  This oscillating behaviour indicates the 
existence of the attractor ${\mathcal T}(VII_h)$} \label{figure(VIIg=1.325b)}
\includegraphics*[scale = 0.60,angle= 0]{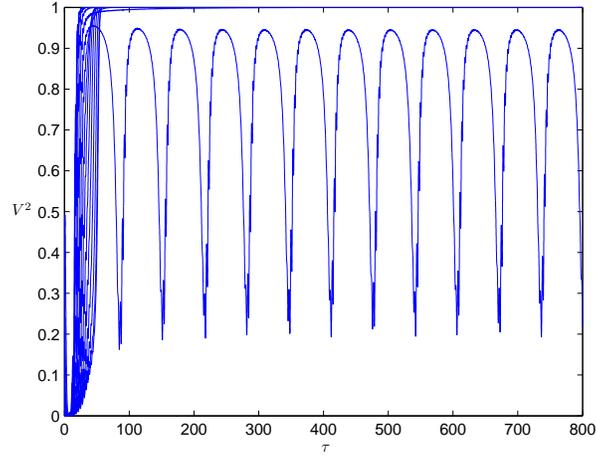}
\end{figure}

\begin{figure} 
\caption{Type VII$_h$:  The figure below displays the dynamical behavior for 
$\gamma=4/3$.  Note how $\Omega\to 0$ and $V^2\to 1$, but $\Sigma_+\not \to 0$.  In this case, the local sinks are $\widetilde{\mathcal E}(VII_h)$ and 
$\widetilde{\mathcal L}_-(VII_h)$.  Note the increased timescale in the graph of 
$\tau$ vs $V^2$.  Only four trajectories were sketched in the first figure to 
show some of the structure of the future asymptotic attractor.} 
\label{figure(VIIg=4/3)}
\includegraphics*[scale = 0.65,angle= 90]{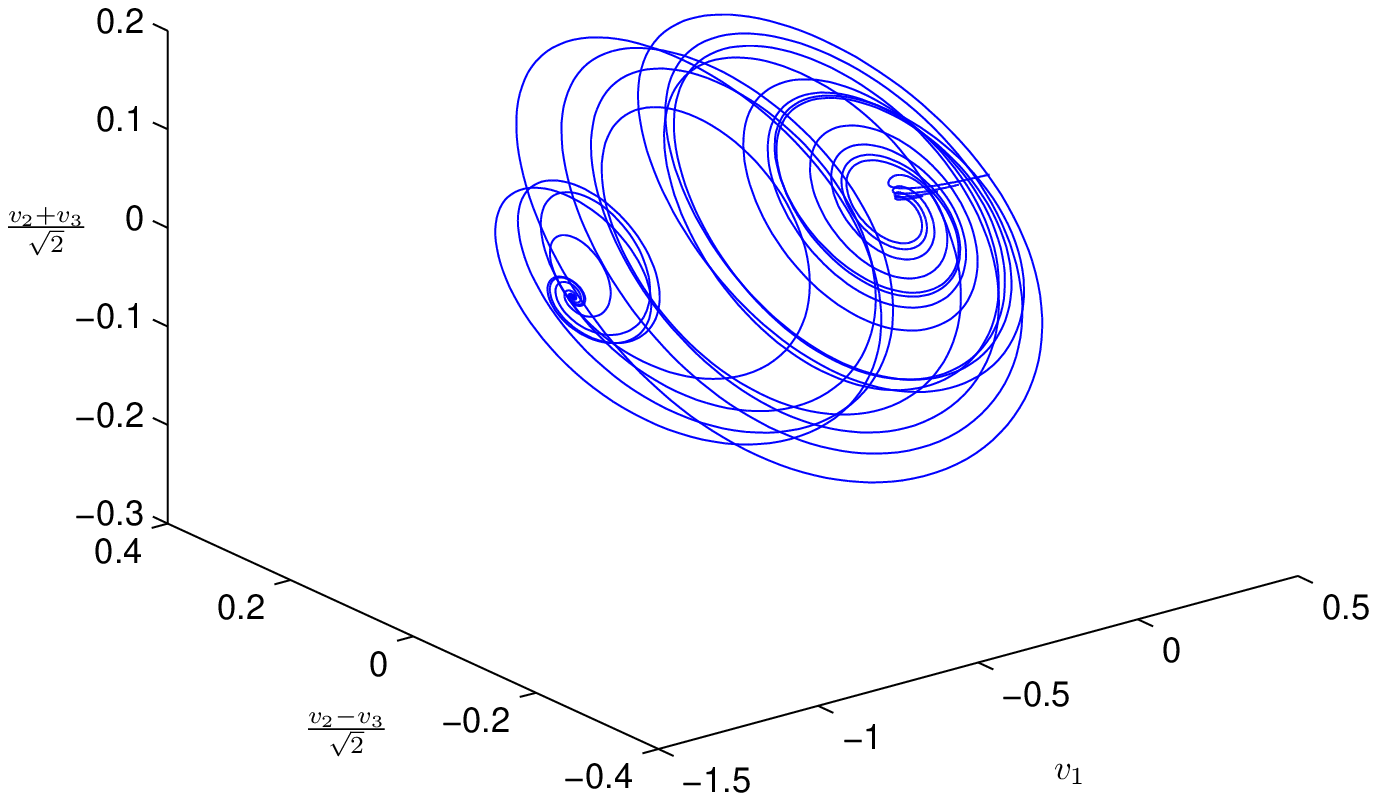}
\includegraphics*[scale = 0.65,angle= 90]{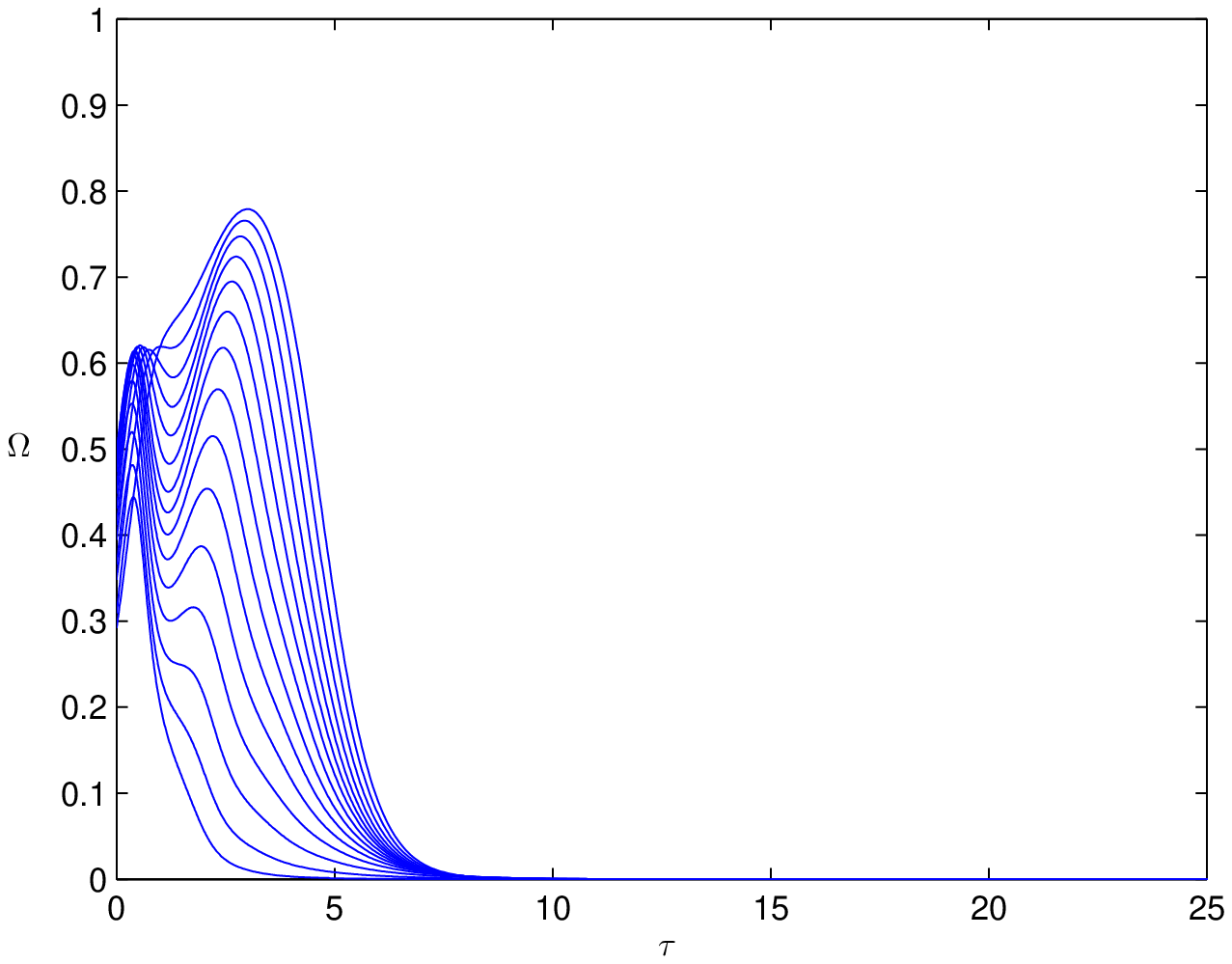}
\linebreak
\linebreak
\includegraphics*[scale = 0.65,angle= 90]{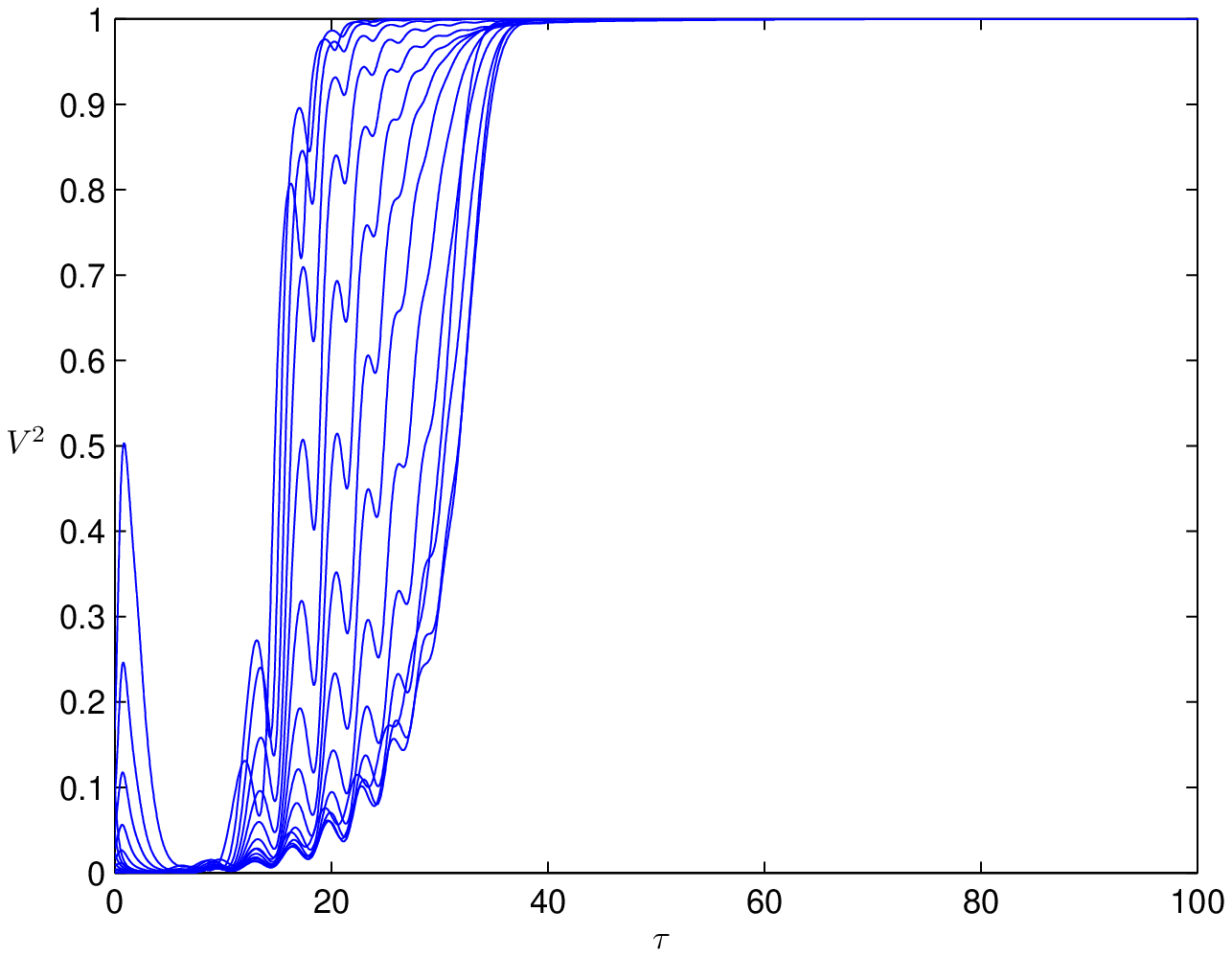}
\includegraphics*[scale = 0.65,angle= 90]{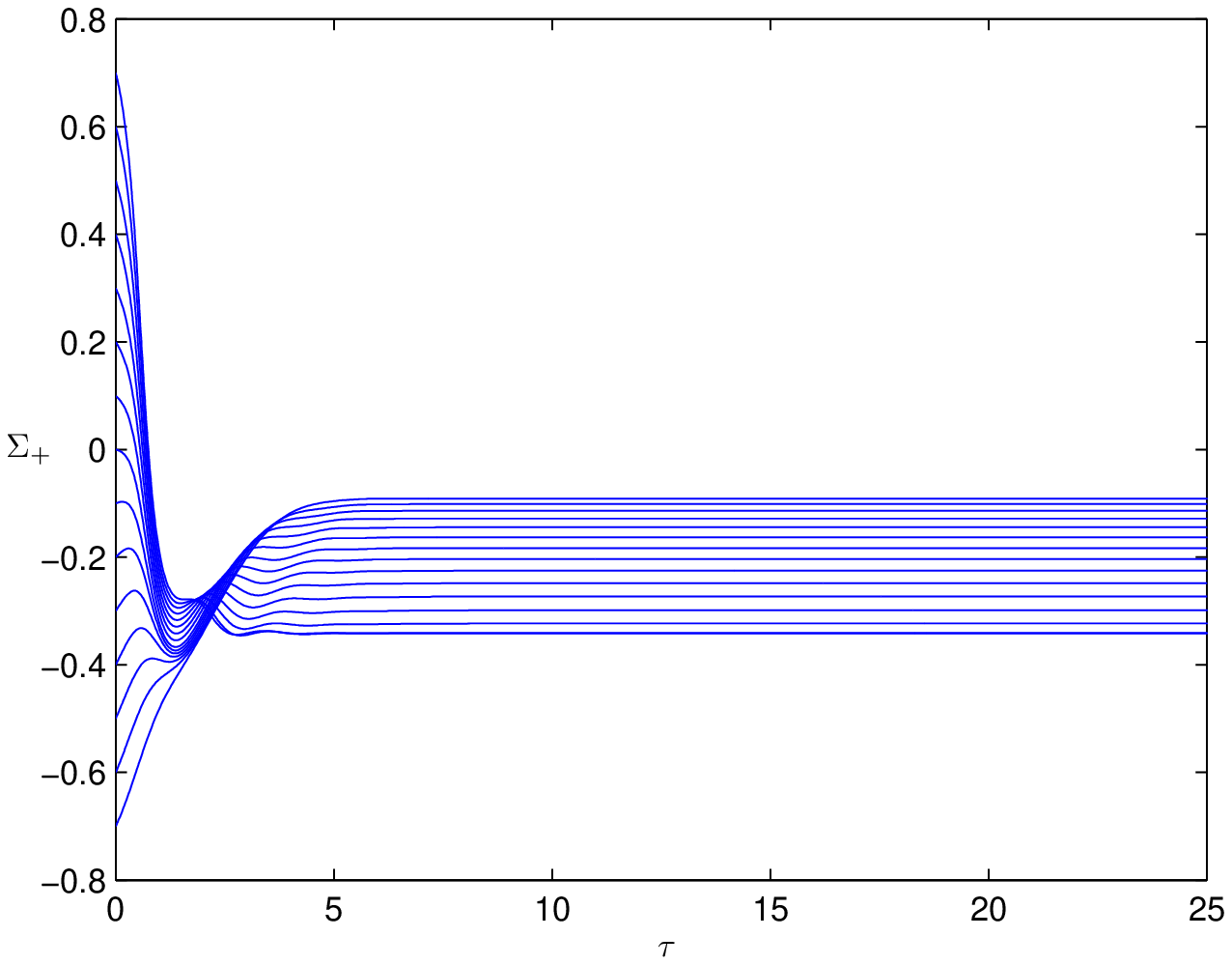}
\end{figure}

For completeness we numerically integrate the dynamical system for the 
cosmologically interesting values of $\gamma$ equal to $1$ ({\em dust}), 
$1.325$ ({\em a dust/radiation mixture}) and $4/3$ ({\em radiation}).  Figures 
\ref{figure(VIIg=1)} to \ref{figure(VIIg=4/3)} depict some of the results of 
this integration. The integrations were done over sufficiently long time 
intervals $[0,800]$ but the plots given here are for time intervals $[0,25]$ or $[0,100]$ if necessary.

\subsection{Bianchi Type  VII$_h$, $\widetilde{\mathcal{F}}(VII_h)$ Throat Attractor}

\begin{figure} 
\caption{ Closed loop $\widetilde{\mathcal{F}}(VII_h)$, $\Sigma_+^*=-0.40$. The picture below is magnified to show additional details of the structure.}\label{figureF}
\includegraphics*[scale = 0.45]{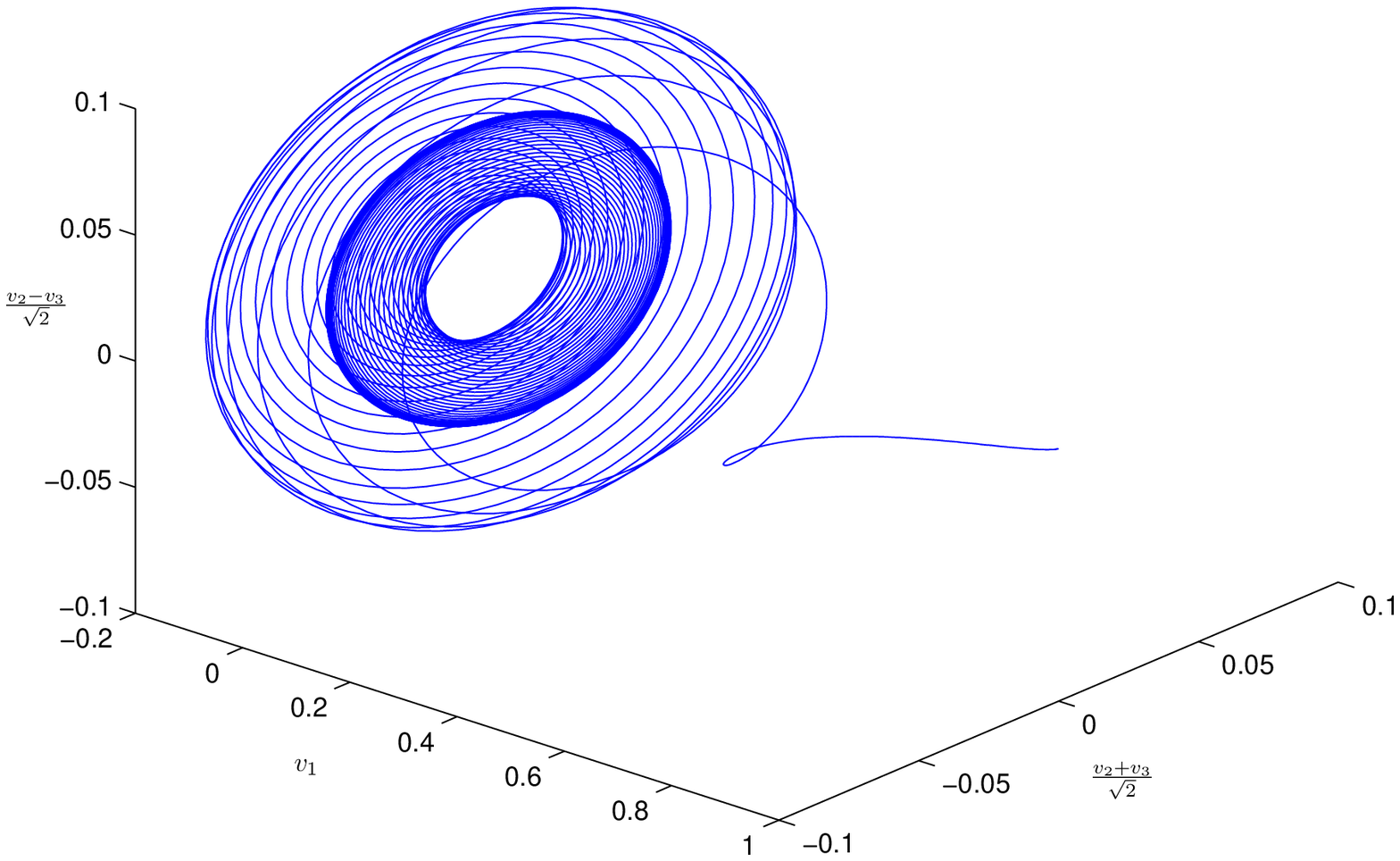}
\includegraphics*[scale = 0.45]{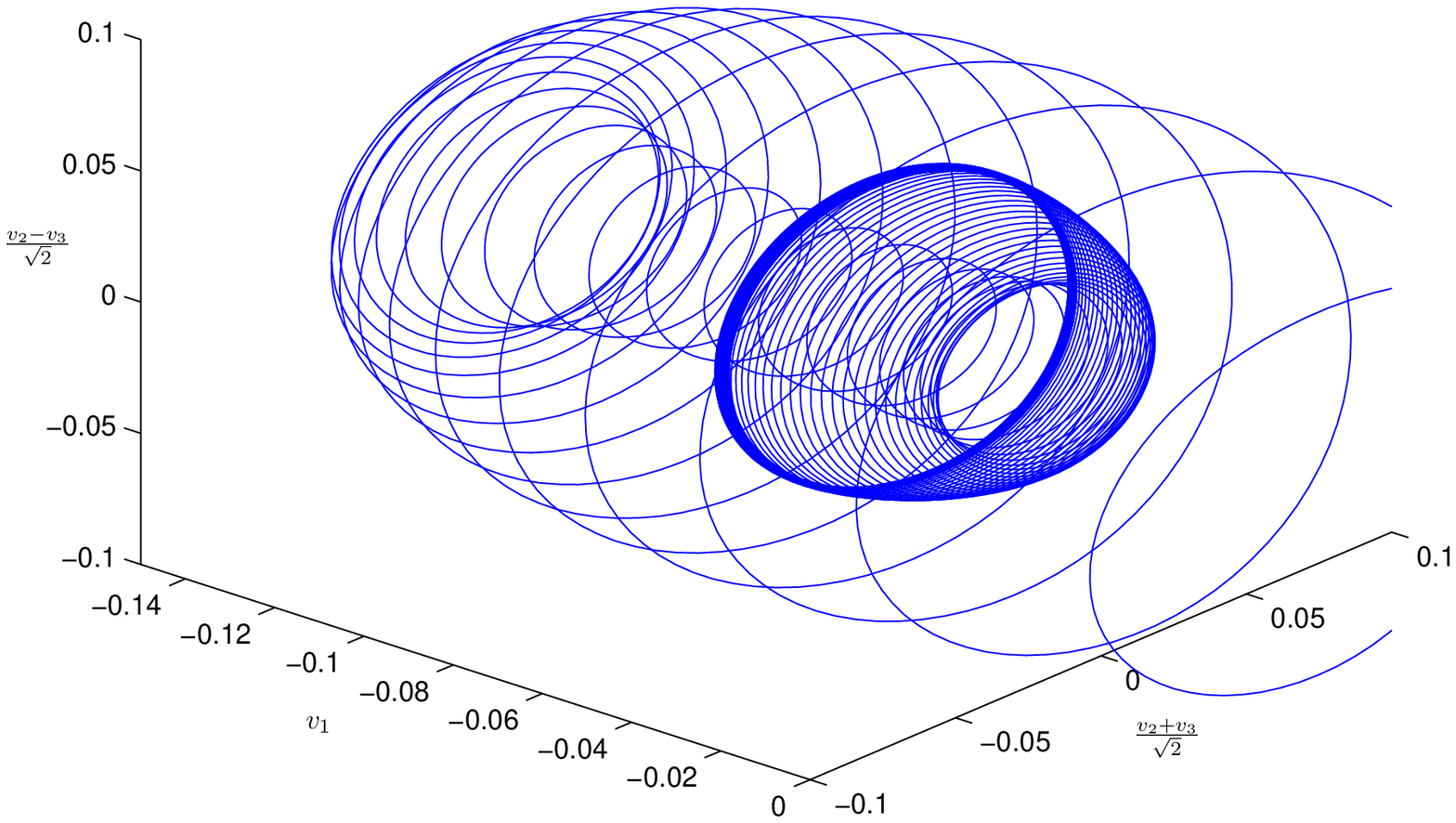}
\end{figure}

Fixing $\gamma=1.325$ and choosing different initial conditions so that the terminal value of $\Sigma_+$ is 'frozen' in between $-0.069773$ and $-0.025$, we are able to observe some rather interesting dynamics.  In this parameter range, we again find that $\widetilde{\mathcal F}(VII_h)$ is a stable attractor. What is perhaps more interesting is the manner in which orbits are attracted to this closed orbit. The attractor $\widetilde{\mathcal F}(VII_h)$ is shown in Figure \ref{figureF}.


\subsection{Bianchi Type VII$_h$, ${\mathcal{T}}(VII_h)$ Torus Attractor}
\begin{figure} 
\caption{The figures below show three orbits with inital conditions ``far away'' from the plane waves solutions (initial condition is $(\Sigma_+)_{0}=0.2$) that end up in the type VII$_h$ loophole. This illustrates the non-zero measure of the attractor basin of the torus attractor. Note how the orbits start to oscillate after the limiting value $\Sigma_+^*$ is reached.}\label{far-torus}\vspace{.5cm}
\includegraphics*[scale = 0.6]{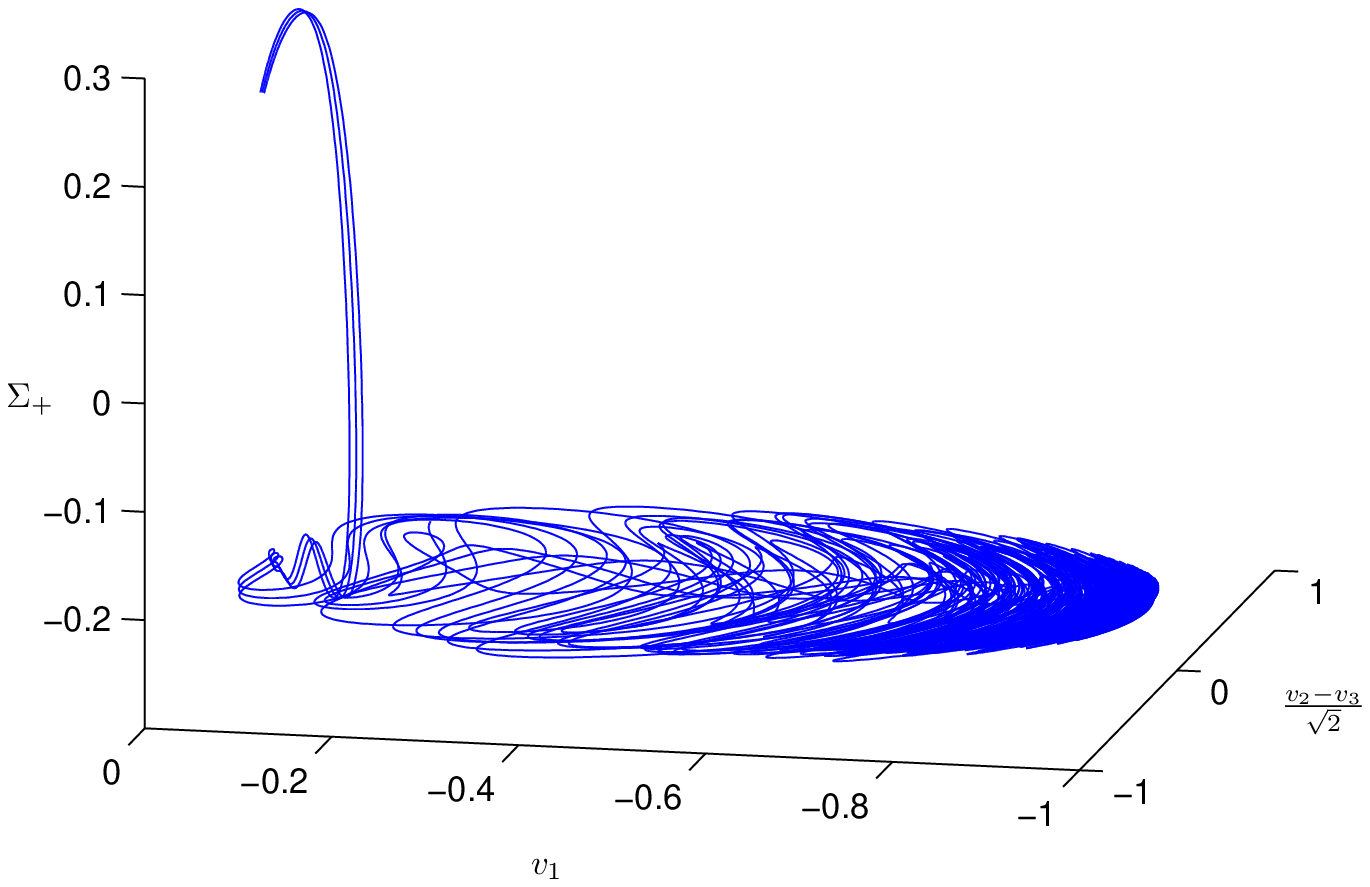}
\includegraphics*[scale = 0.6]{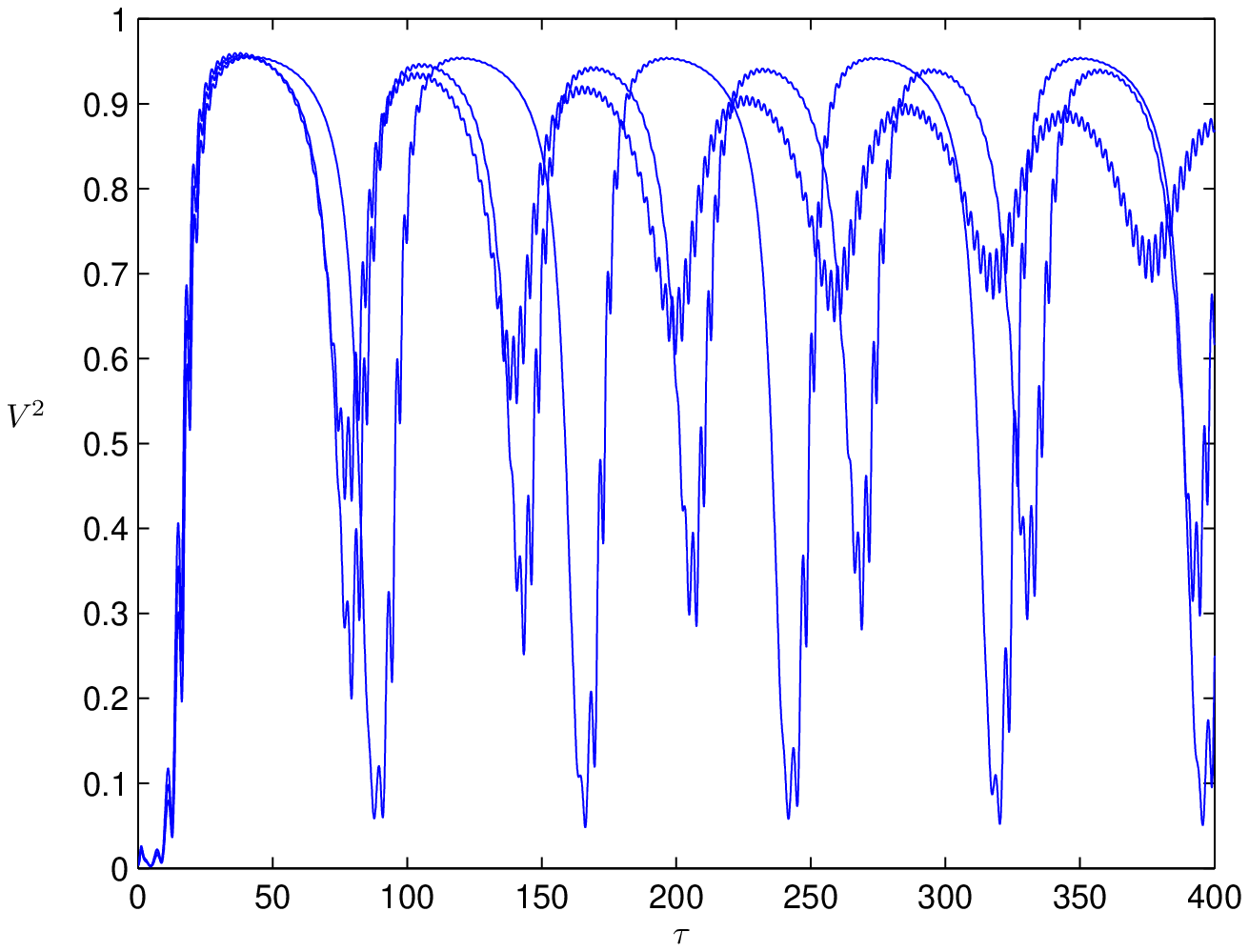}
\end{figure}

Figure \ref{far-torus} shows three orbits with initial conditions ``far away'' from the type VII$_h$ plane wave solutions that approach the torus attractor ${\mathcal{T}}(VII_h)$ at late times. This figure illustrates the fact that there exists an open set of initial conditions for which the orbits are attracted to the Bianchi type VII$_h$ loophole. 

\begin{figure} 
\caption{The figures below show an orbit on its approach to the $\mathcal{T}(VII_h)$ torus attractor, with $h=1$. In the first figure it has almost wound around once. The second and third, one can count almost two and five windings. On the last picture the torus is coming into view.}\label{torus}\vspace{.5cm}
\includegraphics*[scale = 0.6,angle =90]{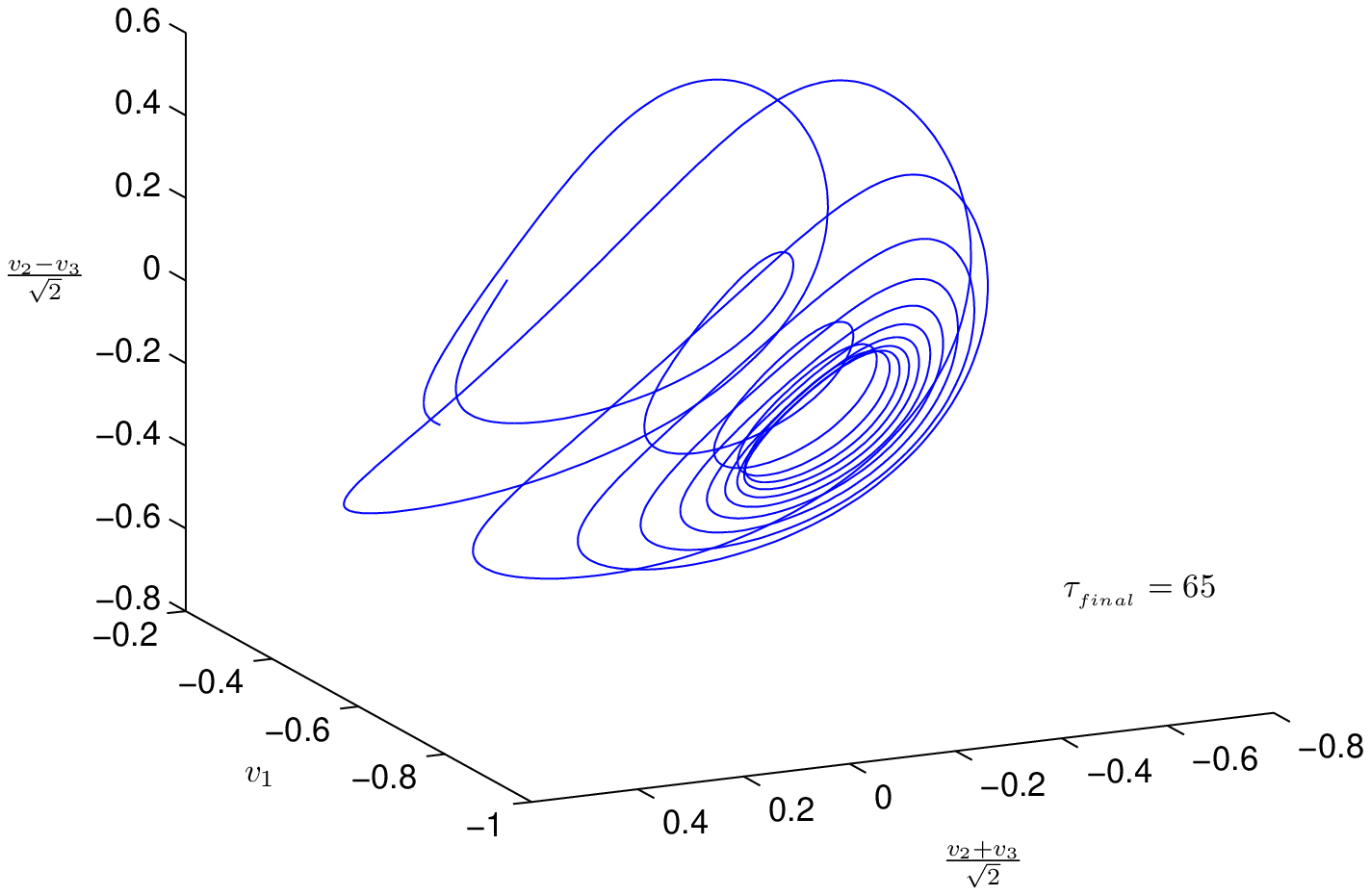}
\includegraphics*[scale = 0.6,angle=90]{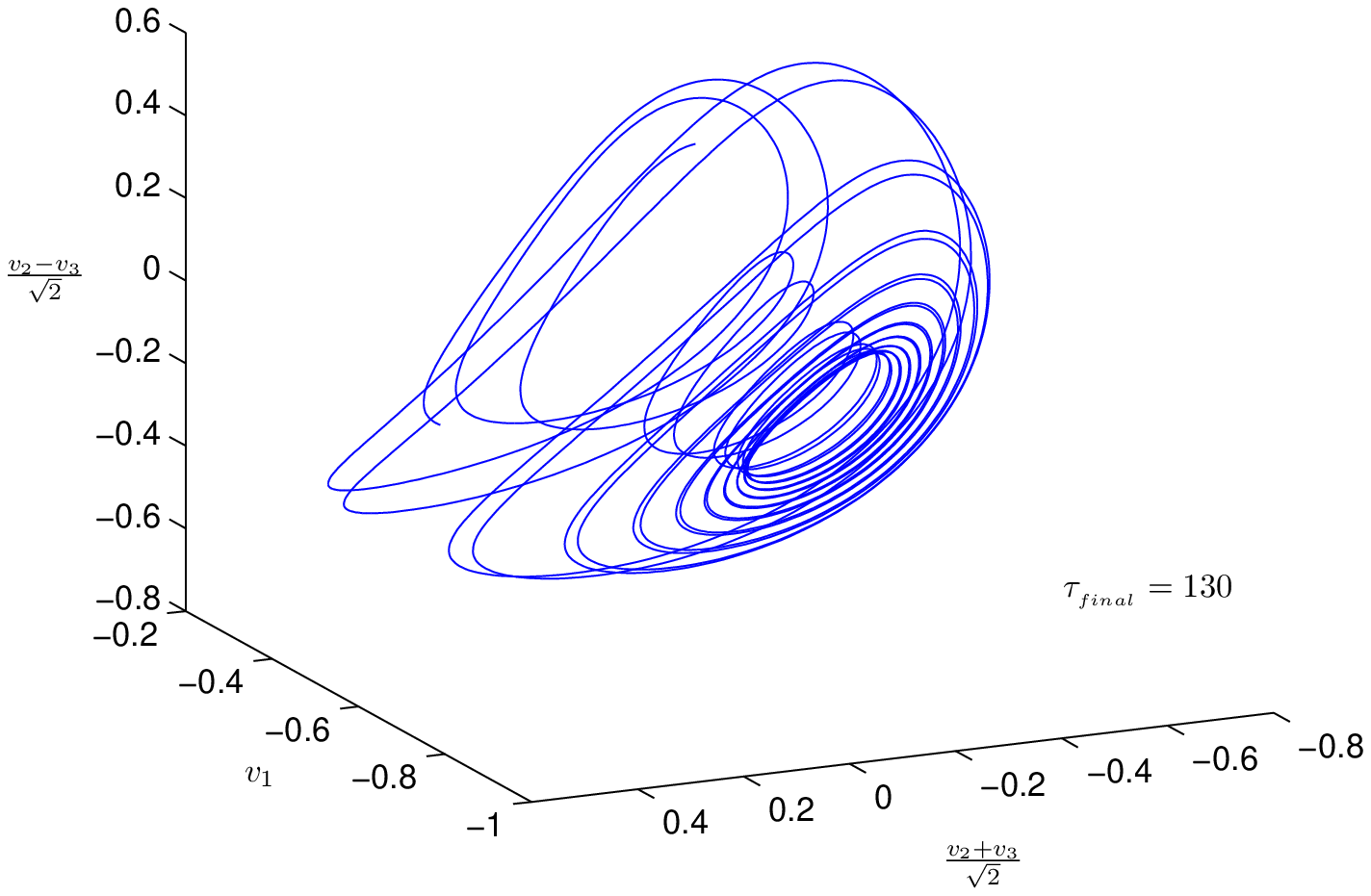}\\
\includegraphics*[scale = 0.6,angle=90]{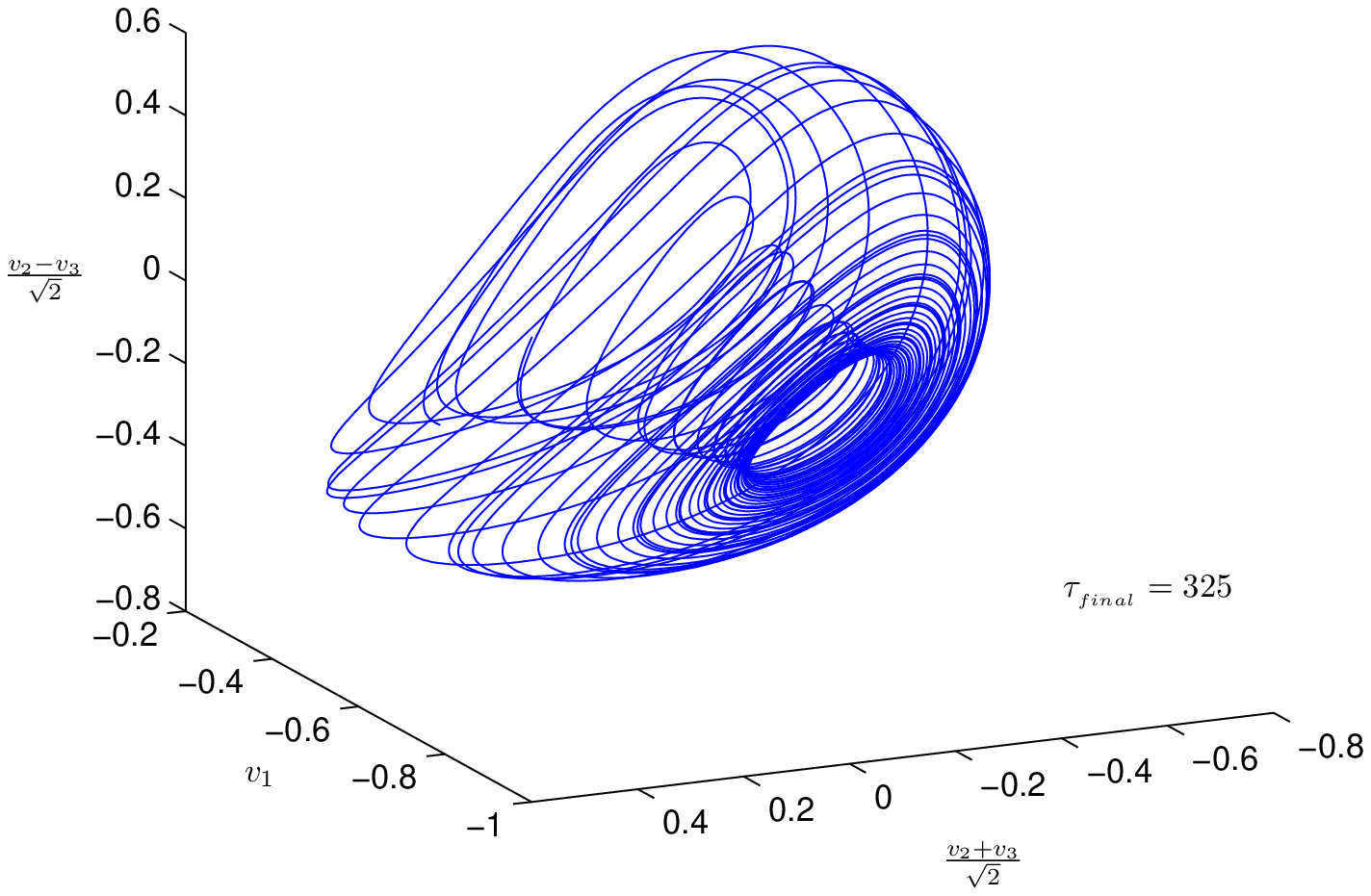}
\includegraphics*[scale = 0.6,angle=90]{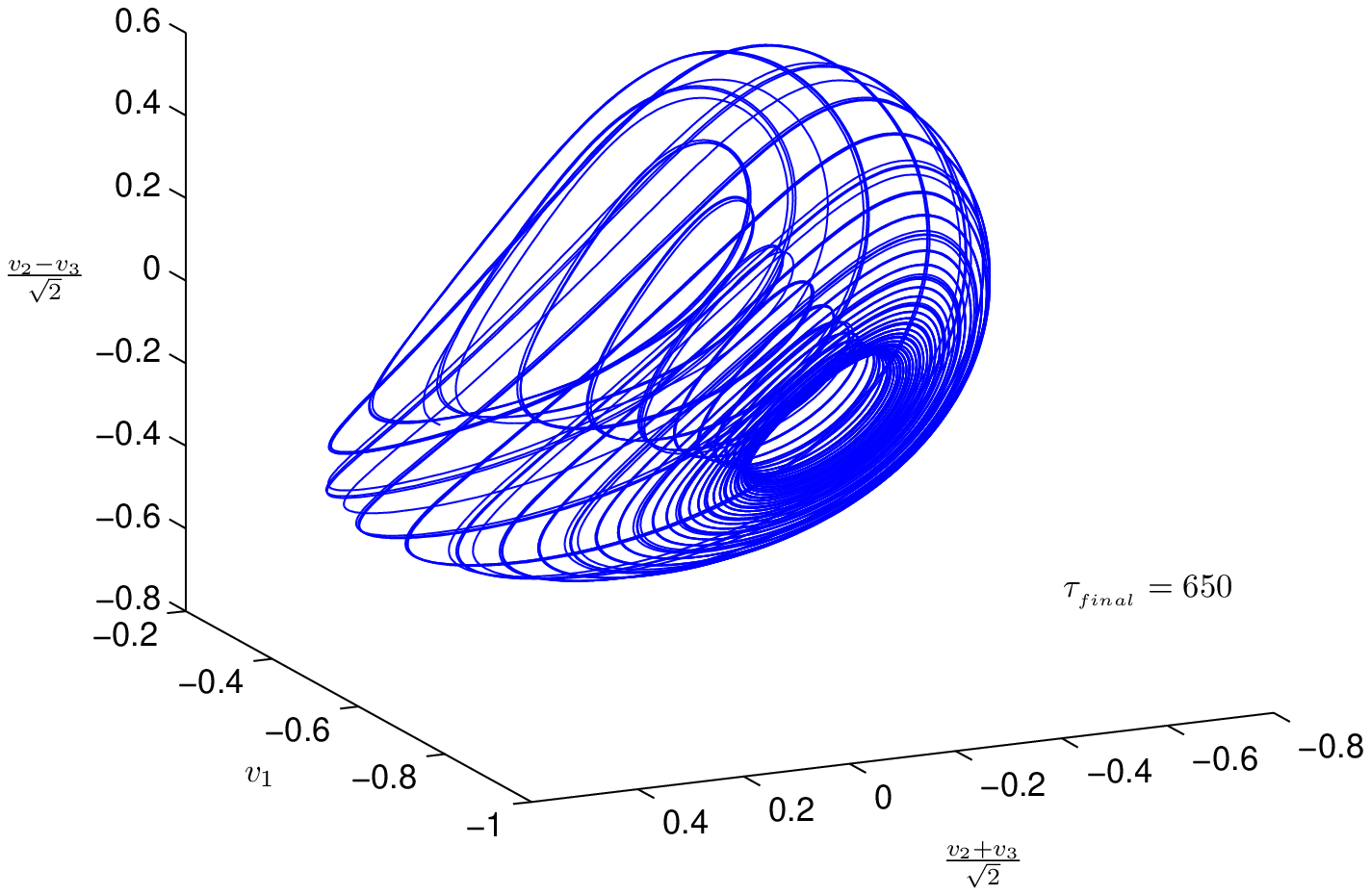}
\end{figure}

Figure \ref{torus} depicts the shape and formation of the attractor $\mathcal{T}(VII_h)$.  We choose different time intervals to illustrate the structure of this attractor. We can see how the curve starts to fill up the torus. Irrational curves will eventually completely fill the torus.

\bibliographystyle{amsplain}

\begin{thebibliography}{99}


\bibitem{DS1}  J. Wainwright and G.F.R. Ellis, \textit{Dynamical Systems
in Cosmology}, Cambridge University Press (1997)


\bibitem{KingEllis} A.R. King and G.F.R. Ellis, \textit{Commun. Math. Phys.}
\textbf{31} (19


\bibitem{rosjan} K.Rosquist  and  R. T. Jantzen,  \textit{Phys. Rep.} {\bf 166} (1988) 89-124.

\bibitem{BN} O. I.  Bogoyavlensky,  (1985)
\textit{Methods in the Qualitative Theory of Dynamical Systems in
Astrophysics and Gas Dynamics} Springer-Verlag.

\bibitem{BS} J.D. Barrow and D.H. Sonoda, \textit{Phys. Reports} \textbf{139}
(1986) 1


\bibitem{Shikin} I.S. Shikin, \textit{Sov. Phys. JETP} \textbf{41} (1976) 794

\bibitem{Collins} C.B. Collins, \textit{Comm. Math. Phys.} \textbf{39}
(1974) 131

\bibitem{CollinsEllis} C. B. Collins and  G. F. R.Ellis, \textit{Phys.Rep.}
{\bf 56} (1979) 65-105.


\bibitem{HWV} C.G. Hewitt and J. Wainwright, \textit{Phys. Rev.} \textbf{D46}
(1992) 4242

\bibitem{HBWII} C.G. Hewitt, R. Bridson, J. Wainwright, \textit{Gen.Rel.Grav.%
} \textbf{33} (2001) 65

\bibitem{BHtilted} J.D. Barrow and S. Hervik, \textit{Class. Quantum
  Grav.} \textbf{20} (2003) 2841


\bibitem{hervik} S. Hervik, \textit{Class. Quantum Grav.} \textbf{21} (2004) 2301

\bibitem{coleyhervik} A. Coley and S. Hervik, \textit{Class. Quantum Grav.} \textbf{21} (2004) 4193-4208


\bibitem{CH2} A.A. Coley and S. Hervik, \textit{Tilted Bianchi models of solvable type}, \texttt{gr-qc/0409100}

\bibitem{carrcoley} B.J. Carr and A.A. Coley, \textit{Class. Quantum Grav.} {\bf 16}  (1999) R31.


\bibitem{Apostolopoulos:04} P. S. Apostolopoulos, \texttt{gr-qc/0407040}

\bibitem{Barrow86} J. D. Barrow, \textit{Can. J. Phys.} {\bf 64} (1986) 152 

\bibitem{wceh} J. Wainwright, A.A. Coley, G.F.R. Ellis and M. Hancock, 
\textit{Class. Quantum Grav.} {\bf 15} (1998) 331.

\bibitem{Novikov} I. D. Novikov, \textit{Sov. Astron.} {\bf 12} (1968) 427 


\bibitem{CollinsHawking} C.B Collins and S.W. Hawking, \textit{Mon. Not. R. Astron Soc.}  {\bf 162}  (1973) 307.



\bibitem{Doreshkevich1973} A. G. Doroshkevich, V. N. Lukash and I. D. Novikov,
\textit{Sov. Phys.-JETP} {\bf 37} (1973) 739.


\bibitem{Barrow1995} J. D. Barrow, \textit{Phys. Rev. D} {\bf 51} (1995) 3113.

\bibitem{JohnChristos}
J.D. Barrow and C. Tsargas, \textit{On the Lukash plane wave solutions}  

\bibitem{DSReza} R. Tavakol in \textit{Dynamical Systems
in Cosmology}, eds: J. Wainwright and G.F.R. Ellis, Cambridge
University Press (1997)

\bibitem{2Ddynsys}
S.Kh. Aranson, G.R. Belitsky, E.V. Zhuzhoma, \textit{Introduction to the Qualitative Theory of Dynamical Sustems on Surfaces}, Trans. Math. Mono. \textbf{153} (1996)

\bibitem{Wald} R.M. Wald, \textit{Phys. Rev. } \textbf{D28} (1983) 2118
\end{thebibliography}

\end{document}